\documentclass[a4paper,12pt]{article}
\usepackage[latin1]{inputenc}
\usepackage[english]{babel}  
\usepackage[authoryear]{natbib}
\usepackage[T1]{fontenc}

\usepackage{graphicx}
\usepackage[titletoc]{appendix}
\usepackage{anysize}
\usepackage{subfig}
\usepackage{float}
\usepackage{hyperref} 
\usepackage{lscape}
\usepackage{afterpage}
\usepackage{pdflscape}
\usepackage{authblk}
\usepackage{tablefootnote}
\usepackage[pass]{geometry}
\usepackage{bm}
\usepackage{xcolor}
\usepackage{ae}
\usepackage{url}

\usepackage{amsmath,amsthm,amssymb,amsfonts,textcomp}
\usepackage{mathtools}
\usepackage{enumerate}
\usepackage{vmargin}
\usepackage{array,multirow}
\usepackage{booktabs}
\usepackage{setspace} 
\usepackage{soul}
\usepackage{indentfirst}
\usepackage{parskip}
\newtheoremstyle{exampstyle}
  {15pt} 
  {15pt} 
  {} 
  {} 
  {\bfseries} 
  {.} 
  {.15em} 
  {} 

\theoremstyle{exampstyle}
\theoremstyle{plain} 

\theoremstyle{exampstyle}

\theoremstyle{exampstyle}
\theoremstyle{exampstyle}
\theoremstyle{exampstyle}


\newcommand{\bo}{\boldsymbol} 

\newcommand{\dd}{{\rm d}}
\newcommand{\diag}{{\text{diag}}}

\title{Robust beta regression through the logit transformation}
\author{Yuri S. Maluf, Silvia L. P. Ferrari\footnote{Corresponding author: e-mail silviaferrari@usp.br.} , Francisco F. Queiroz\\
{\small {\em \textit{Department of Statistics, University of S\~ao Paulo, Brazil}}}}
\date{}                     
\begin{document}
\maketitle
\begin{abstract}
Beta regression models are employed to model continuous response variables in the unit interval, like rates, percentages, or proportions. Their applications rise in several areas, such as medicine, environment research, finance, and natural sciences. The maximum likelihood estimation is widely used to make inferences for the parameters. Nonetheless, it is well-known that the maximum likelihood-based inference suffers from the lack of robustness in the presence of outliers. Such a case can bring severe bias and misleading conclusions. Recently, robust estimators for beta regression models were presented in the literature. However, these estimators require non-trivial restrictions in the parameter space, which limit their application. This paper develops new robust estimators that overcome this drawback. Their asymptotic and robustness properties are studied, and robust Wald-type tests are introduced. Simulation results evidence the merits of the new robust estimators. Inference and diagnostics using the new estimators are illustrated in an application to health insurance coverage data.
\end{abstract} 

\section{Introduction}

Beta regression is a flexible and popular tool for modeling proportions, rates, and other continuous response variables restricted to the open unit interval. Beta regression models are employed to model the relationship between predictors and a continuous response variable that is assumed to follow a beta distribution. The beta regression model with constant precision was introduced by \cite{FerrariCribari} and was extended in various directions. For instance, in \cite{SmithsonVerkuilen} and \cite{SimasSouzaRocha}, both the mean and precision parameters are modeled using predictors. There are numerous applications involving beta regression in different areas such as medicine \citep{GV,Swearingen}, environment research \citep{CHMA}, finance \citep{CKM}, and natural sciences \citep{Geissinger}.

The probability density function of the beta distribution in the mean-precision parameterization is 
\begin{equation}\label{betadensity}
 f(y;\mu,\phi)=\frac{1}{B(\mu\phi,(1-\mu)\phi)} y^{\mu\phi-1}(1-y)^{(1-\mu)\phi-1},\quad 0<y<1,
\end{equation}
where $0<\mu<1$, $\phi>0$, and $B(\cdot,\cdot)$ is the beta function, and we write $y\sim \text{Beta}(\mu,\phi)$. We have $\mathbb{E}\left(y\right)=\mu$  and $\mathbb{V}{\rm ar}\left(y\right)=\mu(1-\mu)/(1+\phi)$, hence $\mu$ is the mean parameter and $\phi$ can be interpreted as a precision parameter. If $\mu \phi>1$ and $(1-\mu) \phi>1$, the beta density \eqref{betadensity} is bounded, has a single mode in $(0,1)$, and decreases to zero as $y \downarrow 0$ or $y \uparrow 1$. When $\mu \phi<1$ or $(1-\mu) \phi<1$, the beta density is unbounded at one or both boundaries. 

The beta regression model considered here is defined as follows. Let $y_1,\ldots,y_n$ be independent random variables such that $y_{i}\sim \text{Beta}(\mu_i,\phi_i)$, for $i=1,\ldots,n$, with
\begin{equation}\label{linearPredictor}
\begin{split}
 g_{\mu}(\mu_{i})&=\bo{X}_{i}^{\top}\bo{\beta}, \,\,\,\,\text{(mean submodel)}\\
 g_{\phi}(\phi_i)&=\bo{Z}_{i}^{\top}\bo{\gamma}, \,\,\,\,\text{(precision submodel)}\\
\end{split}
\end{equation}
where $\bo{\beta}=(\beta_1,\ldots,\beta_{p_{1}})^{\top}$ and $\bo{\gamma}=(\gamma_1,\ldots,\gamma_{p_{2}})^{\top}$ are vectors of unknown regression coefficients ($p=p_1+p_2<n$); $\bo{X}_{i}=(x_{i1},\ldots,x_{ip_{1}})^{\top}$ and $\bo{Z}_{i}=(z_{i1},\ldots,z_{ip_{2}})^{\top}$ are vectors of the covariates, and $\bo{\theta} = (\bo{\beta}^\top, \bo{\gamma}^\top)^\top \in \mathbb{R}^{p}$ is the unknown parameter vector. The link functions $g_\mu:(0,1)\rightarrow \mathbb{R}$ and $g_\phi:(0,\infty)\rightarrow \mathbb{R}$ are strictly increasing and, at least, twice differentiable.

The maximum likelihood approach is usually employed for estimating $\bo{\theta}$. However, the maximum likelihood estimator (MLE) is highly sensitive to outliers. Recently, \cite{GhoshBeta} and \cite{RibeiroFerrari} proposed robust estimators for the beta regression model \eqref{betadensity}-\eqref{linearPredictor}. These estimators require suitable restrictions in the parameter space. If all the beta densities in model (\ref{betadensity})-(\ref{linearPredictor}) are bounded, the robust estimators and their respective asymptotic covariance matrices are well-defined. Note that the boundedness of all the beta densities assumption requires implicit, non-trivial restrictions in the parameter space of the regression parameters $\bo{\beta}$ and $\bo{\gamma}$. Moreover, these restrictions depend on the covariate vectors $\bo{X}_{i}$ and $\bo{Z}_{i}$ for all $i=1,\ldots,n$. If the restrictions are not satisfied, relevant numerical problems may arise when employing the robust estimators in empirical applications.    

This paper introduces two new robust estimators for the beta regression model \eqref{betadensity}-\eqref{linearPredictor}. They are derived using methods similar to those employed by \cite{GhoshBeta} and \cite{RibeiroFerrari}, with the advantage of not requiring restrictions in the parameter space. 

The remaining of this paper is organized as follows. Section \ref{Sec.BetaReg} briefly describe the estimators developed by \cite{GhoshBeta} and \cite{RibeiroFerrari}. Section \ref{Sec.robL} presents two new robust estimators that overcome the limitations of the current robust estimators. Robustness and asymptotic properties of the new estimators are also presented in Section \ref{Sec.robL}. Section \ref{Sec.SimStudy} shows simulation results that evidence the merits of the new estimators over the MLE and the current robust estimators. An application of the proposed methods is discussed in Section \ref{Sec.App}. The paper closes with some remarks and directions for future works.

\section{Current robust estimators}\label{Sec.BetaReg}

\cite{GhoshBeta} proposed the minimum density power divergence estimator (MDPDE) for the beta regression model, a robust estimator based on the density power divergence that involves a tuning constant, $\alpha \geq 0$ \citep{Basu98, GhoshBasu}. It solves the estimating equation
\begin{align*}
\begin{split}
\displaystyle \sum_{i=1}^n \left[\bm{U}(y_i; \bm{\theta})f_{\bm{\theta}}(y_i; \mu_i, \phi_i)^\alpha  - \mathcal{E}_{i,1-\alpha}(\bm{\theta})\right]& = \bm{0},
\end{split}
\end{align*}
where $f_{\bm{\theta}}(y_i; \mu_i, \phi_i)$ denotes the beta density \eqref{betadensity} with $\mu_i$ and $\phi_i$ given by \eqref{linearPredictor}, $\bm{U}(y_i; \bm{\theta})=\nabla_{\bm{\theta}}\log(f_{\bm{\theta}}(y_i; \mu_i, \phi_i))$, and $\mathcal{E}_{i,1-\alpha}(\bm{\theta}) = \mathbb{E}\left[\bo{U}(y_i; \bm{\theta})f_{\bm{\theta}}(y_i; \mu_i, \phi_i)^\alpha \right]$. The factor $f_{\bm{\theta}}(y_i; \mu_i, \phi_i)^\alpha$ acts as the weight of the $i$-th observation in the estimation procedure. If $\alpha=0$, we have the maximum likelihood estimator. Choices of $\alpha \in (0,1)$ leads to a robust procedure because observations that are inconsistent with the postulated model receive smaller weights. If $\alpha\geq 1$, the estimator is highly robust but severely inefficient. Hence, from now on, we will restrict $\alpha \in [0,1)$. The role of $\mathcal{E}_{i,1-\alpha}(\bm{\theta})$ is to center the weighted score, ensuring Fisher-consistency. However, $\mathcal{E}_{i,1-\alpha}(\bm{\theta})$ is not well-defined unless $\mu_i \phi_i>\alpha/(1+\alpha)$ and $(1-\mu_i) \phi_i>\alpha/(1+\alpha)$. Moreover, the asymptotic covariance matrix of the MDPDE is not well-defined unless $\mu_i \phi_i>2\alpha/(1+2\alpha)$ and $(1-\mu_i) \phi_i>2\alpha/(1+2\alpha)$ (see \citet{RibeiroFerrari} for details). 


\cite{RibeiroFerrari} proposed an estimator based on the maximization of a reparameterized L$_q$-likelihood. The L$_q$-likelihood \citep{FerrariYang} is 
\begin{align}\label{Lqlikelihood}
\ell_q(\bm{\theta}) = \displaystyle \sum_{i=1}^n L_q\left( f_{\bm{\theta}}(y_i; \mu_i, \phi_i) \right),
\end{align}
where $q=1-\alpha \in (0,1]$ is the tuning constant and $L_q(u) = (u^{1-q} - 1)/(1-q)$, for $q \in (0,1)$, and $L_q(u) = \log(u)$, for $q=1$. The estimator that comes from the maximization of \eqref{Lqlikelihood} solves the estimating equation 
\[
\displaystyle \sum_{i=1}^n \bm{U}(y_i; \bm{\theta})f_{\bm{\theta}}(y_i; \mu_i, \phi_i)^\alpha = \bm{0}.
\]
Note that the estimating function is not unbiased unless $\alpha=0$, hence the resulting estimator is not Fisher-consistent. In \cite{GhoshBeta}, the Fisher-consistency is achieved by centering the weighted score. \cite{RibeiroFerrari} obtained a Fisher-consistent estimator through a reparametrization of the L$_q$-likelihood, named surrogate maximum likelihood estimator (SMLE). The estimating equation is given in \citet[eq. (10)]{RibeiroFerrari}; it is not well-defined unless $\mu_i \phi_i>\alpha$ and $(1-\mu_i) \phi_i>\alpha$. Also, the validity of its asymptotic covariance matrix requires that $\mu_i \phi_i>2\alpha/(1+\alpha)$ and $(1-\mu_i) \phi_i>2\alpha/(1+\alpha)$.


A sufficient condition for the MDPDE and the SMLE and their respective asymptotic covariance matrices to be well-defined is that all the beta densities in model (\ref{betadensity})-(\ref{linearPredictor}) are bounded. Under such assumption, the MDPDE and the SMLE have good properties such as B-robustness, V-robustness, and asymptotic normality. A crucial issue for the use of the proposed estimators is the choice of the tuning constant. Higher values of $\alpha$ increase robustness and decrease efficiency. \cite{RibeiroFerrari} developed an effective data-driven algorithm for selecting the optimal $\alpha$. Simulation results and real data applications in \cite{RibeiroFerrari} evidence the superior performance of these estimators relative to the MLE for datasets containing outlier observations. 

The findings in \cite{GhoshBeta} and in \cite{RibeiroFerrari} are guaranteed for bounded beta densities but not necessarily otherwise. As we will show later, simulations for unbounded beta densities reveal serious numerical problems of the MDPDE and the SMLE. In the next section we propose alternative robust estimators which have the advantage of being well-defined for all beta densities.

\section{Robust estimators through the logit transformation}\label{Sec.robL}

The limitation of the MDPDE and the SMLE discussed in the previous section comes from the fact that the beta densities are not closed under power transformations. Given a density $v$ and a constant $\xi\in(0,\infty)$, the power transformations is
\begin{equation*}
 v^{\left(\xi\right)}(y)=\frac{v(y)^{\xi}}{\int v(y)^{\xi} \dd y} \varpropto v(y)^\xi, \quad \forall y \text{~in the support,}
\end{equation*}
provided that $\int v(y)^{\xi} \dd y<\infty$. For the beta density \eqref{betadensity},
\[
f(y; \mu, \phi)^\xi \varpropto y^{\xi (\mu \phi -1) } (1-y)^{\xi [(1-\mu) \phi -1] },
\]
which is integrable for all $\xi \in (0, \infty)$ if and only if $\mu \phi \geq 1$ and $(1-\mu) \phi\geq 1$. Hence, the class of the bounded beta densities is closed under power transformations, unlike the complete class of the beta densities.

To overcome this problem, consider the logit transformation $y^\star=\log[y/(1-y)]$. If $y \sim \text{Beta}(\mu, \phi)$, the density function of $y^\star$ is given by 
\begin{equation}\nonumber
\begin{split}
  h(y^\star;\mu,\phi) = \dfrac{1}{B(\mu\phi,(1-\mu)\phi)}\dfrac{e^{-y^\star(1-\mu)\phi}}{(1+e^{-y^\star})^{\phi}},\quad y^\star\in\mathbb{R}.
\end{split}
\end{equation}
The distribution of $y^\star$ is called exponential generalized beta of the second type \citep{McDonald} and we write $y^\star\sim \text{EGB}(\mu,\phi)$. Note that $h(y^\star;\mu,\phi)^\xi \propto h(y^\star;\mu,\xi \phi)$, for all $y^\star\in\mathbb{R}$, $\mu \in (0,1)$, and $\phi, \xi >0$. That is, the class of the EGB densities is closed under power transformations. 

We will construct robust estimators for the parameters of the beta regression model \eqref{betadensity}-\eqref{linearPredictor} using the density function of the logit transformed response variable. These estimators are based on \cite{GhoshBeta} and  \cite{RibeiroFerrari} methods and will be described in the following.

Let $y_i^\star=\log[y_i/(1-y_i)]$, where $y_i$, for $i=1, \ldots, n$, follow the postulated beta regression model \eqref{betadensity}-\eqref{linearPredictor}. We denote the density function of $y_i^\star$ by $h_{\bm{\theta}}(\cdot; \mu_i, \phi_i)$. The first proposed robust estimator, named logit minimum density power divergence estimator (LMDPDE), minimizes the empirical version of the density power divergence given by 
\begin{align*}
\mathcal{H}_n(\bm{\theta}) = \dfrac{1}{n} \displaystyle\sum_{i=1}^n \mathcal{V}_i(y_i^\star; \bm{\theta}),
\end{align*}
where
\[
\mathcal{V}_i(y_i^\star; \bm{\theta}) = \mathcal{K}_{i, 1+\alpha}(\bm{\theta}) - \dfrac{1 + \alpha}{\alpha} h_{\bm{\theta}}(y_i^\star; \mu_i, \phi_i)^{\alpha},
\]
and
$$\mathcal{K}_{i, 1+\alpha}(\bm{\theta}) = \displaystyle\int_{-\infty}^\infty h_{\bm{\theta}}(y^\star; \mu_i, \phi_i)^{1+\alpha} \dd y^\star = \dfrac{B(\mu_i \phi_i(1+\alpha), (1-\mu_i) \phi_i (1+\alpha))}{B(\mu_i \phi_i, (1-\mu_i) \phi_i)^{1+\alpha}},$$
for $0\leq\alpha<1$. Note that the integral is finite for all $0\leq\alpha<1$. The estimating equation $\nabla_{\bm{\theta}} \mathcal{H}_n(\bm{\theta}) = \bm{0}$ is given by
\begin{align}\label{eqEstimacaoEGBMDPDE}
\displaystyle \sum_{i=1}^n [\bm{U}(y_i; \bm{\theta})h_{\bm{\theta}}(y_i^\star; \mu_i, \phi_i)^\alpha  - E_{i,1-\alpha}(\bm{\theta})]& = \bm{0},
\end{align}
in which 
\[
\bm{U}(y_i; \bm{\theta}) = \left( \phi_i \dfrac{(y_i^\star - \mu_i^\star)}{g'_\mu (\mu_i)}\bm{{X}}_i^\top, \quad \dfrac{\mu_i(y_i^\star - \mu_i^\star) + (y_i^\dagger - \mu_i^\dagger)}{g'_\phi (\phi_i)} \bm{{Z}}_i^\top   \right)^\top, 
\]
$$E_{i,1-\alpha}(\bm{\theta}) = \mathbb{E}\left[\bm{U}(y_i; \bm{\theta})h_{\bm{\theta}}(y_i^\star; \mu_i, \phi_i)^\alpha 
 \right] = \left( \gamma_{1, i}^{(1+\alpha)}\bm{\bm{X}}_i^\top, \gamma_{2, i}^{(1+\alpha)}\bm{\bm{Z}}_i^\top \right)^\top,$$
where $y_i^\dagger = \log(1-y_i)$,
\begin{align*}
\gamma_{1, i}^{(\alpha)} = \dfrac{\phi_i \mathcal{K}_{i, \alpha}(\bm{\theta}) }{g'_\mu(\mu_i)}(\mu_{i, \alpha}^{\star} -\mu^\star_{i} ), \quad \quad \gamma_{2, i}^{(\alpha)} = \dfrac{ \mathcal{K}_{i, \alpha}(\bm{\theta}) }{g'_\phi(\phi_i)}[\mu_i(\mu_{i, \alpha}^{ \star} -\mu^\star_{i}) + (\mu_{i, \alpha}^{ \dagger} -\mu^\dagger_{i})],
\end{align*}
with $\mu_i^\star = \mathbb{E}({y}_i^\star )  =  \psi(\mu_{i}\phi_i)-\psi((1-\mu_{i})\phi_i)$, $\mu_i^\dagger = \mathbb{E}(y_i^\dagger) =  \psi((1-\mu_{i})\phi_i)-\psi(\phi_i)$, $\mu_{i, \alpha}^{ \star} = \psi(\mu_{i} \phi_{i, \alpha}) - \psi((1-\mu_{i}) \phi_{i, \alpha})$, $\mu_{i, \alpha}^{ \dagger} = \psi((1-\mu_{i}) \phi_{i, \alpha}) - \psi( \phi_{i,\alpha})$, $\psi(\cdot)$ denoting the digamma function,  and $\phi_{i, \alpha} = \phi_i \alpha$. 

The weight of the $i$-th observation in the estimating equation \eqref{eqEstimacaoEGBMDPDE} of the LMDPDE is $h_{\bm{\theta}}(y_i^\star; \mu_i, \phi_i)^\alpha$. In contrast, the corresponding weight for the MDPDE is $f_{\bm{\theta}}(y_i; \mu_i, \phi_i)^\alpha$. If $\alpha=0$, $h_{\bm{\theta}}(y_i^\star; \mu_i, \phi_i)^\alpha=1$ and $E_{i,1-\alpha}(\bm{\theta}) = 0$, for all $i=1, \ldots, n$; hence the LMDPDE coincides with the MLE. Unlike the estimating function of the MDPDE, that of the LMDPDE is well-defined for all $\alpha \in [0,1)$ and $\bm{\theta} \in \mathbb{R}^p$.

The second robust estimator is based on \cite{RibeiroFerrari} method, and is named logit surrogate maximum likelihood estimator (LSMLE). The L$_q$-likelihood based on the density $h_{\bm{\theta}}(\cdot; \mu_i, \phi_i)$ is given by \eqref{Lqlikelihood} with $f_{\bm{\theta}}(y_i; \mu_i, \phi_i)$ replaced by $h_{\bm{\theta}}(y_i^\star; \mu_i, \phi_i)$. As expected, the estimator that comes from the maximization of the L$_q$-likelihood is not Fisher-consistent. In other words, the estimating function is biased. Since the class of the EGB densities is closed under power transformations, Fisher-consistency can be achieved by maximizing the L$_q$-likelihood in the parametrization $\tau_{1/(1-\alpha)}(\bm{\theta})$ \citep{FerrariLaVecchia, LaVecchiaetal}, where $\tau_\omega(\bm{\theta}): \bm{\Theta} \longmapsto \bm{\Theta}$ is a continuous function satisfying $h_{\tau_\omega(\bm{\theta})}(y^\star;\mu, \phi) = h_{\bm{\theta}}^{(\omega)}(y^\star;\mu, \phi)$, for all $y^\star \in \mathbb{R}$. The LSMLE is the maximizer of 
\[
\ell_{1-\alpha}^*(\bm{\theta}) = \displaystyle \sum_{i=1}^n L_{1-\alpha}\left(h_{\bm{\theta}}^{\left(\frac{1}{1-\alpha}\right)}(y_i^\star; \mu_i, \phi_i)\right), 
\]
where $h_{\bm{\theta}}^{\left(\frac{1}{1-\alpha}\right)}(y_i^\star; \mu_i, \phi_i) = h_{\bm{\theta}}\left(y_i^\star; \mu_i, {\phi}_{i, (1-\alpha)^{-1}}\right)$, with $0\leq\alpha<1$, and $\mu_i$ and $\phi_i$ satisfying \eqref{linearPredictor}. Note that $h_{\bm{\theta}}\left(y_i^\star; \mu_i, {\phi}_{i, (1-\alpha)^{-1}}\right)$ is the density function of the logit transformation of a variable that follows a modified beta regression model with mean and precision submodels given respectively by
\[
g^*_\mu(\mu_i) = g_\mu(\mu_i) = \bm{{X}}_i^{\top} \bm{\beta}, \quad \quad g^*_\phi(\phi_i) = g_\phi({\phi}_{i, 1-\alpha}) = \bm{{Z}}_i^{\top} \bm{\gamma},
\]
which will be denoted by $h^\ast_{\bm{\theta}}(y_i^\star; \mu_{i}, \phi_{i})$. Thus, the LSMLE is the maximizer of
\[
\ell_{1-\alpha}^*(\bm{\theta}) = \displaystyle \sum_{i \in \wp} L_{1-\alpha}\left(h^\ast_{\bm{\theta}}(y_i^\star; \mu_{i}, \phi_{i})\right).
\]
It solves the estimating equation 
\begin{align}\label{estimatingequationLSMLE}
\begin{split}
\displaystyle \sum_{i=1}^n \bm{U}^\ast(y_i^\star; \bm{\theta})h^\ast_{\bm{\theta}}(y_i^\star; \mu_i, \phi_i)^\alpha & = \bm{0},
\end{split}
\end{align}
where $\bm{U}^\ast(y_i^\star; \bm{\theta}) = \nabla_{\bm{\theta}}\log h^\ast_{\bm{\theta}}(y_i^\star; \mu_i, \phi_i)$ is the modified score vector for the $i$-th observation given by 
\[
\bm{U}^\ast(y_i^\star; \bm{\theta}) = \left( \phi_{i} \dfrac{(y_i^\star - {\mu}_i^\star)}{g'_\mu (\mu_{i})}\bm{{X}}_i^\top, \quad (1-\alpha)^{-1} \dfrac{\mu_{i}(y_i^\star - {\mu}_i^\star) + (y_i^\dagger - {\mu}_i^\dagger)}{g'_\phi ({\phi}_{i, 1-\alpha})} \bm{{Z}}_i^\top   \right)^\top.
\]

In the Supplementary Material (Section 1) we show that the LSMLE is Fisher-consistent.

\paragraph{Asymptotic normality.} Let $\widehat{\bo{\theta}}_\alpha = (\widehat{\bo{\beta}}_\alpha^\top, \widehat{\bo{\gamma}}_\alpha^\top)^\top$ and $\widetilde{\bo{\theta}}_\alpha = (\widetilde{\bo{\beta}}_\alpha^\top, \widetilde{\bo{\gamma}}_\alpha^\top)^\top$ be the LMDPDE and the LSMLE, respectively, for fixed $\alpha \in [0,1)$. Since they are M-estimators, we have that $\widehat{\bo{\theta}}_\alpha \stackrel{a}{\sim} \text{N}(\bo{\theta}, \bo{V}_{1, \alpha}(\bo{\theta})) $ and $\widetilde{\bo{\theta}}_\alpha \stackrel{a}{\sim} \text{N}(\bo{\theta}, \bo{V}_{2, \alpha}(\bo{\theta})) $, where $\stackrel{a}{\sim}$ denotes asymptotic distribution,
\[
\bo{V}_{1, \alpha}(\bo{\theta}) = \bo{\Lambda}_{1, \alpha}^{-1}(\bo{\theta}) \bo{\Sigma}_{1, \alpha}(\bo{\theta})\bo{\Lambda}_{1, \alpha}^{-1}(\bo{\theta}), \quad \quad \bo{V}_{2, \alpha}(\bo{\theta}) = \bo{\Lambda}_{2, \alpha}^{-1}(\bo{\theta}) \bo{\Sigma}_{2, \alpha}(\bo{\theta})\bo{\Lambda}_{2, \alpha}^{-1}(\bo{\theta}),
\]
and the expressions for $\bo{\Lambda}_{j, \alpha}(\bo{\theta})$ and $\bo{\Sigma}_{j, \alpha}(\bo{\theta})$, $j=1,2$, are given in the Appendix; see the Supplementary Material (Section 2) for details. The covariance matrices $\bo{V}_{1, \alpha}(\bo{\theta})$ and $\bo{V}_{2, \alpha}(\bo{\theta})$ are well-defined for all $\alpha \in [0,1)$ and $\bm{\theta} \in \mathbb{R}^p$ unlike those of the SMLE and MDPDE \citep{RibeiroFerrari}. 
In addition, the asymptotic covariance matrices are equal to the asymptotic covariance matrix of the MLE for $\alpha=0$. 

\paragraph{Robustness properties.} In the context of robust estimators, the influence function plays an important role. Introduced by \cite{Hampel}, the influence function represents the first-order measure of the effect on the asymptotic bias caused by a slight contamination in a data point. Since the LMDPDE and the LSMLE are M-estimators, their influence functions are respectively given by
\[
\text{IF}(y^\star; \widehat{\bo{\theta}}_\alpha) = \bo{\Lambda}_{1, \alpha}^{-1}(\bo{\theta})  [\bm{U}(y; \bm{\theta})h_{\bm{\theta}}(y^\star; \mu, \phi)^\alpha  - E_{1-\alpha}(\bm{\theta})],
\]
\[
\text{IF}(y^\star; \widetilde{\bo{\theta}}_\alpha) = \bo{\Lambda}_{2, \alpha}^{-1}(\bo{\theta})\bm{U}^\ast(y^\star; \bm{\theta})h^\ast_{\bm{\theta}}(y^\star; \mu, \phi)^\alpha,
\]
where $E_{1-\alpha}(\bm{\theta}) = \mathbb{E}\left[\bm{U}(y; \bm{\theta})h_{\bm{\theta}}(y^\star; \mu, \phi)^\alpha  \right]$. The influence functions of the LMDPDE and the LSMLE are bounded, that is they are B-robust. We also extend the robustness analysis to the change-of-variance function, which measures the bias on the covariance matrix due to an infinitesimal contamination in a data point. We show that the change-of-variance functions of the LMDPDE and the LSMLE are bounded, that is they are V-robust \citep[Section 2.5]{Hampel2011}; see the Supplementary Material, Section 3, for details. These robustness properties do not hold for the MLE and are guaranteed for the MDPDE and SMLE for bounded beta densities; see \cite{RibeiroFerrari}. 

\paragraph{Robust Wald-type tests.} Let $m:\mathbb{R}^p\rightarrow \mathbb{R}^{d}$, with $d\leq p$, be a continuously differentiable function of $\bo{\theta}$. Assume that its Jacobian $d \times p$ matrix, $\bo{J}_{m}(\bo{\theta})$, has rank $d$. Consider the null hypothesis $m(\bo{\theta}) = \bo{\eta}_0$, for a fixed $\bo{\eta}_0 \in \mathbb{R}^d$, to be tested against a two sided alternative. Let
\begin{equation}\nonumber
W_{j,\alpha}(\bo{\theta})=(m(\bo{\theta})-\bo{\eta}_{0})^{\top}\left[\bo{J}_{m}(\bo{\theta})\bo{V}_{j, \alpha}(\bo{\theta})\bo{J}_{m}(\bo{\theta})^{\top}\right]^{-1}(m(\bo{\theta})-\bo{\eta}_{0}),
\end{equation}
for $j=1,2$. The Wald-type test statistics that use the LMDPDE and the LSMLE are, respectively, given by $W_{1,\alpha}(\widehat{\bo{\theta}}_\alpha)$ and $W_{2,\alpha}(\widetilde{\bo{\theta}}_\alpha)$. Under the null hypothesis, both statistics are asymptotically $\chi_d^2$-distributed. They coincide with the usual Wald test statistic if $\alpha=0$.

\paragraph{Selecting the tuning constant.} \citet[Section 3]{RibeiroFerrari} proposed a data-driven algorithm to select the tuning constant for the MDPDE and the SMLE. The idea is to select $\alpha$, in an ordered grid from $\alpha=0$ to $\alpha=\alpha_{\text{max}}$, that is closest to zero such that the estimates of the parameters are sufficiently stable, ensuring full efficiency for non-contaminated data. If the algorithm does not reach stability up to $\alpha_{\text{max}}$, it returns the MLE ($\alpha=0$). The authors suggest setting $\alpha_{\text{max}}=0.5$. Here, the algorithm is implemented for selecting $\alpha$ for the LMDPDE and the LSMLE.

\section{Simulation studies}\label{Sec.SimStudy}

In this section, we evaluate the performance of the robust estimators and the MLE for the beta regression model \eqref{betadensity}-\eqref{linearPredictor}. We employ the logit and the logarithmic link functions for the mean and precision submodels, respectively. Both submodels include an intercept, i.e., $x_{i,1} = z_{i,1} = 1$, $i=1, \ldots, n$. The sample sizes are set at $n=40,~80, ~160$, and $320$. The covariate values for the mean submodel are set for the sample size $n=40$ as random draws from a standard uniform distribution and replicated twice, four times and eight times for the other values of $n$. For the non-constant precision scenario, the covariate values for the mean submodel are used in the precision submodel. All the covariate values are kept constant over all the simulated samples. We consider non-contaminated and contaminated samples with a fixed contamination rate, namely 5\%. All simulations were carried out using the R software \citep{R}, and the results are based on $1000$ Monte Carlo replications.

\begin{figure}[H]
\captionsetup[subfigure]{labelformat=empty}
\centering
 \subfloat[Scenario A]{\includegraphics[width=0.32\textwidth]{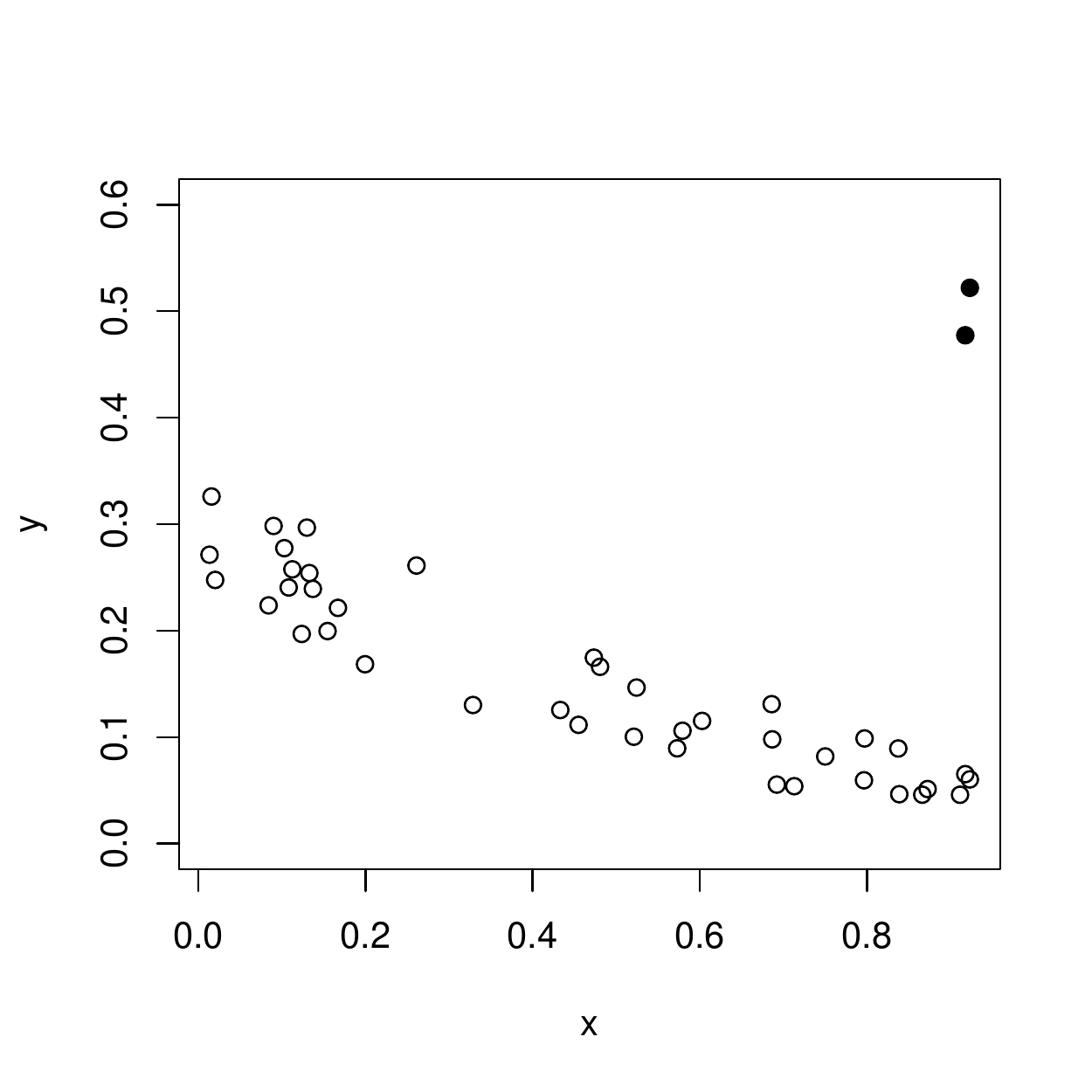}}
 \subfloat[Scenario B]{\includegraphics[width=0.32\textwidth]{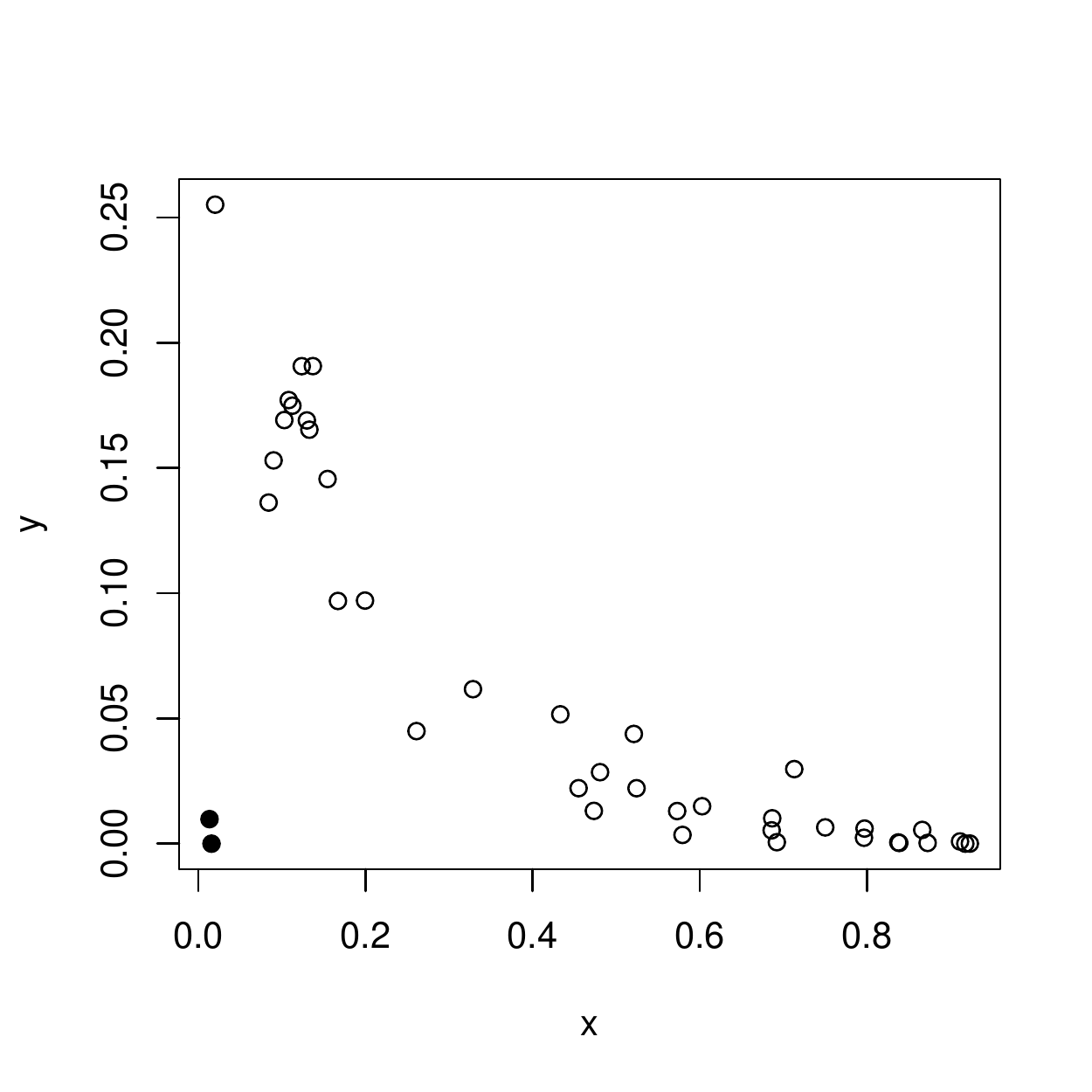}}
 \subfloat[Scenario C]{\includegraphics[width=0.32\textwidth]{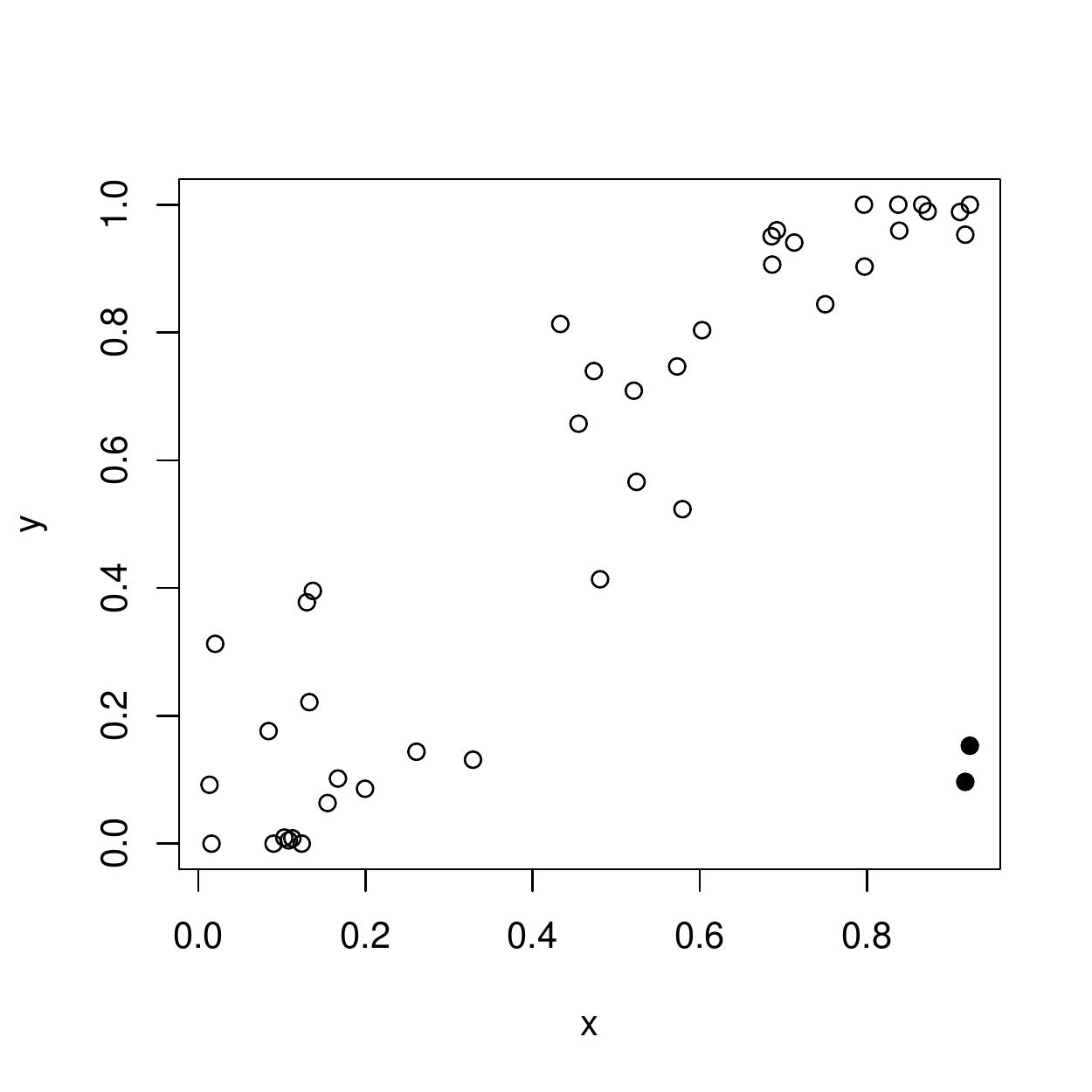}}\\
\caption{Scatter plots of contaminated samples generated as in Scenarios A, B, and C with $n=40$. The contaminated observations are represented by solid black circles.}
     \label{Fig.scatter.sim}
\end{figure}

We consider three different scenarios. Figure \ref{Fig.scatter.sim} shows scatter plots of a sample generated under each scenario for $n=40$.
\newline\newline
\textit{Scenario A: bounded beta densities; constant precision.} The parameters are set at $\beta_1=-1$, $\beta_2=-2$ and $\gamma_1=5$. The possible values for $\mu$ range in $(0.05,0.27)$ and $\phi=\exp(5) \cong148$. For the contaminated samples, we replace the observations generated with the 5\% smallest means by observations generated with mean $\mu'_{i}=(1+\mu_i)/2$. All the beta densities in this scenario are bounded.
\newline\newline
\textit{Scenario B: unbounded beta densities; constant precision.} The parameters are set at $\beta_1=-1$, $\beta_2=-5.5$ and $\gamma_1=5$. The possible values for $\mu$ range in $(0.001,0.27)$ and $\phi=\exp(5)\cong1 48$. For the contaminated samples, the observations generated with the 5\% highest means are replaced by observations generated with mean $\mu'_{i}=0.002$. Some beta densities in this scenario are unbounded.
\newline\newline
\textit{Scenario C: unbounded beta densities; varying precision.} The parameters are set at $\beta_1=-3$, $\beta_2=7.5$, $\gamma_1=1$, and $\gamma_2=2$. The possible values for $\mu$ and $\phi$ range in $(0.05,0.98)$ and $\phi\in(2.7,20.1)$, respectively. For the contaminated samples, the observations generated with the $5\%$ highest means are replaced by observations generated with $\mu'_{i}=\exp{(\beta_1)}/(1+\exp{(\beta_1)}) \cong 0.05$ and $\phi'_{i}=\exp (\gamma_1+\gamma_2) \cong 20.1$. Some beta densities in this scenario are unbounded. 

\begin{figure}[!h]
\captionsetup[subfigure]{labelformat=empty}
\centering
\includegraphics[scale=0.6]{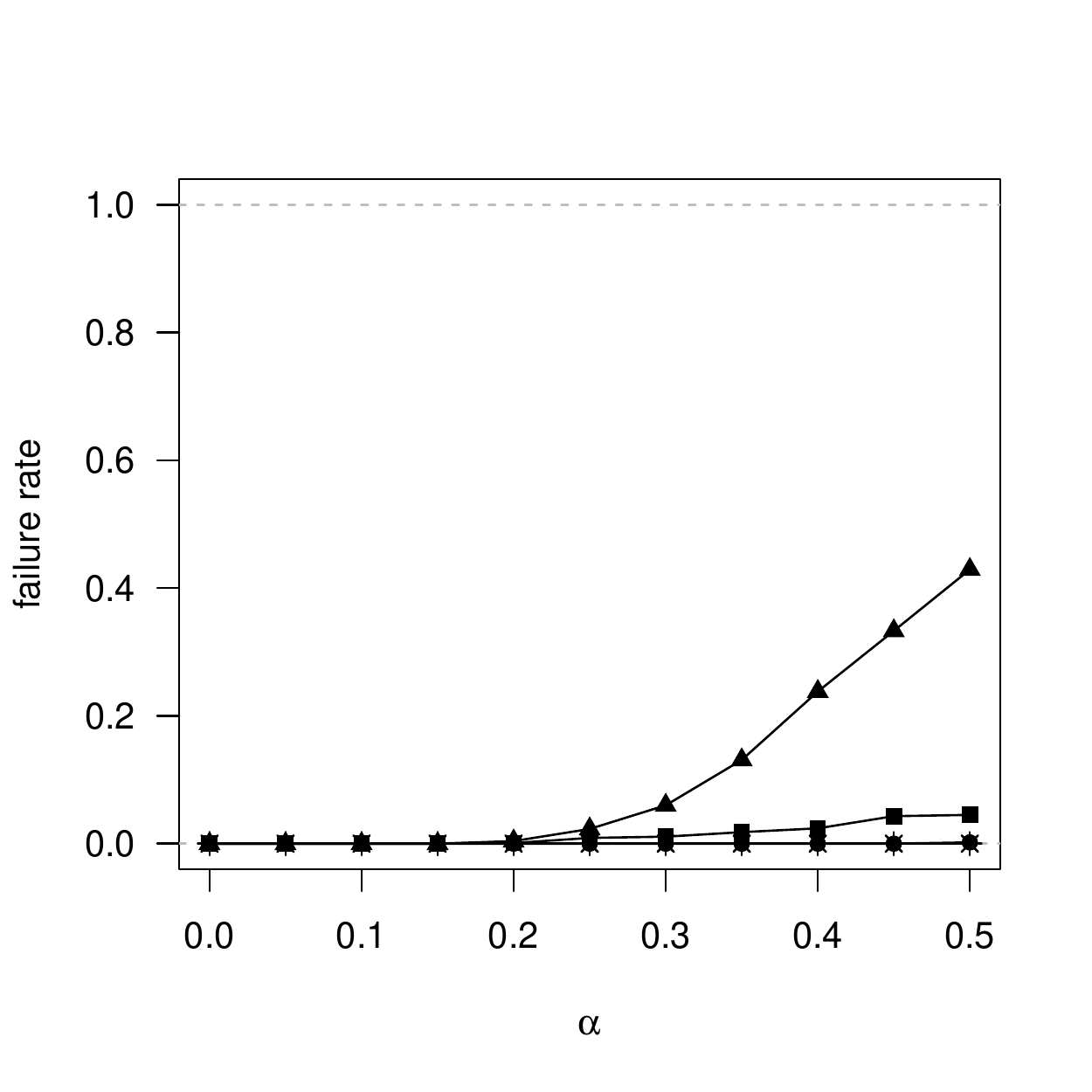}
\includegraphics[scale=0.6]{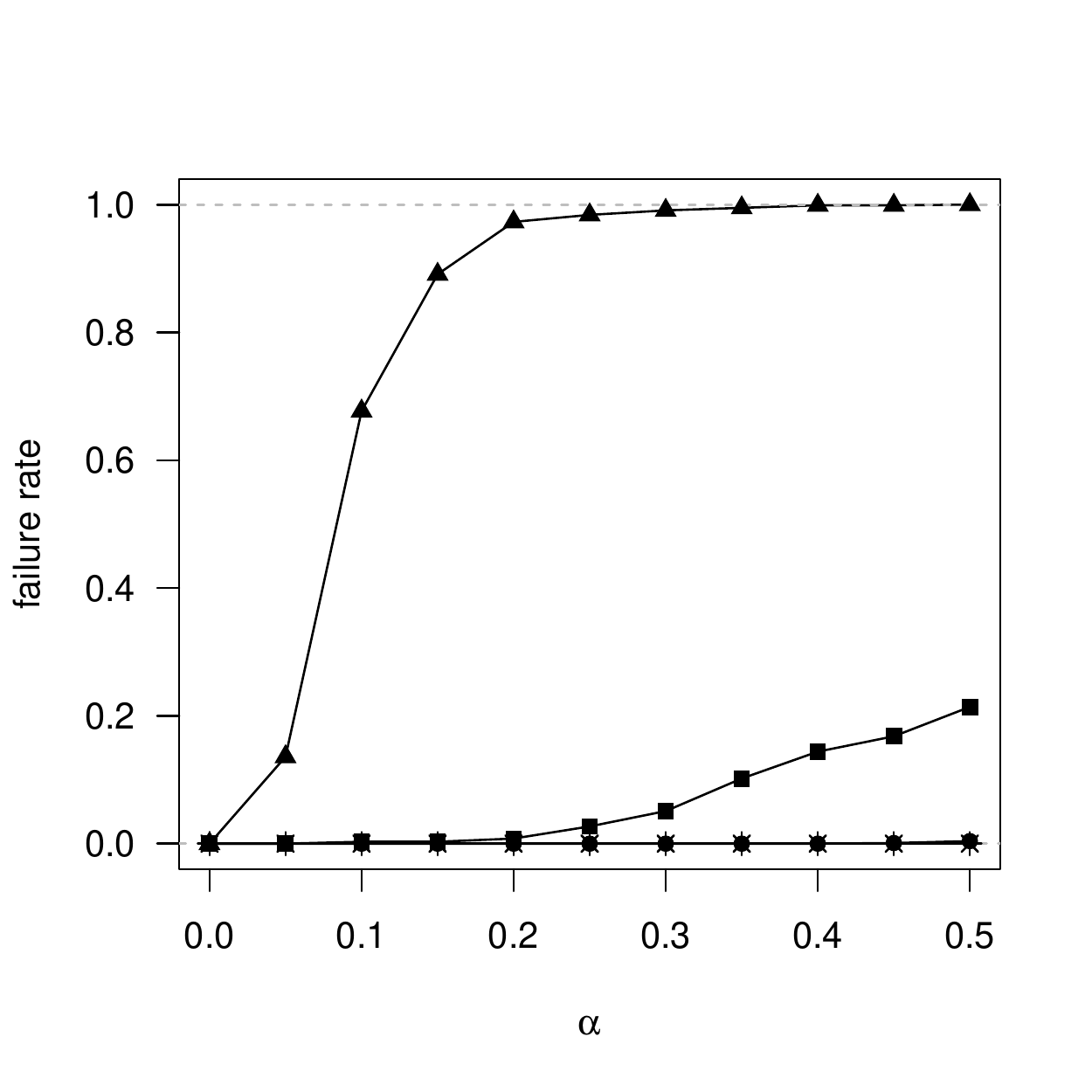}
\includegraphics[scale=0.6]{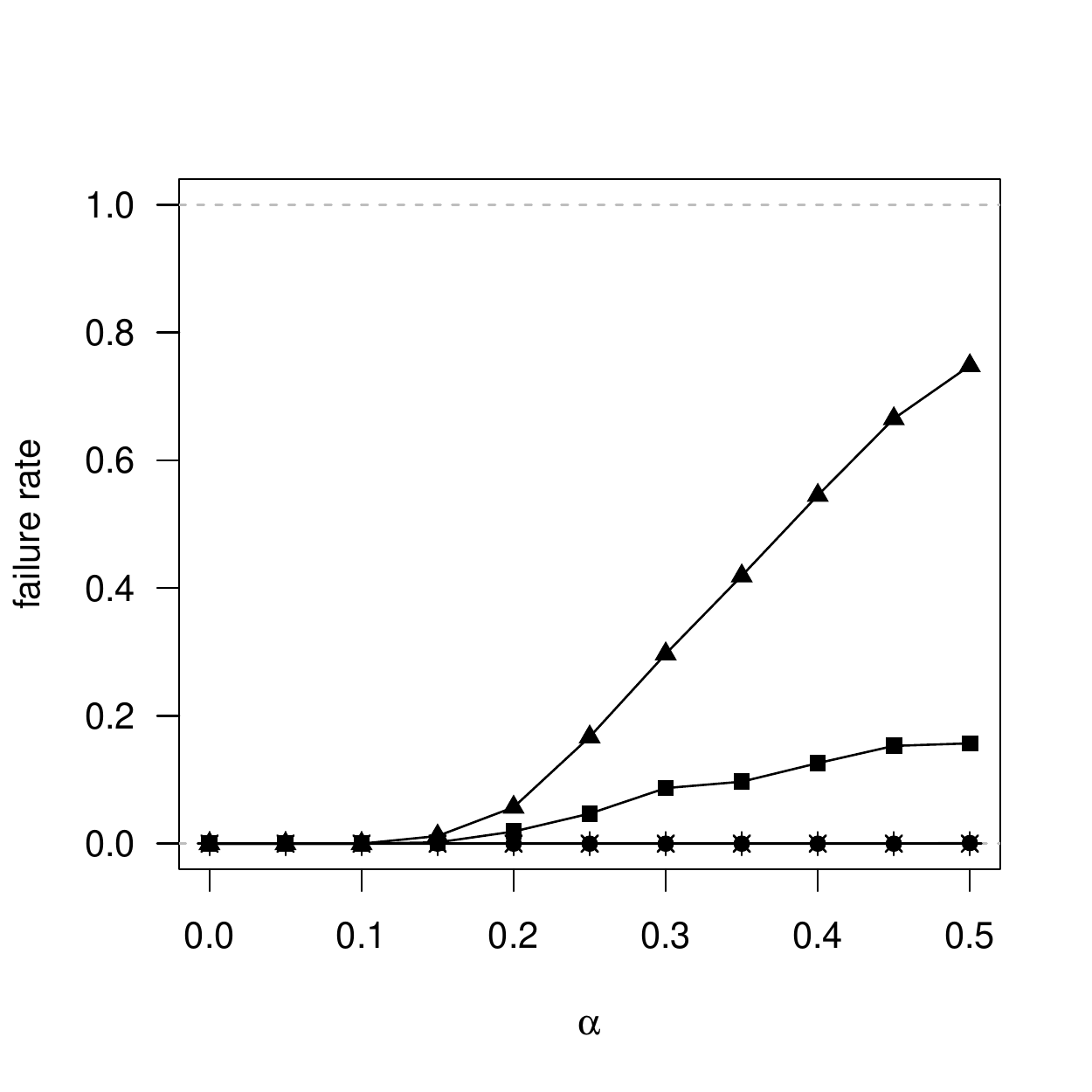}
\includegraphics[scale=0.6]{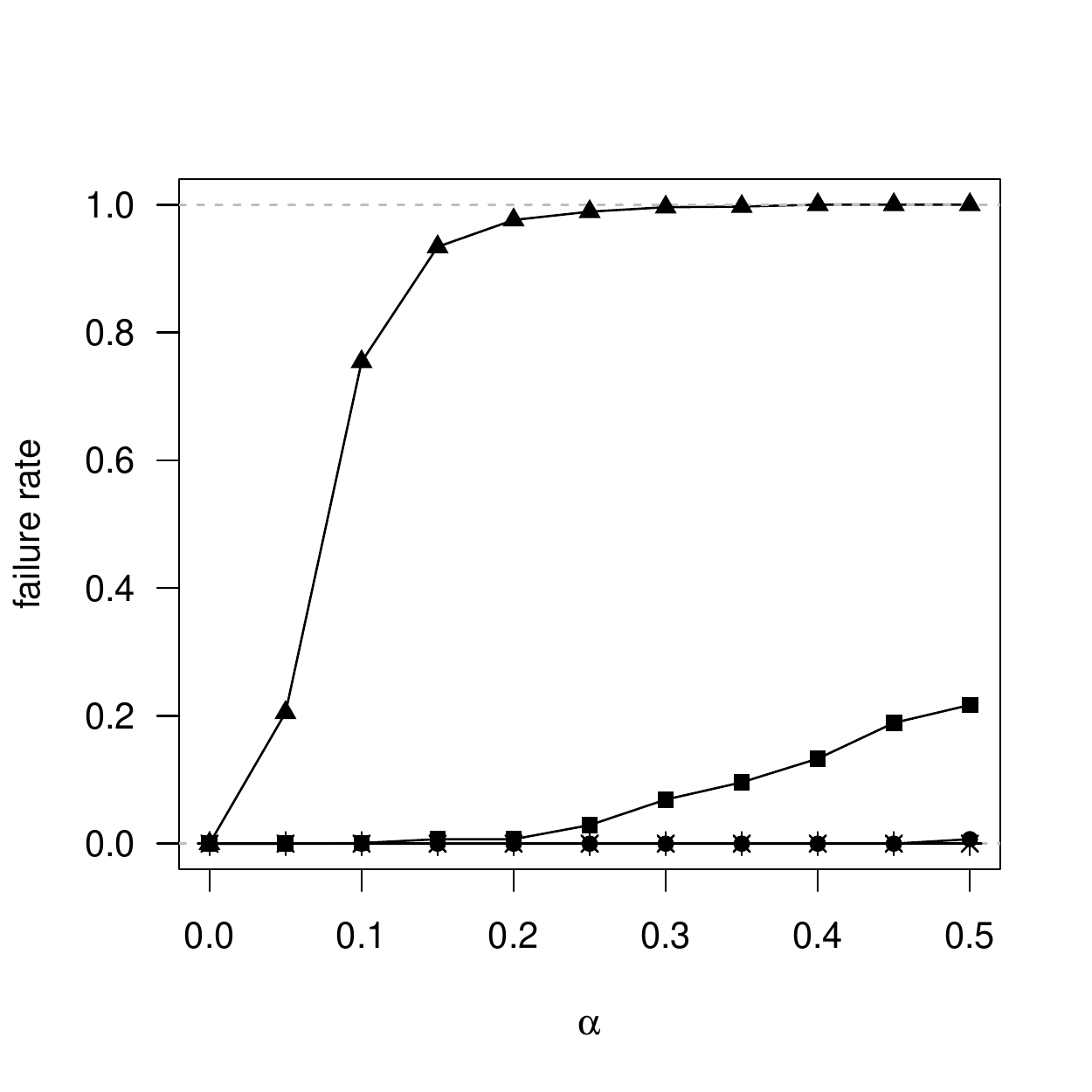}
 \caption{Plots of the failure rate versus the tuning constant $\alpha$ for Scenario B (left) and Scenario C (right) for non-contaminated samples (first row) and contaminated samples (second row) : SMLE (triangle), MDPDE (square), LSMLE (circle), and LMDPDE (star).}
     \label{Fig.CR.BC}
\end{figure}

First, we run simulations for the three scenarios with $n=40$ for fixed values of the tuning constant $\alpha$ ranging from $0$ to $0.05$ incremented by $0.05$. For each $\alpha$ value, we compute the failure rate over the $1000$ simulated samples for the robust estimates: the MDPDE, the SMLE, the LMDPDE, and the LSMLE. We consider a failure whenever the optimization algorithm for computing the estimate does not reach convergence or the asymptotic standard error can not be calculated. In Scenario A, no sample resulted in failure for any of the estimators. Recall that all the beta densities in this scenario are bounded. Figure \ref{Fig.CR.BC} displays plots of the failure rates for Scenarios B and C. The failure rate of SMLE and the MDPDE tend to increase as $\alpha$ grows, more so in Scenario C. In contrast, for the new robust estimators, the failure rate is equal (or close) to zero for all values of $\alpha$, both under non-contaminated or contaminated data. Simulations for the other sample sizes reveal a similar pattern.

We now report simulation results using the data-driven algorithm for selecting the optimum value of $\alpha$ proposed by \cite{RibeiroFerrari}. Figures \ref{Fig.BPA}, \ref{Fig.BPB}, and \ref{Fig.BPCb}-\ref{Fig.BPCg} display the boxplots of the parameter estimates using the MLE, the SMLE, the MDPDE, the LSMLE, and the LMDPDE under Scenarios A, B, and C, respectively. 

The MLE is highly affected by contaminated observations for all the scenarios and presents a severe bias. For instance, the maximum likelihood estimates of $\gamma_1$ in Scenario B are around its true value, $5$, for the non-contaminated data and around $2$ for the contaminated data. In Scenario A, all the robust estimators present good performances for both contaminated and non-contaminated data. They behave similarly to the MLE under non-contaminated samples. Recall that all the beta densities in this scenario are bounded. In Scenarios B and C, which include unbounded beta densities, the MDPDE and the SMLE do not behave well for the non-contaminated data. In these cases, for almost all the samples, the selected optimum tuning constants are zero, resulting in non-robust estimates. It happens because the optimization fails; hence, the algorithm does not reach stability in the estimates and returns the MLE. In contrast, the LMDPDE and the LSMLE perform well in the presence and absence of contamination. Also, these estimators have similar behavior. For samples of moderate and large sizes ($n=80, ~160, ~360$), the performances of the LMDPDE and LSMLE are excellent. The simulation results indicate that the proposed estimators (LMDPDE and LSMLE) are robust in the presence of outliers, unlike the MLE. Also, the MDPDE and SMLE may not be useful in scenarios involving unbounded beta densities. In general, in the presence of contamination, the robustness obtained through the new proposed estimators comes at the cost of a slight increase in variability, especially in small sample sizes ($n=40$).

\begin{figure}[!h]
\captionsetup[subfigure]{labelformat=empty}
\centering
 \subfloat{\includegraphics[width=0.25\textwidth]{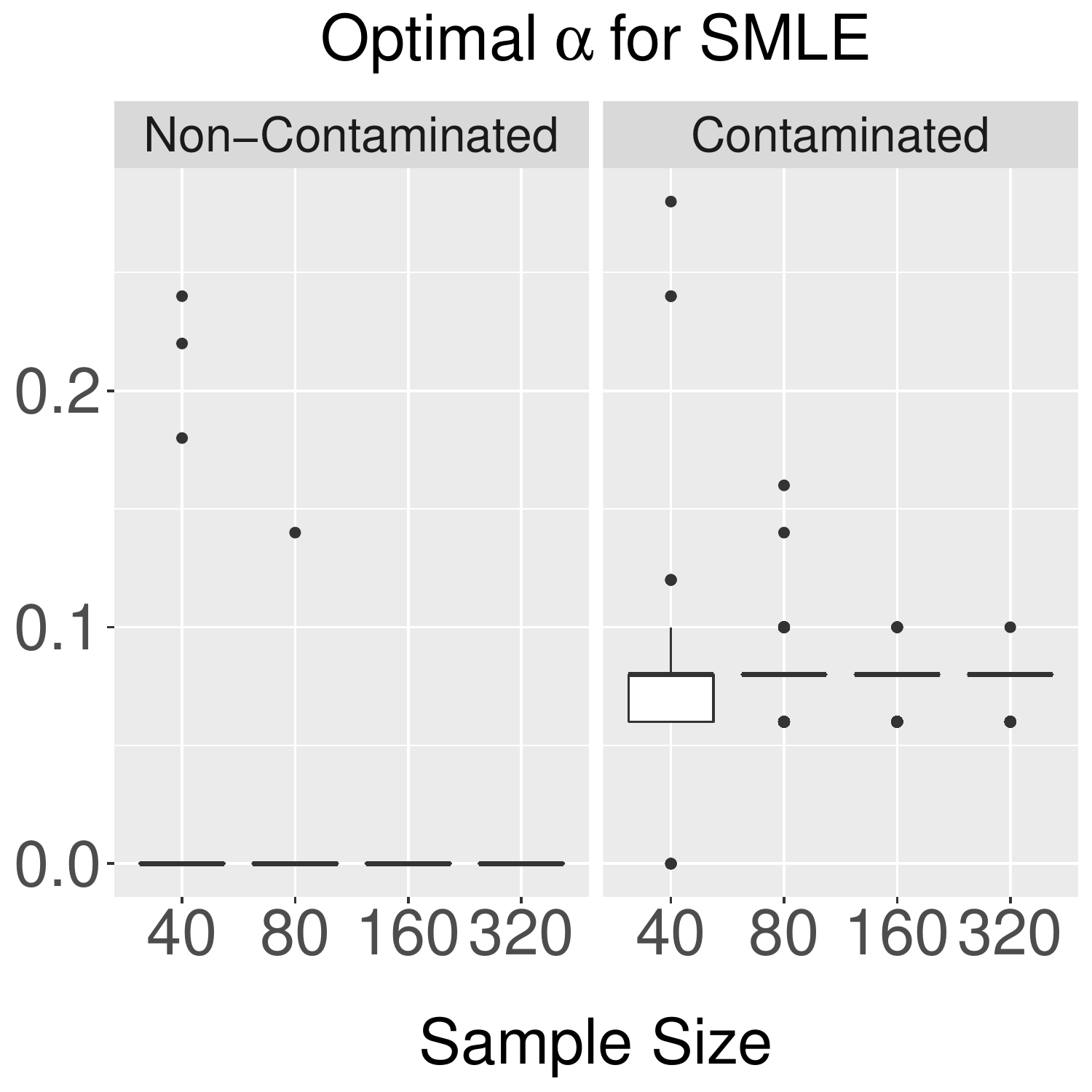}}
 \subfloat{\includegraphics[width=0.25\textwidth]{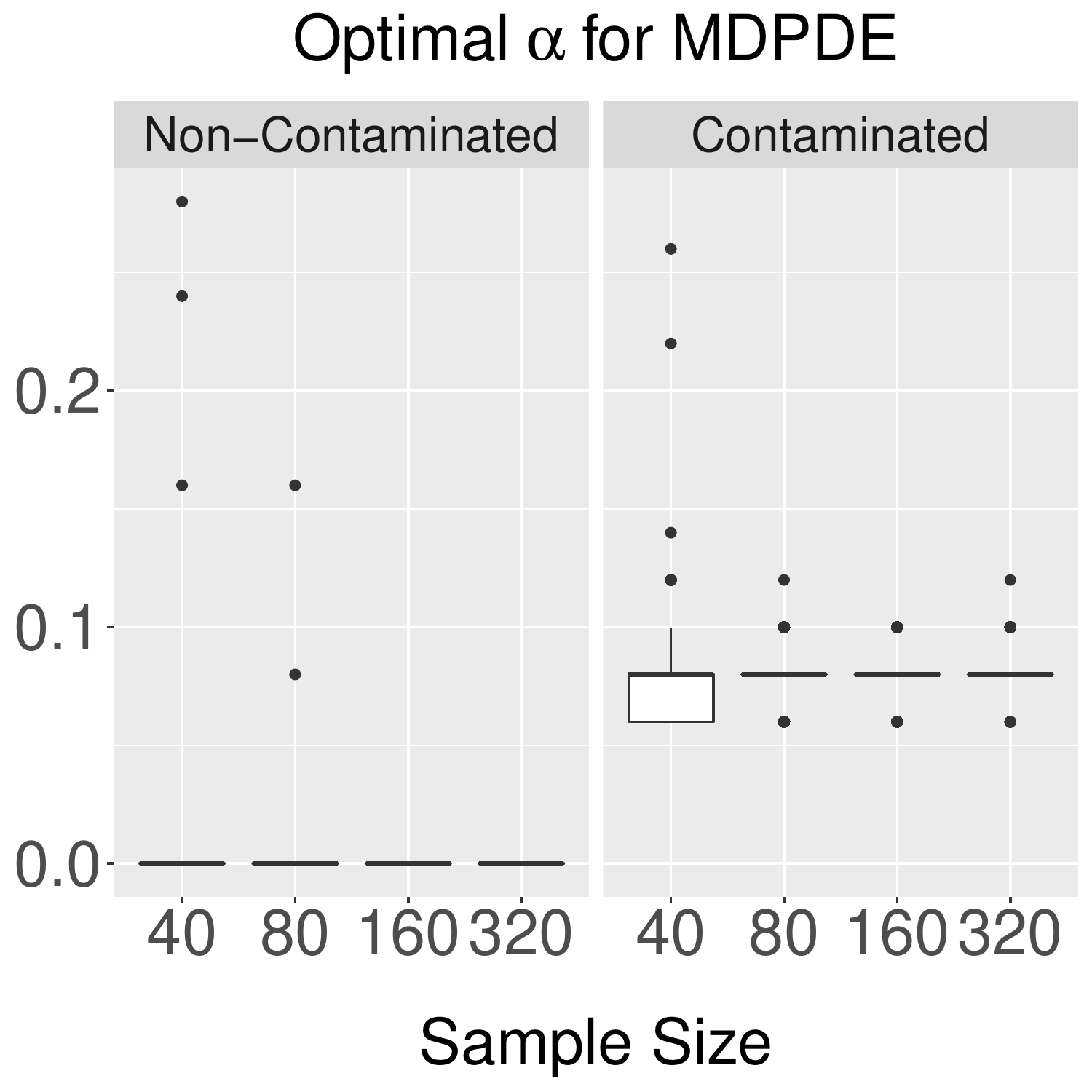}}
 \subfloat{\includegraphics[width=0.25\textwidth]{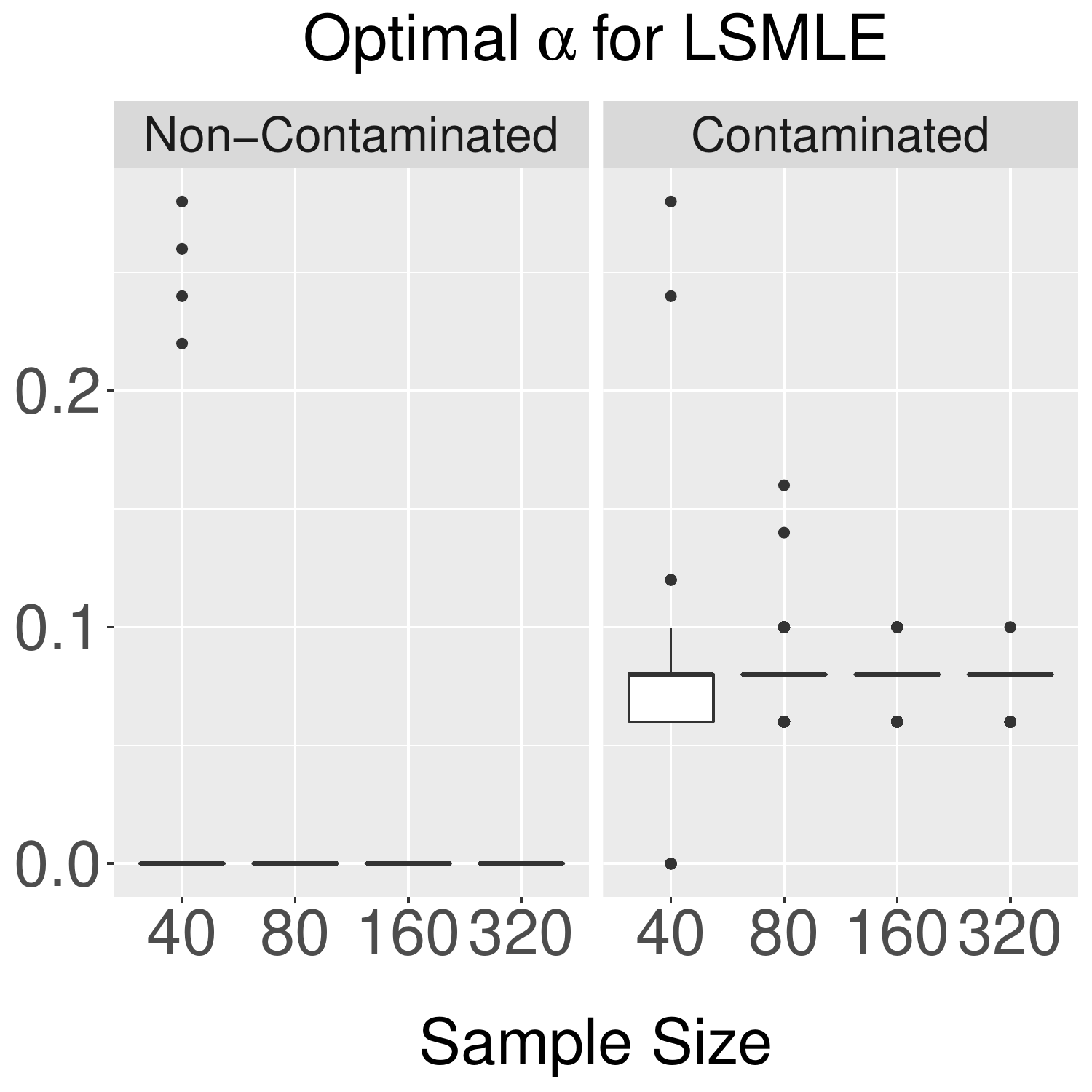}}
 \subfloat{\includegraphics[width=0.25\textwidth]{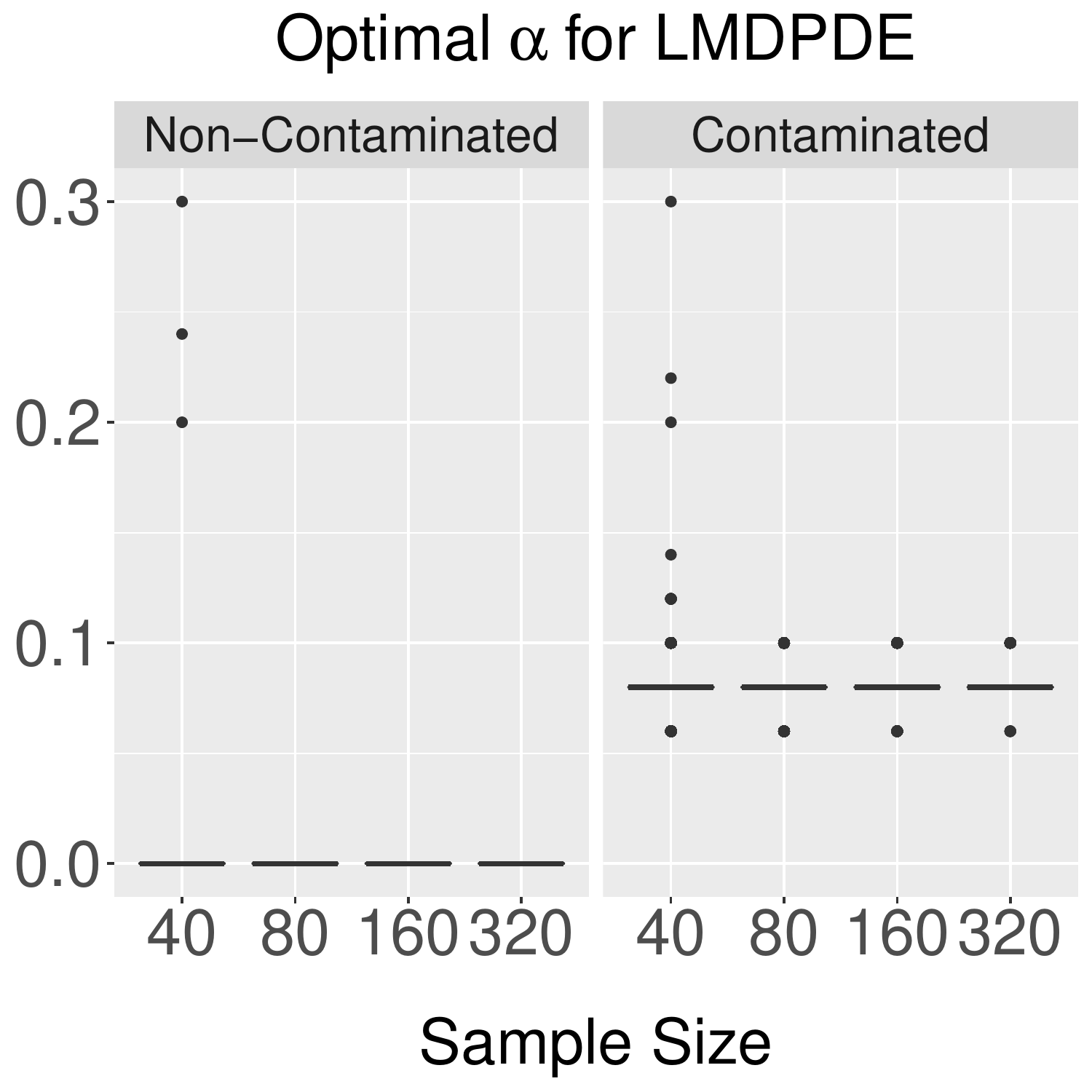}}\\
 \subfloat{\includegraphics[width=0.25\textwidth]{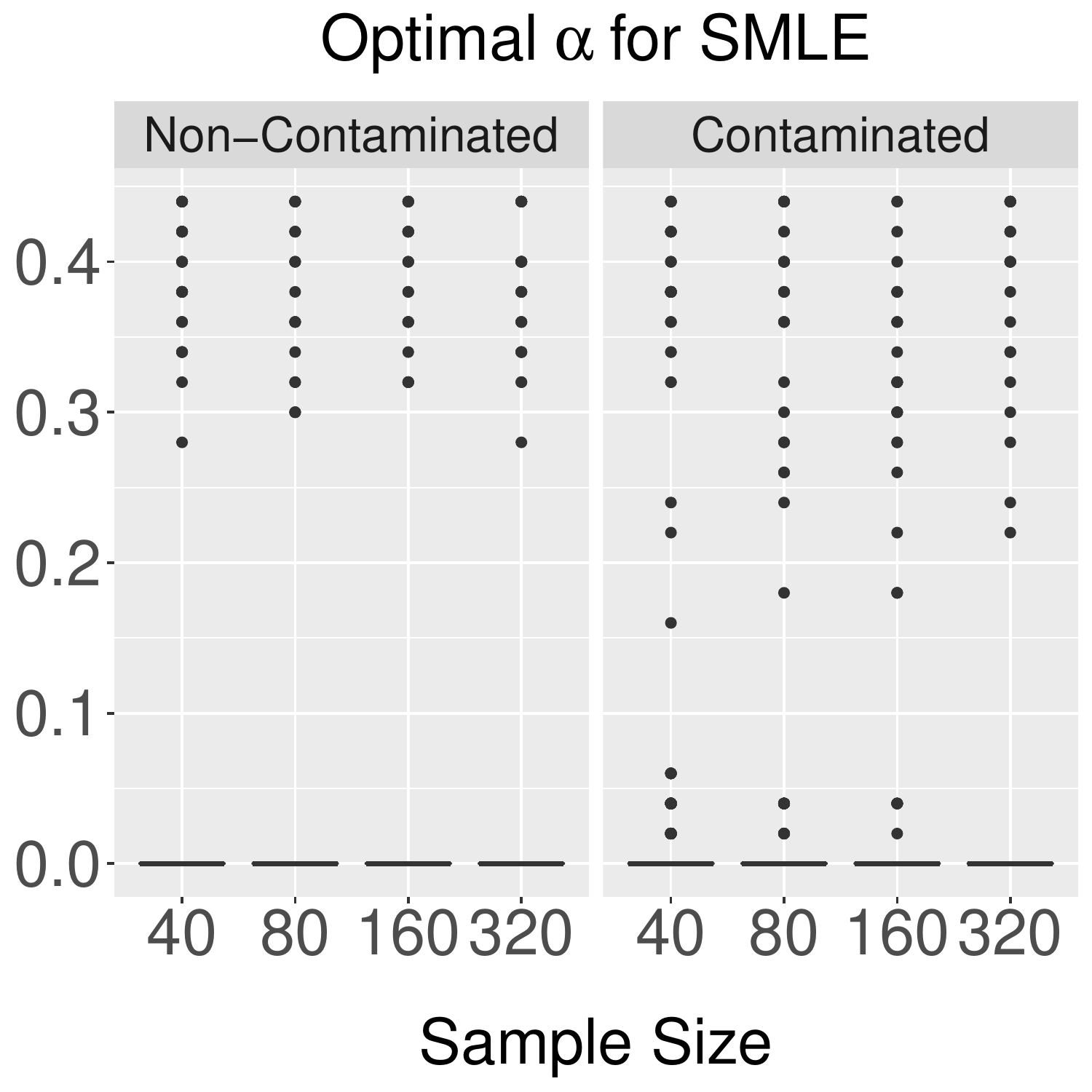}}
 \subfloat{\includegraphics[width=0.25\textwidth]{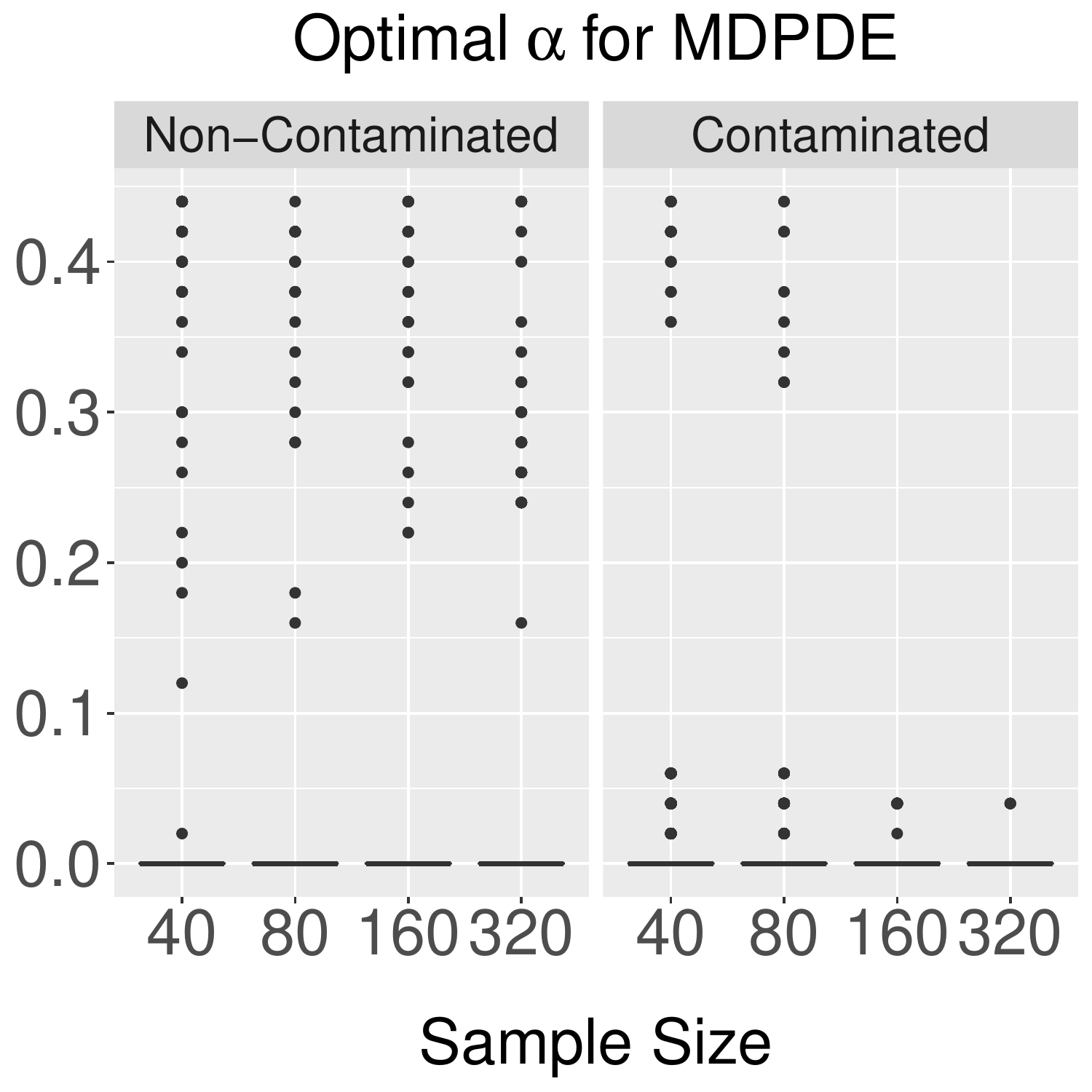}} 
 \subfloat{\includegraphics[width=0.25\textwidth]{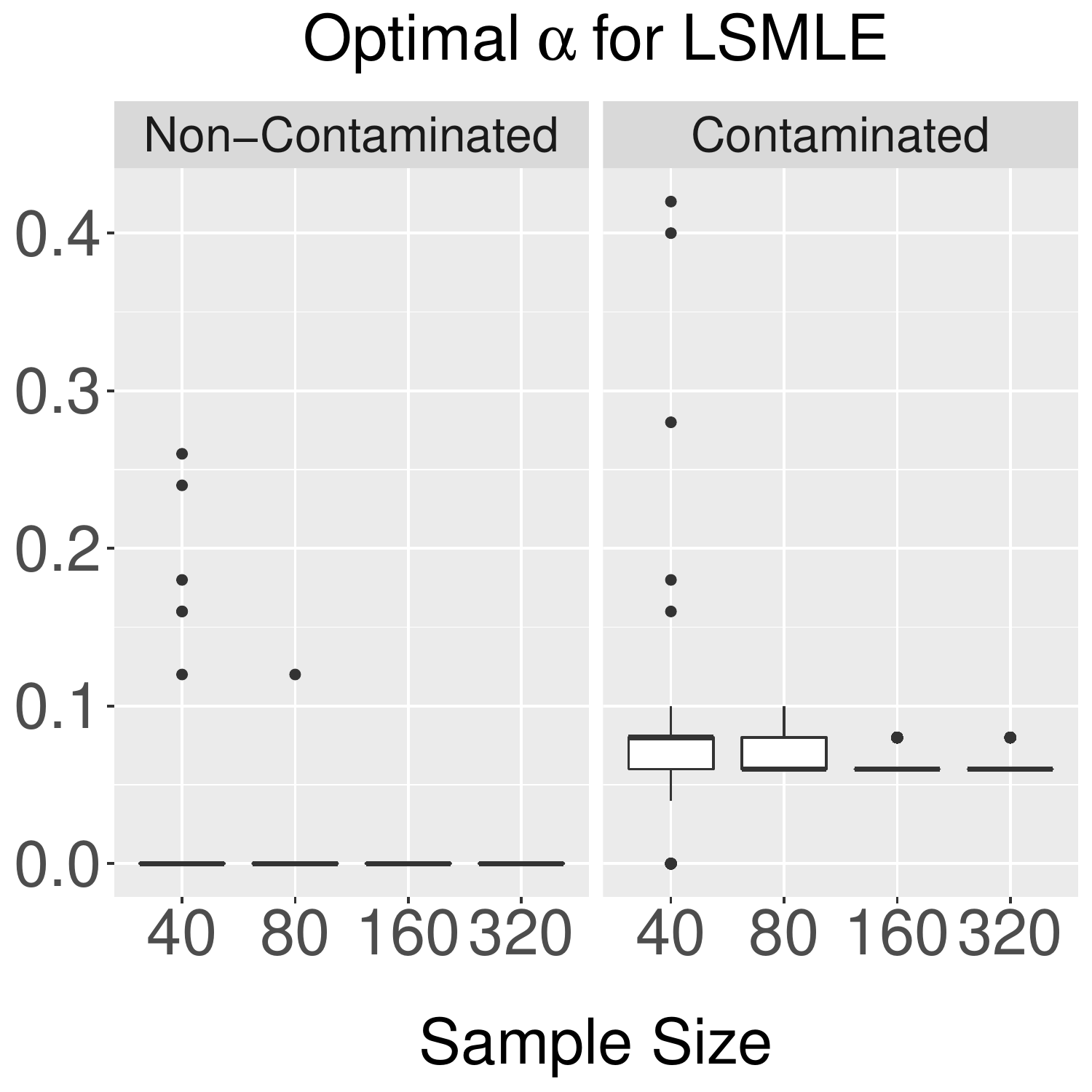}}
 \subfloat{\includegraphics[width=0.25\textwidth]{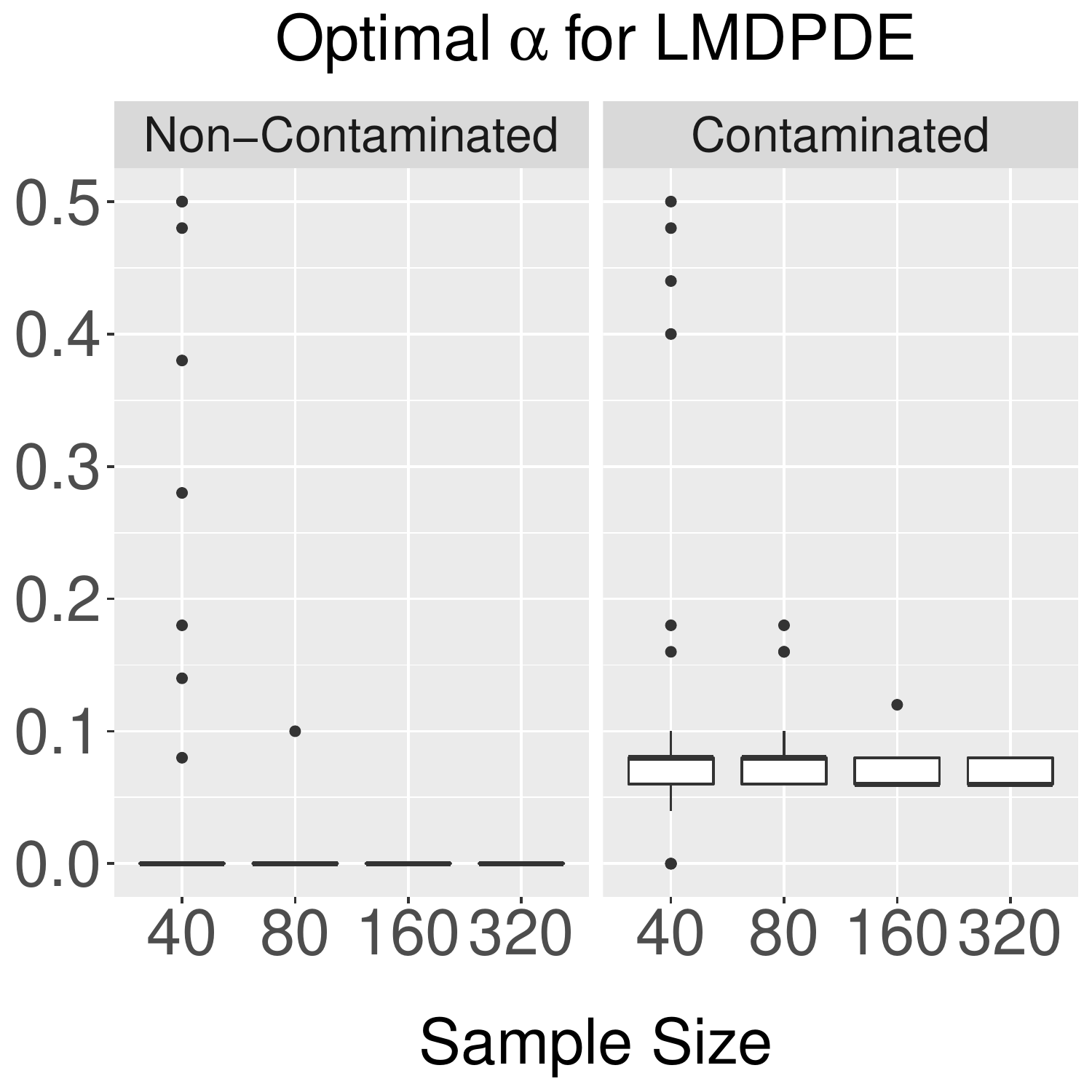}}\\
  \subfloat{\includegraphics[width=0.25\textwidth]{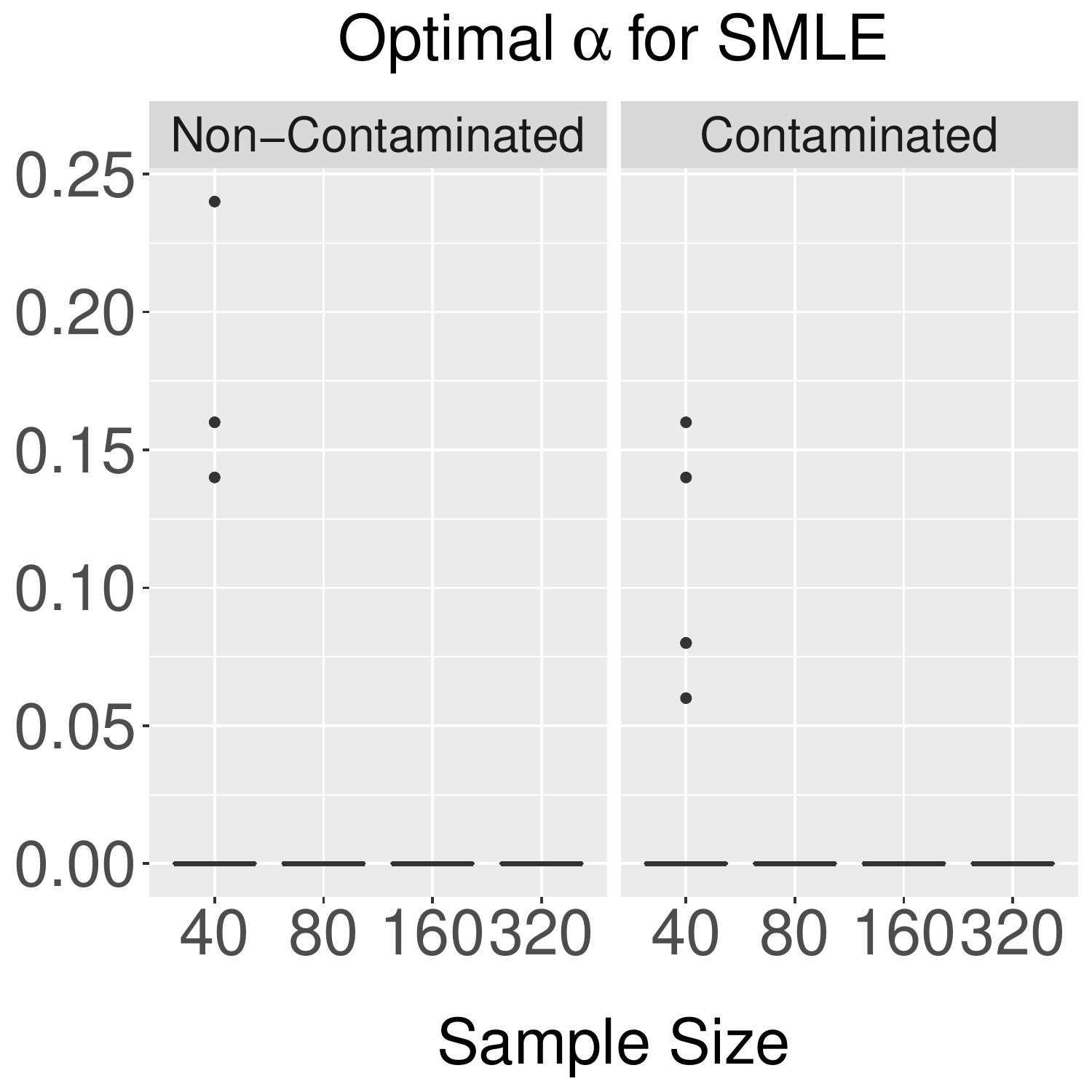}}
 \subfloat{\includegraphics[width=0.25\textwidth]{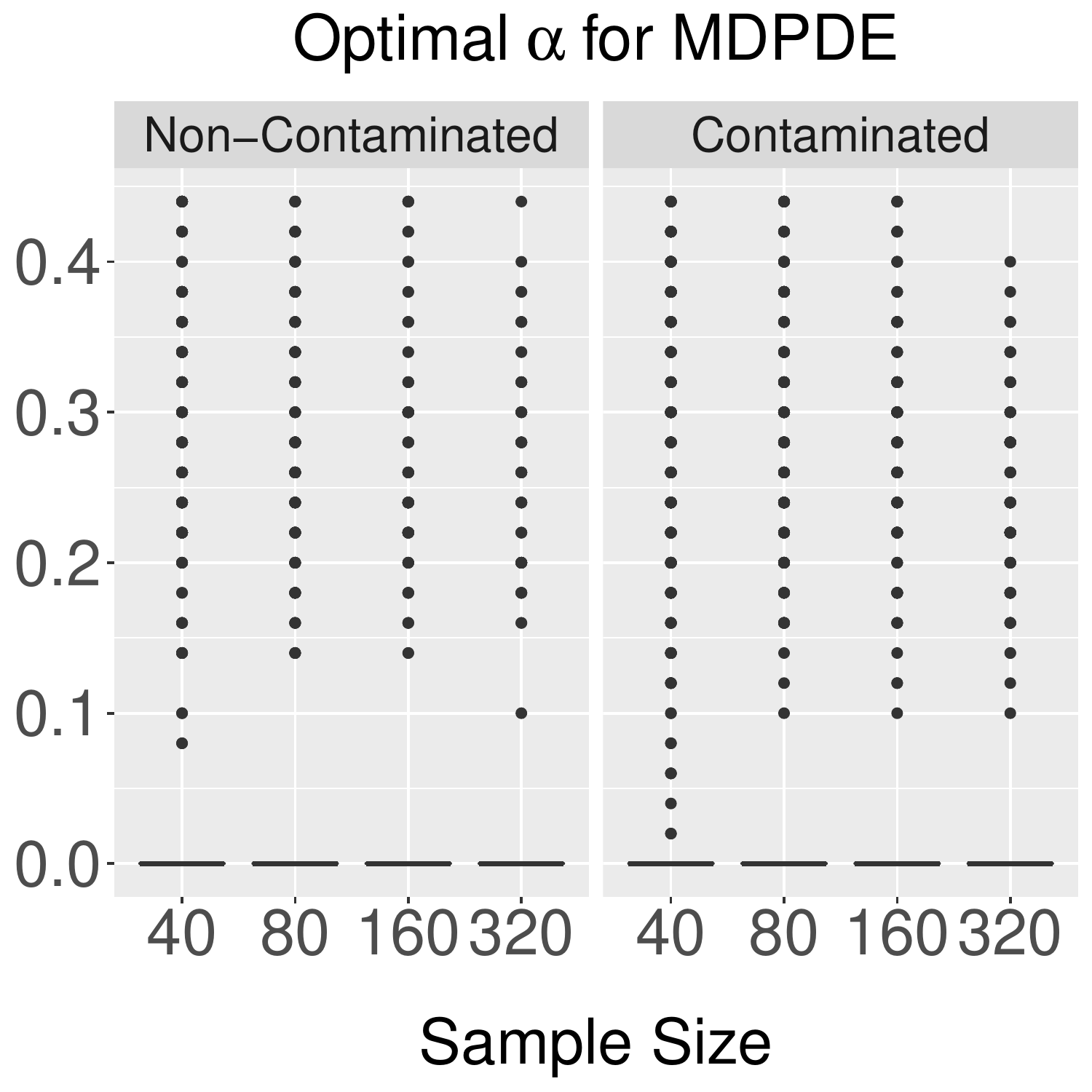}} 
 \subfloat{\includegraphics[width=0.25\textwidth]{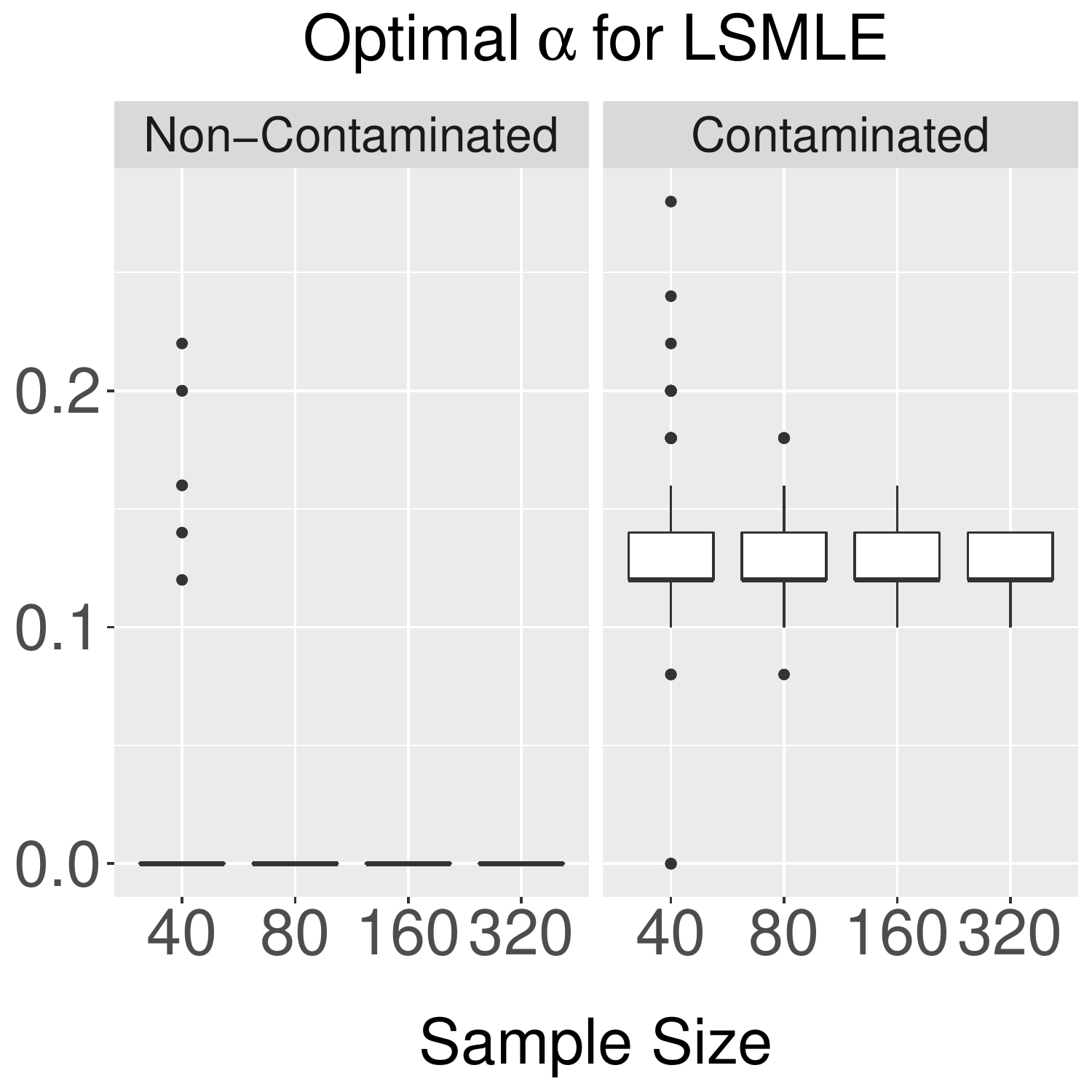}}
 \subfloat{\includegraphics[width=0.25\textwidth]{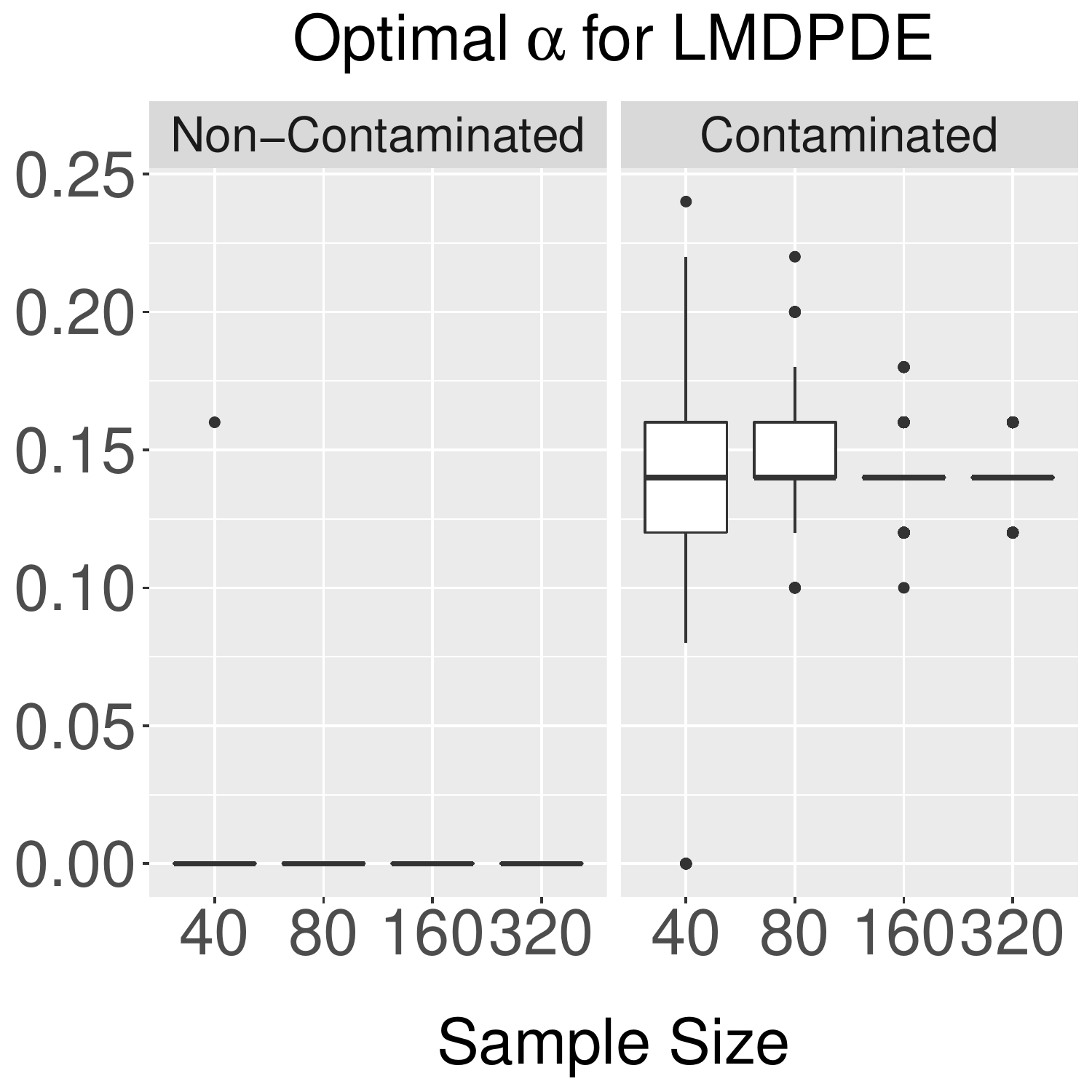}}
     \caption{Boxplots of the optimal values for the tuning parameter $\alpha$ for the robust estimators under Scenario A (first row), B (second row), and C (third row).}
     \label{Fig.BP-Tuning}	
\end{figure}

The data-driven algorithm to select the tuning constant $\alpha$ had an excellent performance for the LMDPDE and the LSMLE (see Figure \ref{Fig.BP-Tuning}). The selected optimum $\alpha$ is zero for the LMDPDE and the LSMLE for non-contaminated data, except for a few samples when $n=40$. Recall that $\alpha=0$ corresponds to the MLE. For the contaminated data, the selected optimum values of $\alpha$ are around $0.1$ for Scenarios A and B and around $0.15$ for Scenario C. Hence, the algorithm can identify the need to use a robust procedure. In Scenario A, the MDPDE and the SMLE behave like the LMDPDE and the LSMLE. On the other hand, for Scenarios B and C, the selected optimum $\alpha$ values for the MDPDE and the SMLE are close to zero for both non-contaminated and contaminated data. This happens because the algorithm does not achieve stability under contaminated data for most of the samples.

We now report the empirical levels of the Wald test (that uses the MLE) and the robust Wald-type test based on the LMDPDE and the LSMLE. The considered nominal level is $5\%$. The null hypotheses considered for Scenarios A and B are $\text{H}_{0}^{1}:\beta_2=\beta_2^0$, $\text{H}_{0}^{2}:(\beta_1,\beta_2)=(\beta_{1}^0,\beta_{2}^0)$, and $\text{H}_{0}^{3}:(\beta_1,\beta_2,\gamma_1)=(\beta_{1}^0,\beta_{2}^0,\gamma_{1}^0)$. For Scenario C, we set $\text{H}_{0}^{4}:(\beta_2,\gamma_2)=(\beta_{2}^0,\gamma_{2}^0)$, $\text{H}_{0}^{5}:(\beta_1,\beta_2,\gamma_2)=(\beta_{1}^0,\beta_{2}^0,\gamma_{2}^0)$, and $\text{H}_{0}^{6}:\gamma_2=\gamma_{2}^0$. The values of the parameters fixed at the null hypotheses are those used in the simulations above. The results are shown in Table \ref{Tab.Emp-Level-5}. For non-contaminated data, the empirical levels of all the tests are close to the nominal levels. For contaminated data, the usual Wald test presents a type I error close to $100\%$, being highly unreliable. In contrast, the robust Wald-type tests show to be reasonably reliable, with only slight inflation in the type I error relatively to non-contaminated situations.

Overall, our simulations suggest that the new robust estimators proposed in this paper, namely the LSMLE and the LMDPDE, exhibited the same performance as the SMLE and the MDPDE for bounded beta densities (Scenario A). For unbounded beta densities, as in Scenarios B and C, the MDPDE and the SMLE are unreliable, presenting severe bias for contaminated data. For all the scenarios, the new estimators behave as the MLE for non-contaminated data and prove to be robust in the presence of contamination. Hence, practitioners should employ the new proposed estimators in real data applications.

\afterpage{

\begin{landscape}
\begin{figure}[!h]
\captionsetup[subfigure]{labelformat=empty}
\centering
 \subfloat{\includegraphics[scale=0.32]{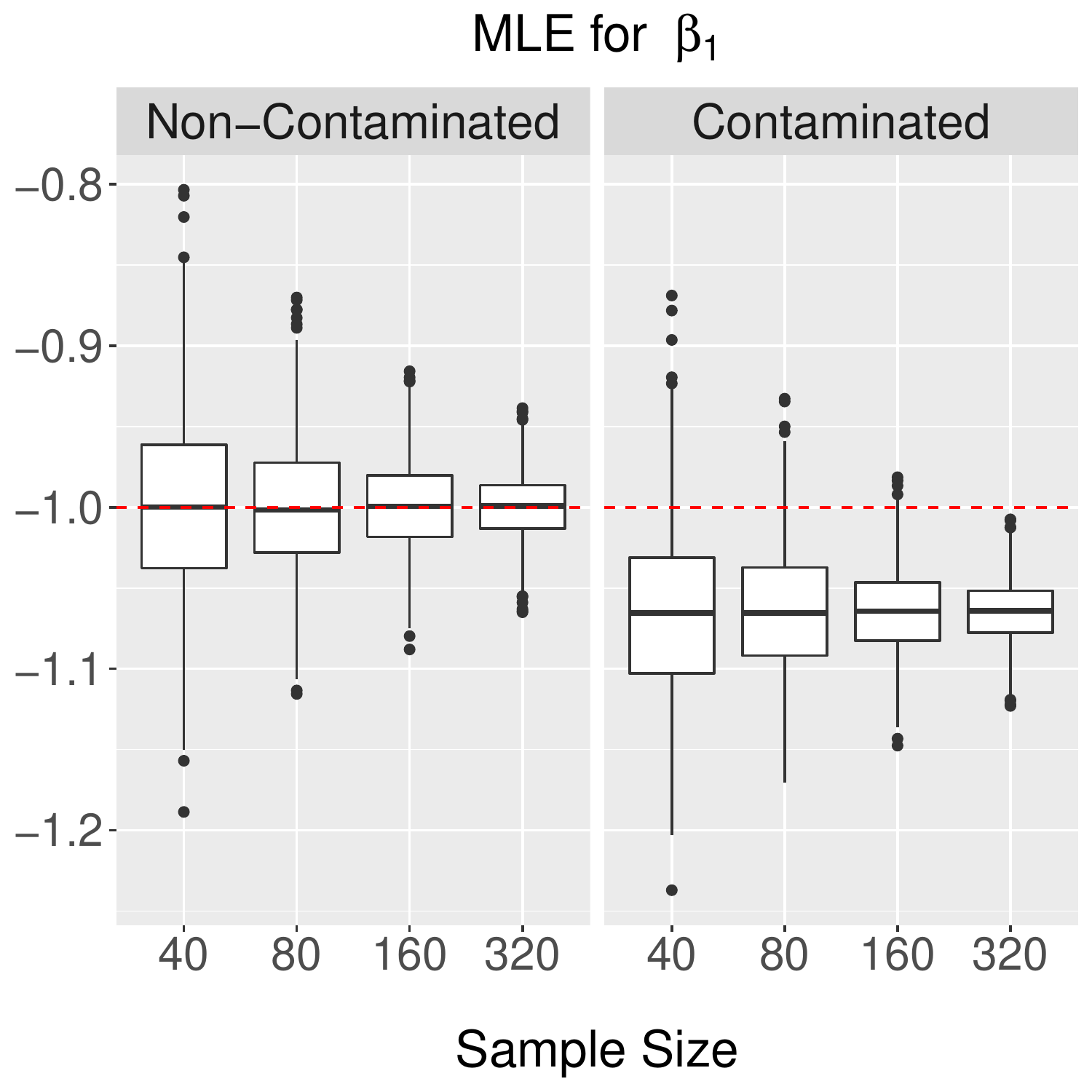}}
 \subfloat{\includegraphics[scale=0.32]{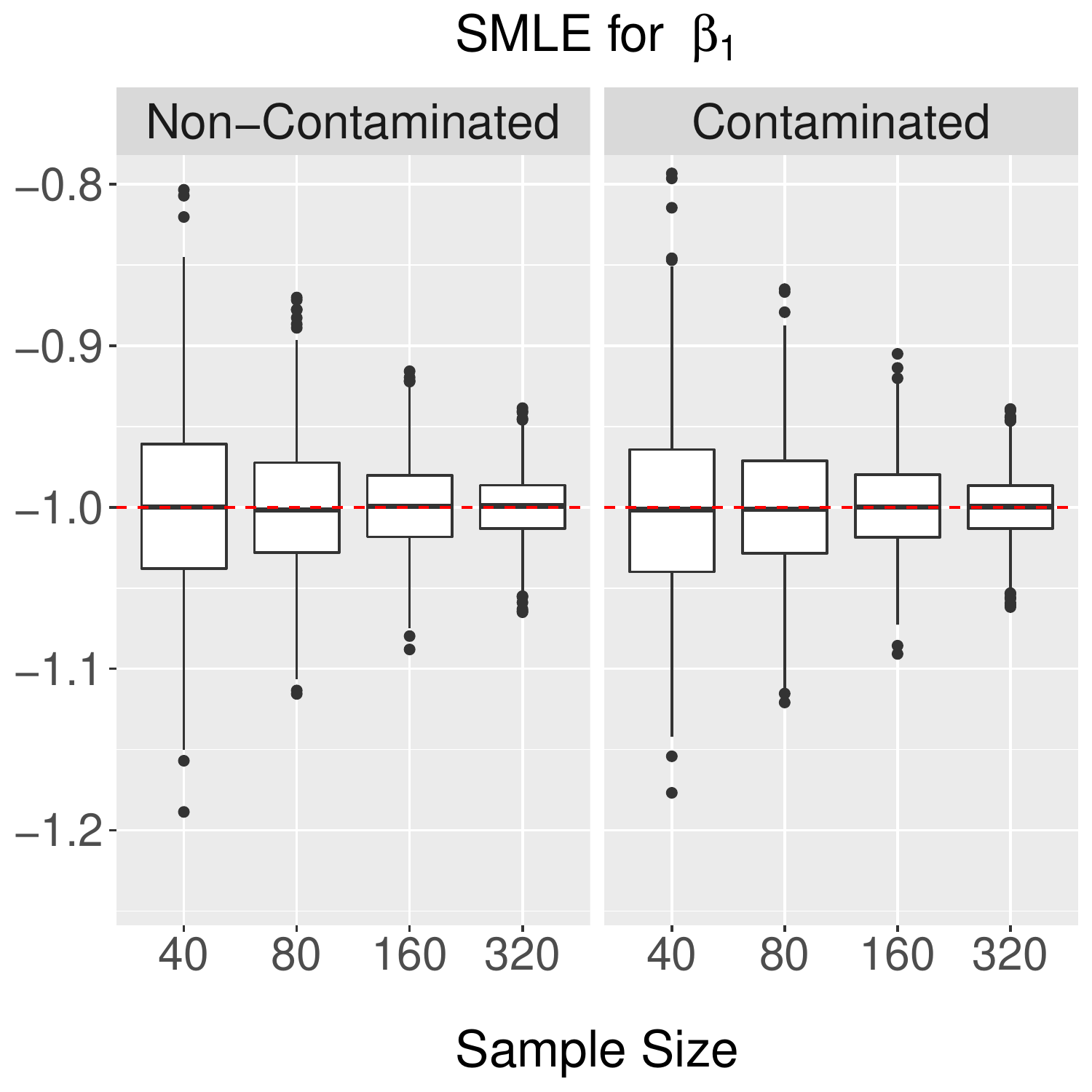}}
 \subfloat{\includegraphics[scale=0.32]{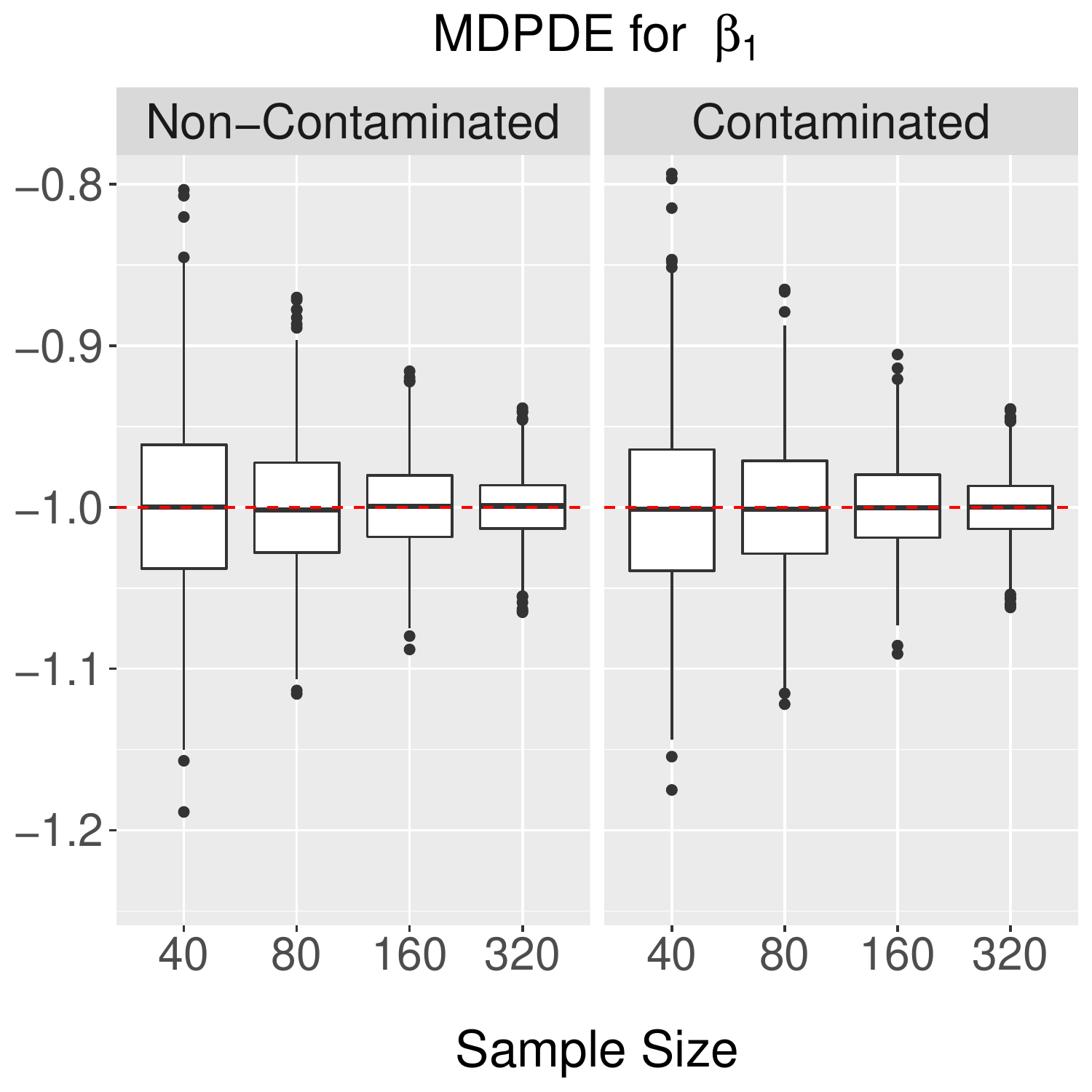}}
 \subfloat{\includegraphics[scale=0.32]{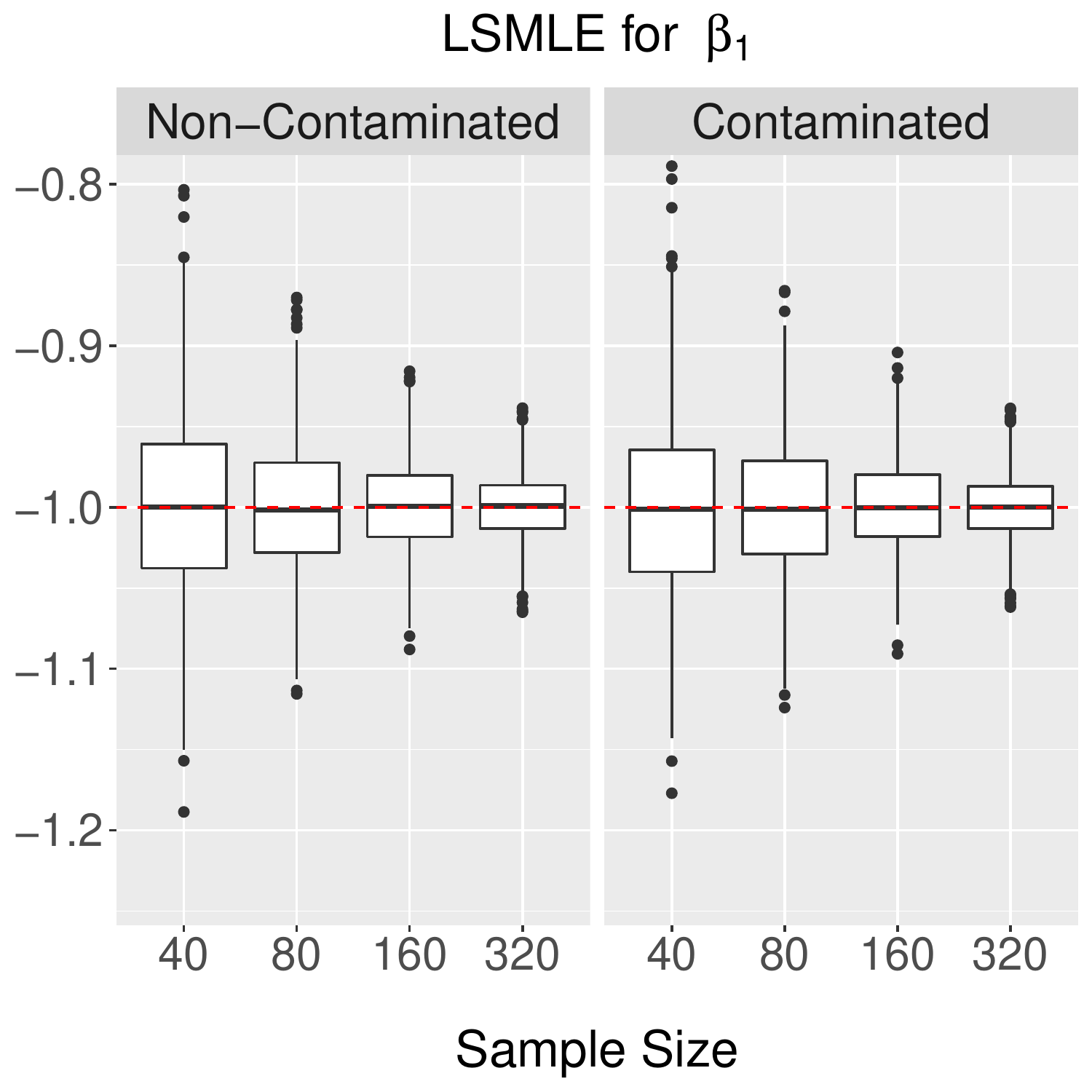}}
 \subfloat{\includegraphics[scale=0.32]{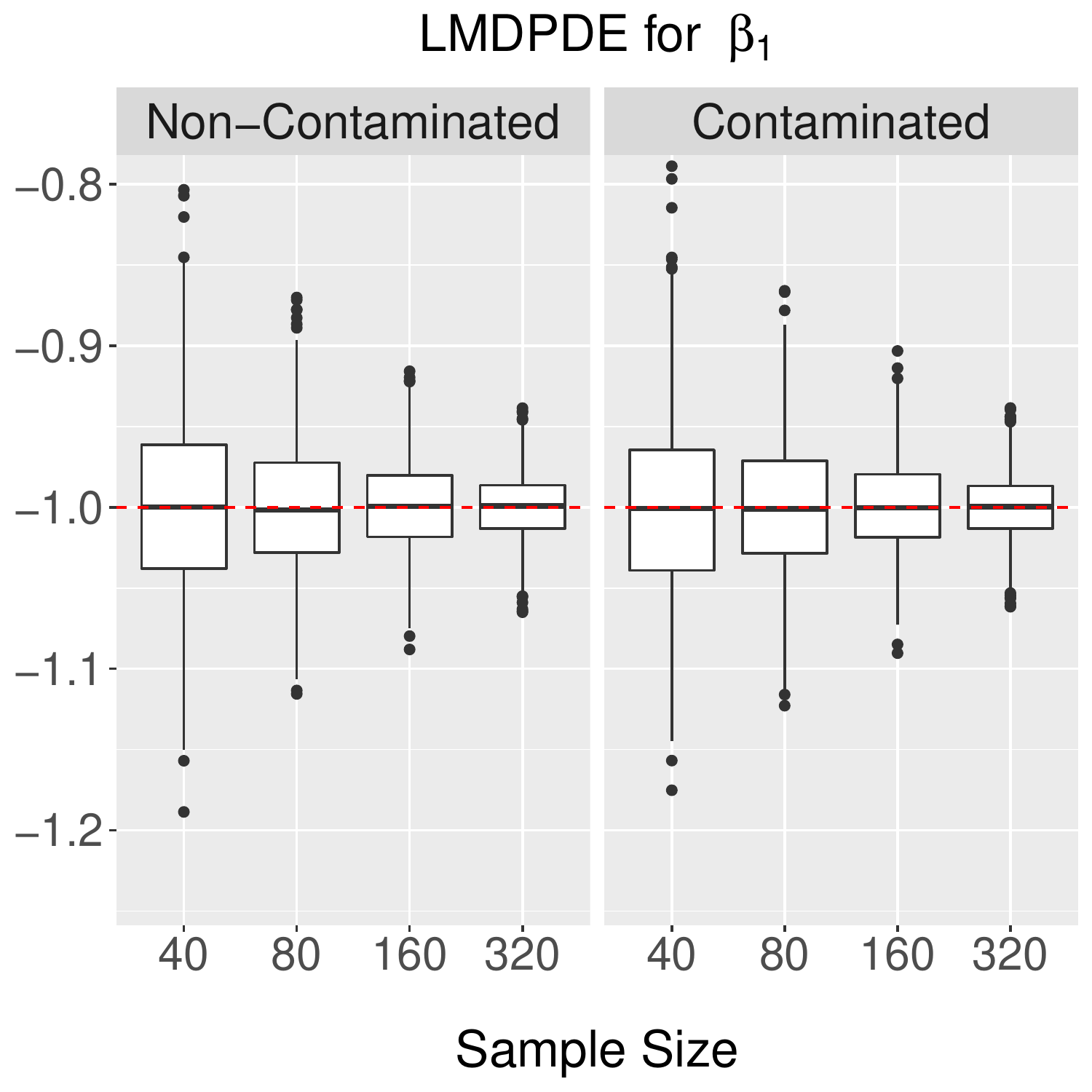}}\\
 \subfloat{\includegraphics[scale=0.32]{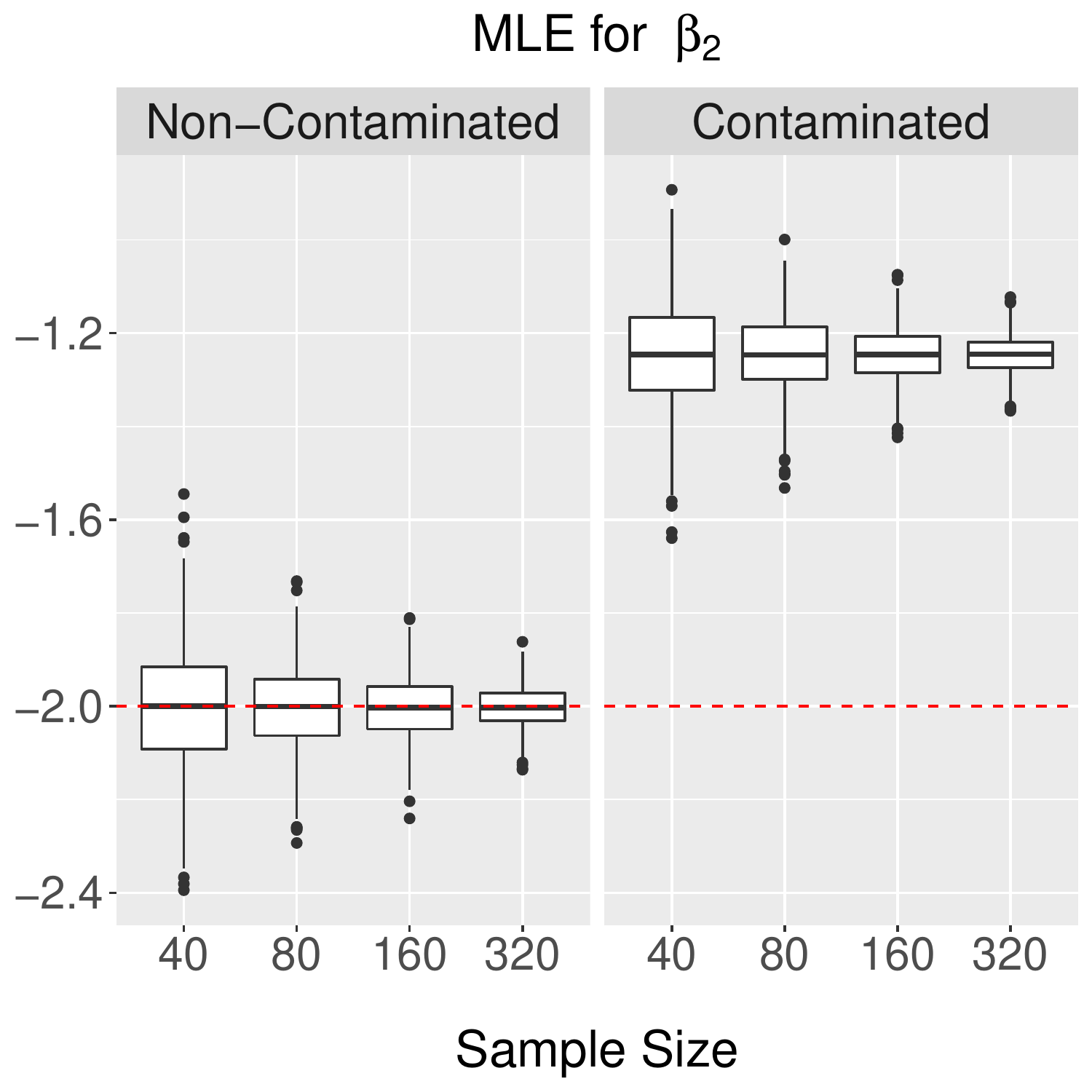}}
 \subfloat{\includegraphics[scale=0.32]{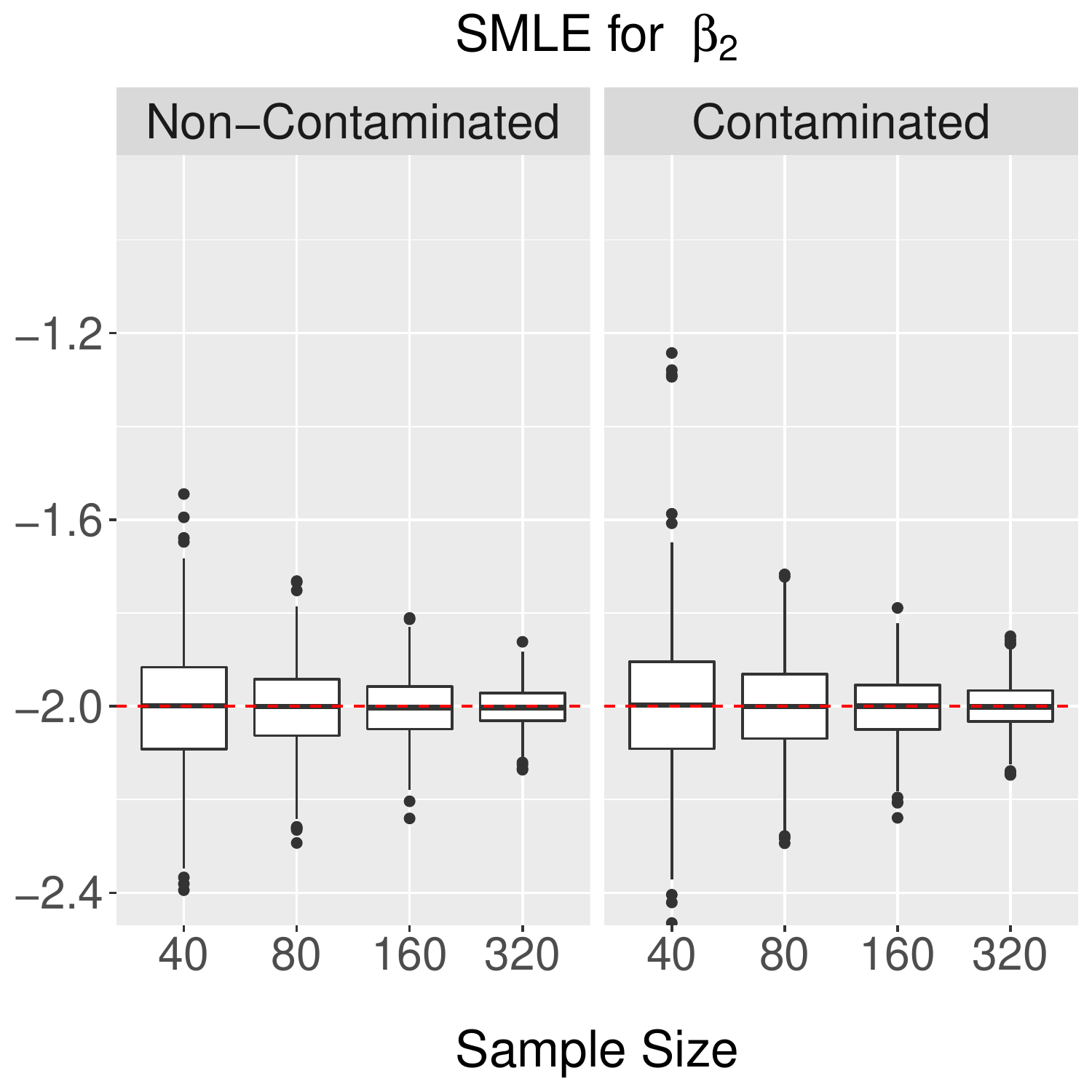}}
 \subfloat{\includegraphics[scale=0.32]{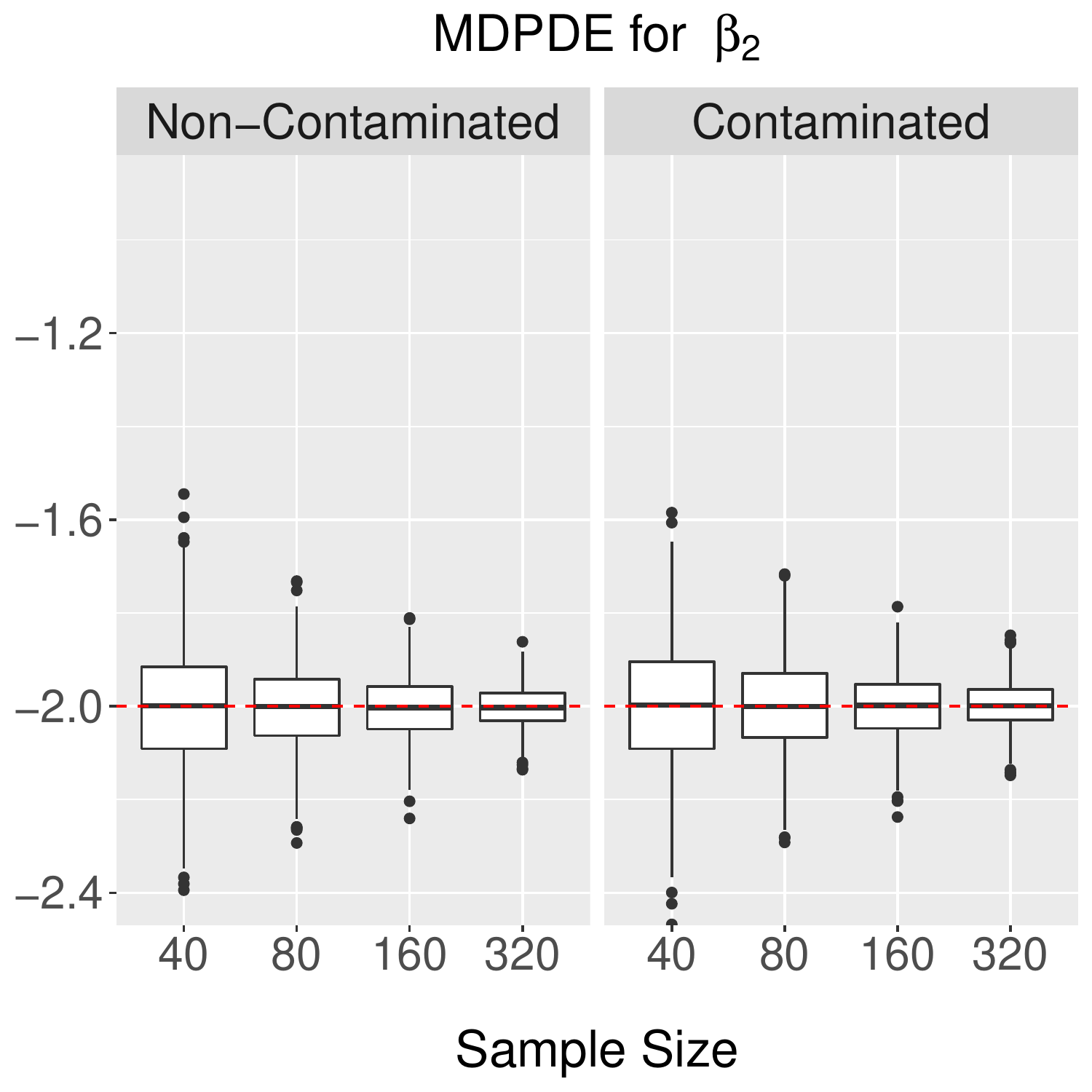}}
 \subfloat{\includegraphics[scale=0.32]{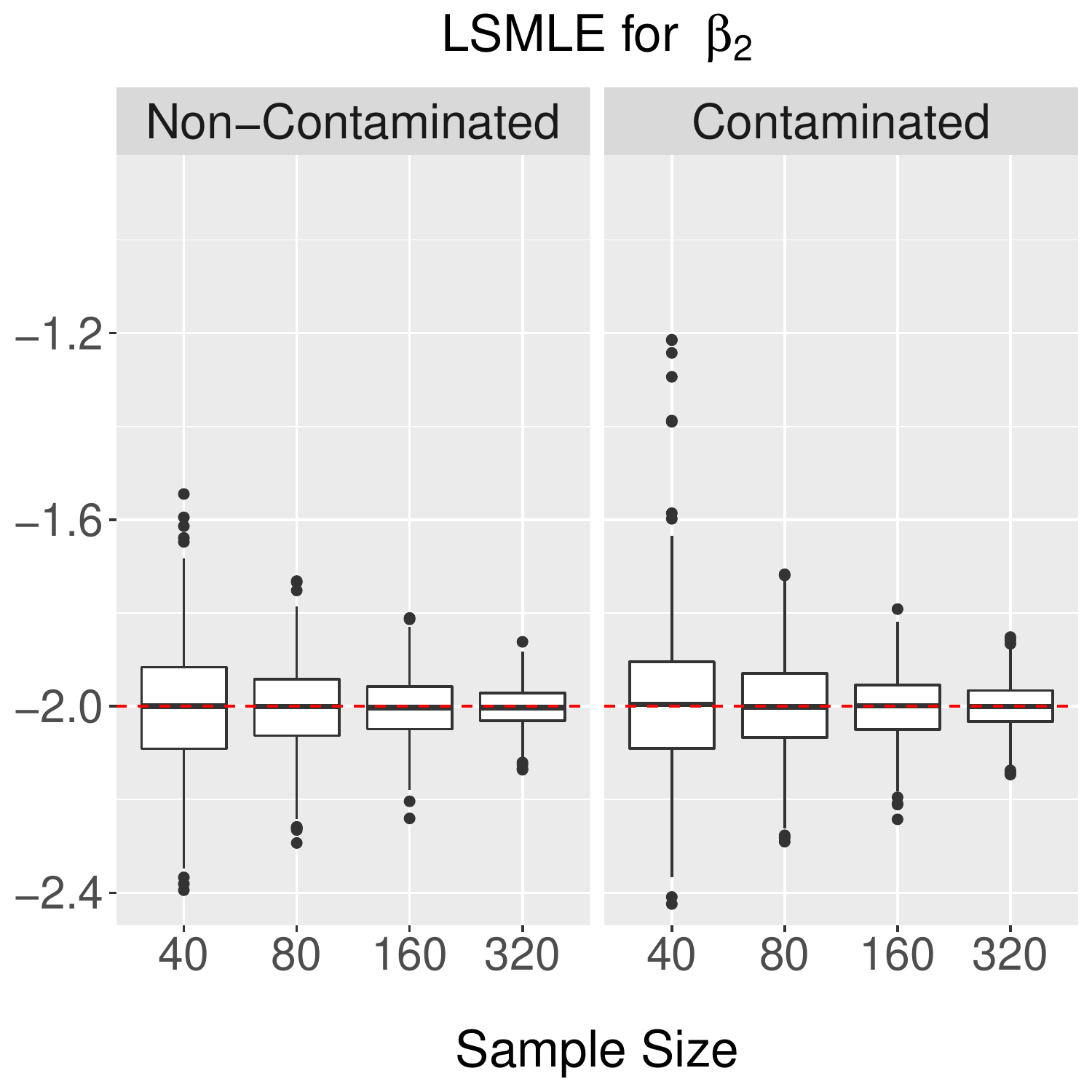}}
 \subfloat{\includegraphics[scale=0.32]{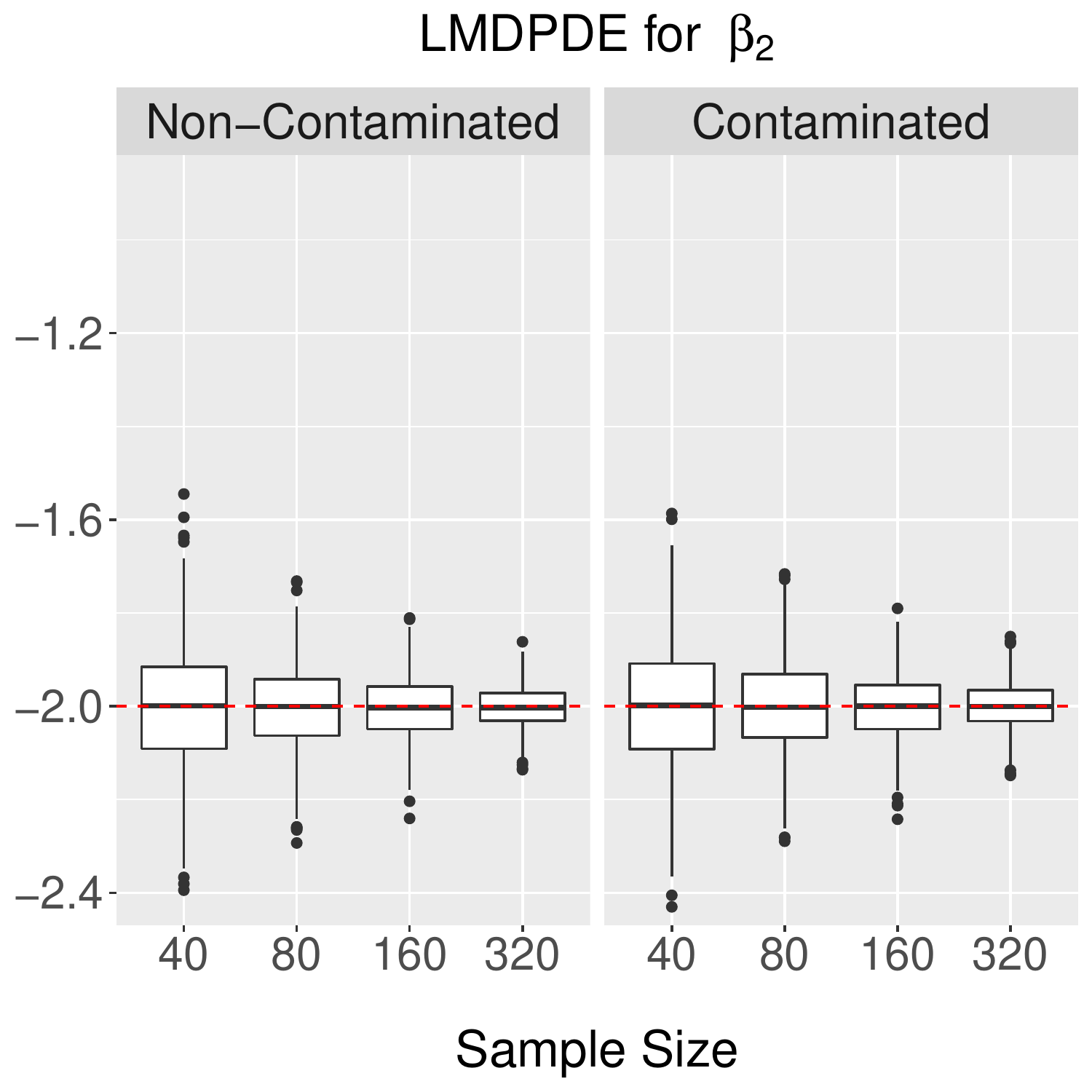}}\\
 \subfloat{\includegraphics[scale=0.32]{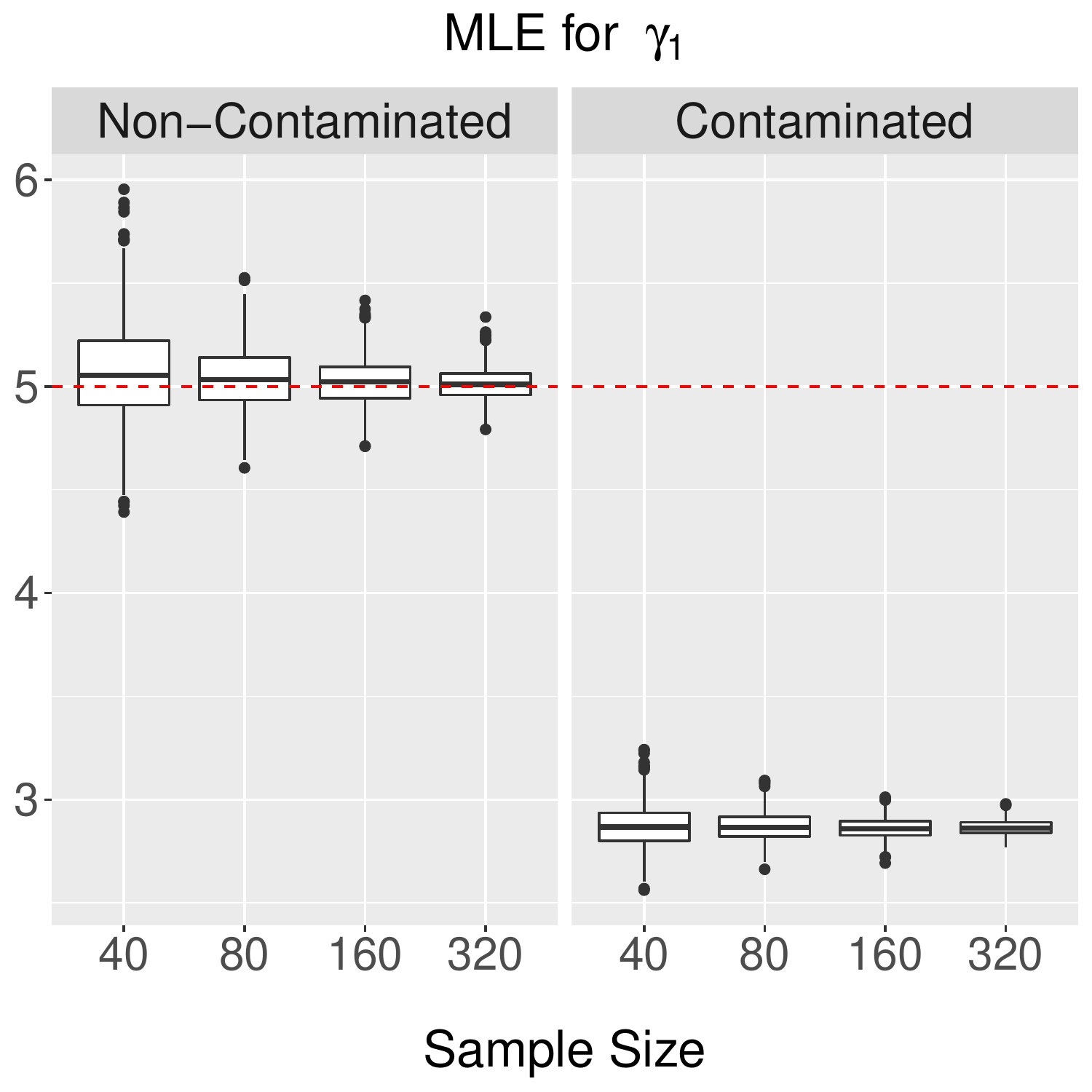}}
 \subfloat{\includegraphics[scale=0.32]{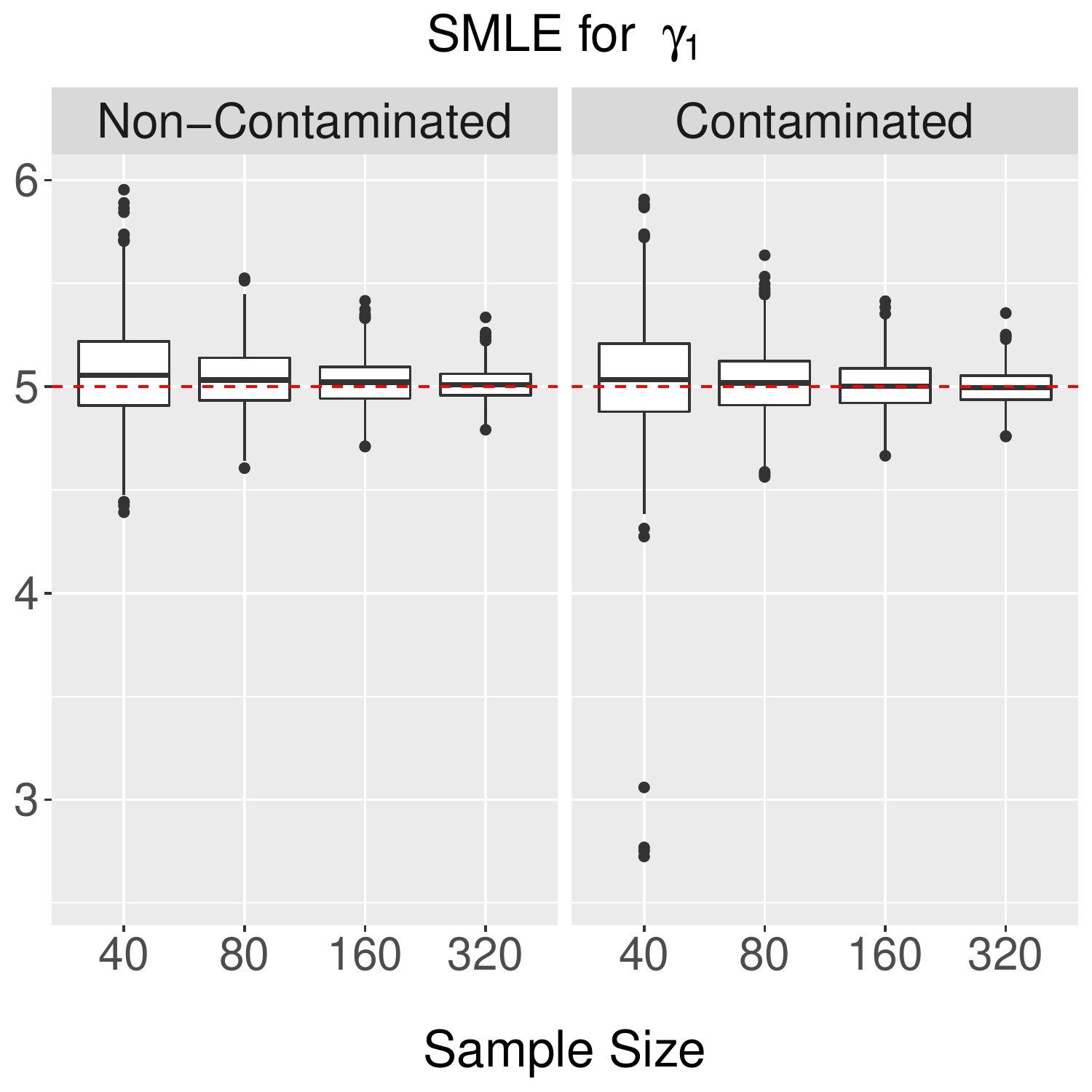}}
 \subfloat{\includegraphics[scale=0.32]{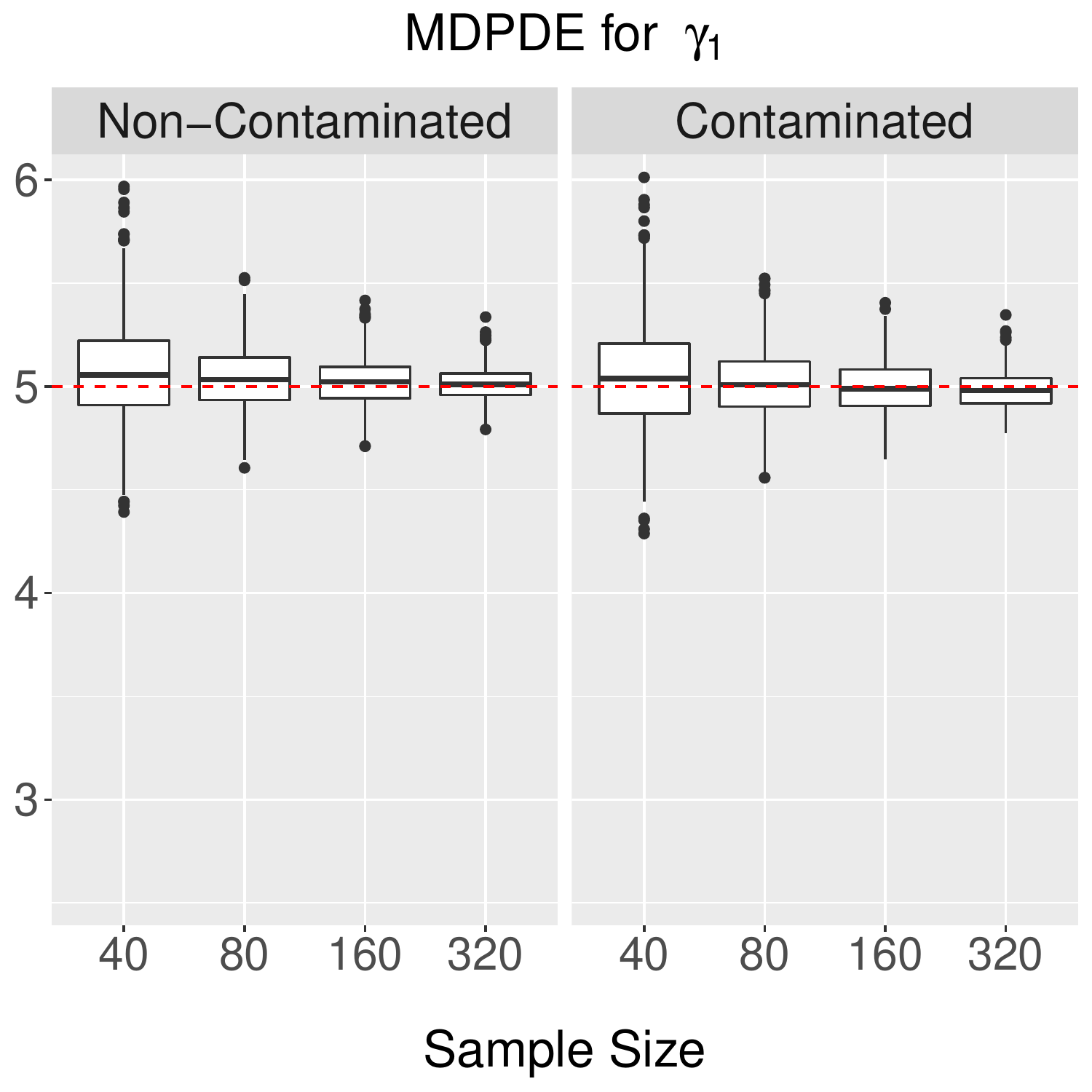}}
 \subfloat{\includegraphics[scale=0.32]{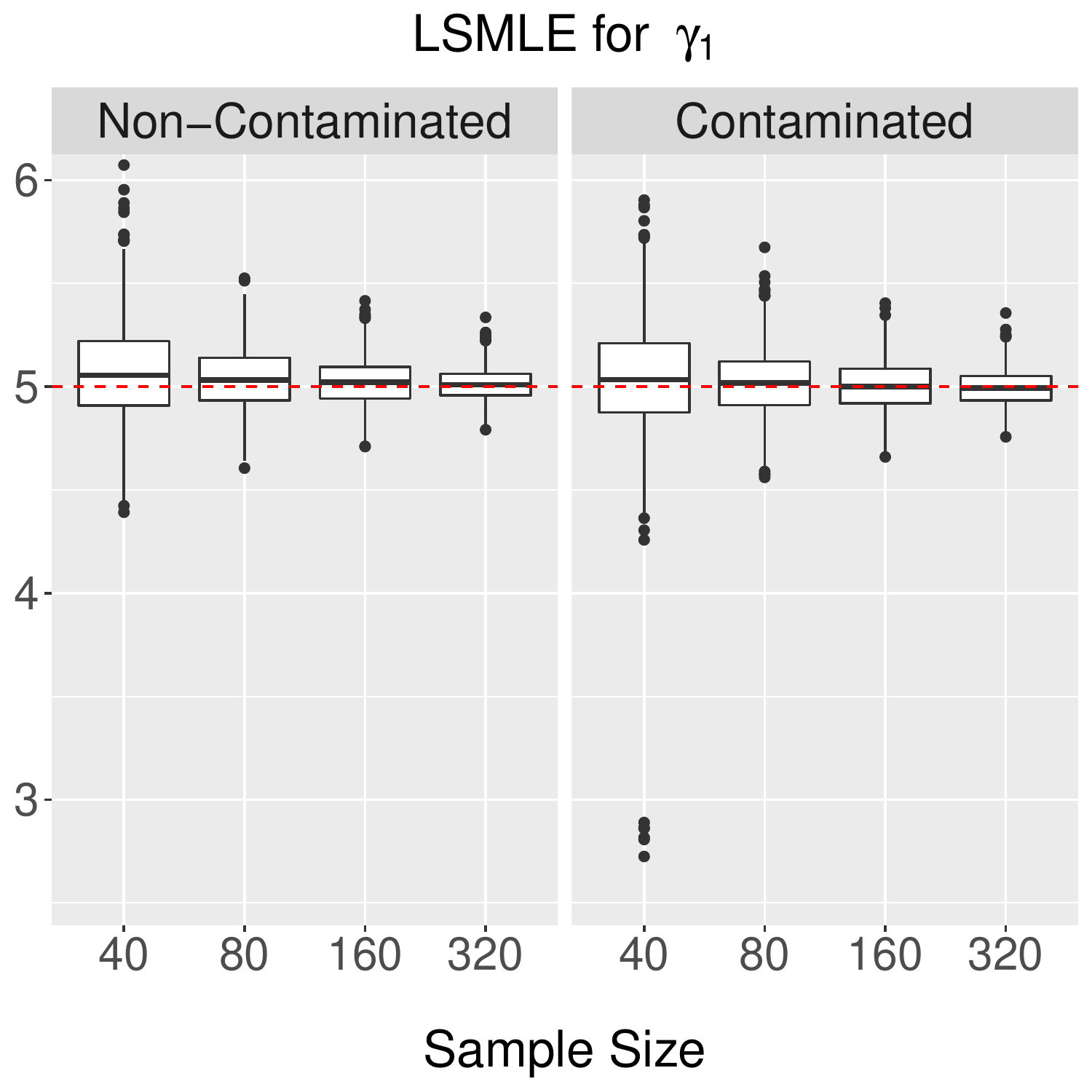}}
 \subfloat{\includegraphics[scale=0.32]{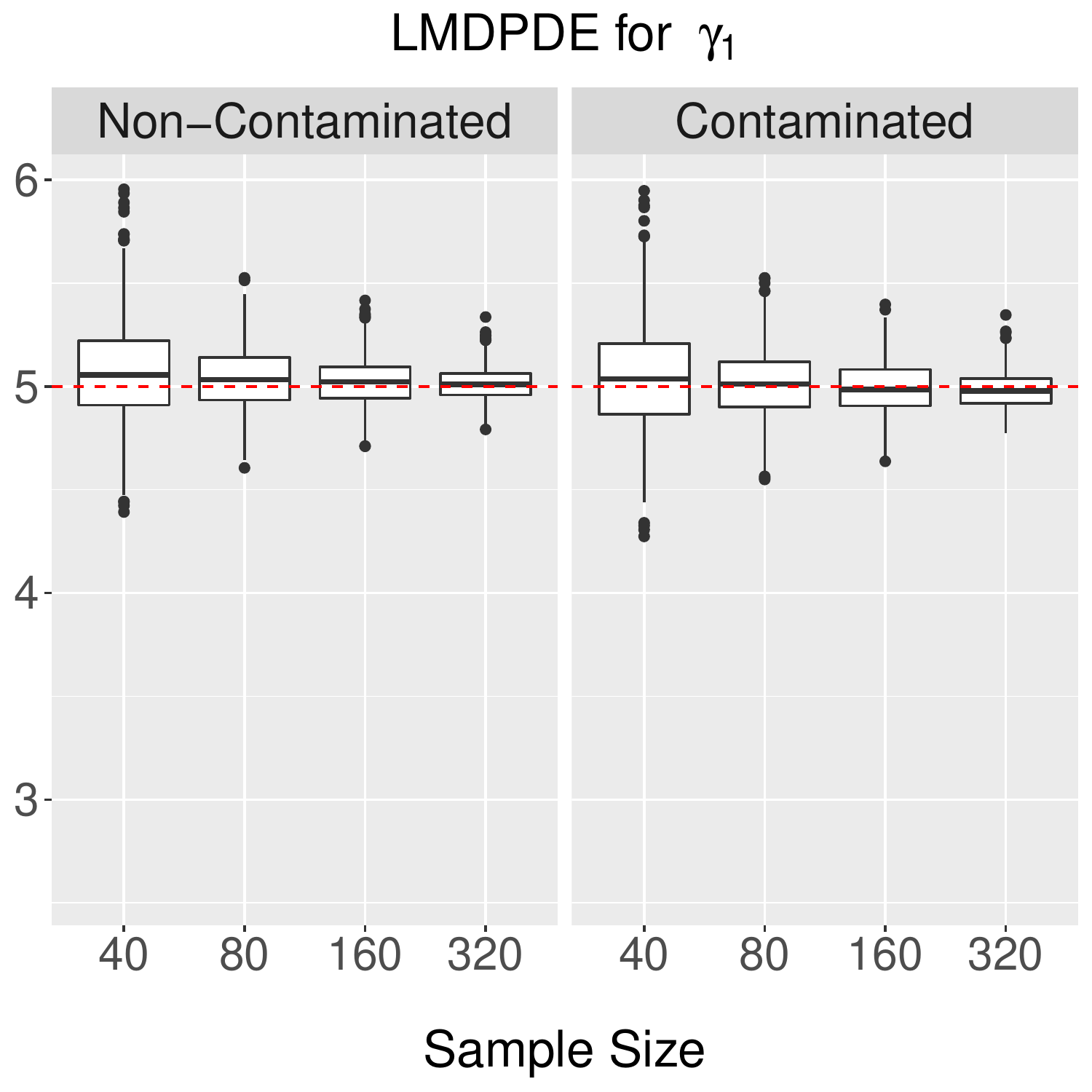}}\\
     \caption{Boxplots of estimates of $\beta_1$ (first row), $\beta_2$ (second row), and $\gamma_1$(third row) for the MLE and the robust estimators. The red dashed line represents the true parameter value.}
 \label{Fig.BPA}
\end{figure}
\begin{figure}[!h]
 \captionsetup[subfigure]{labelformat=empty}
 \centering
 \subfloat{\includegraphics[scale=0.32]{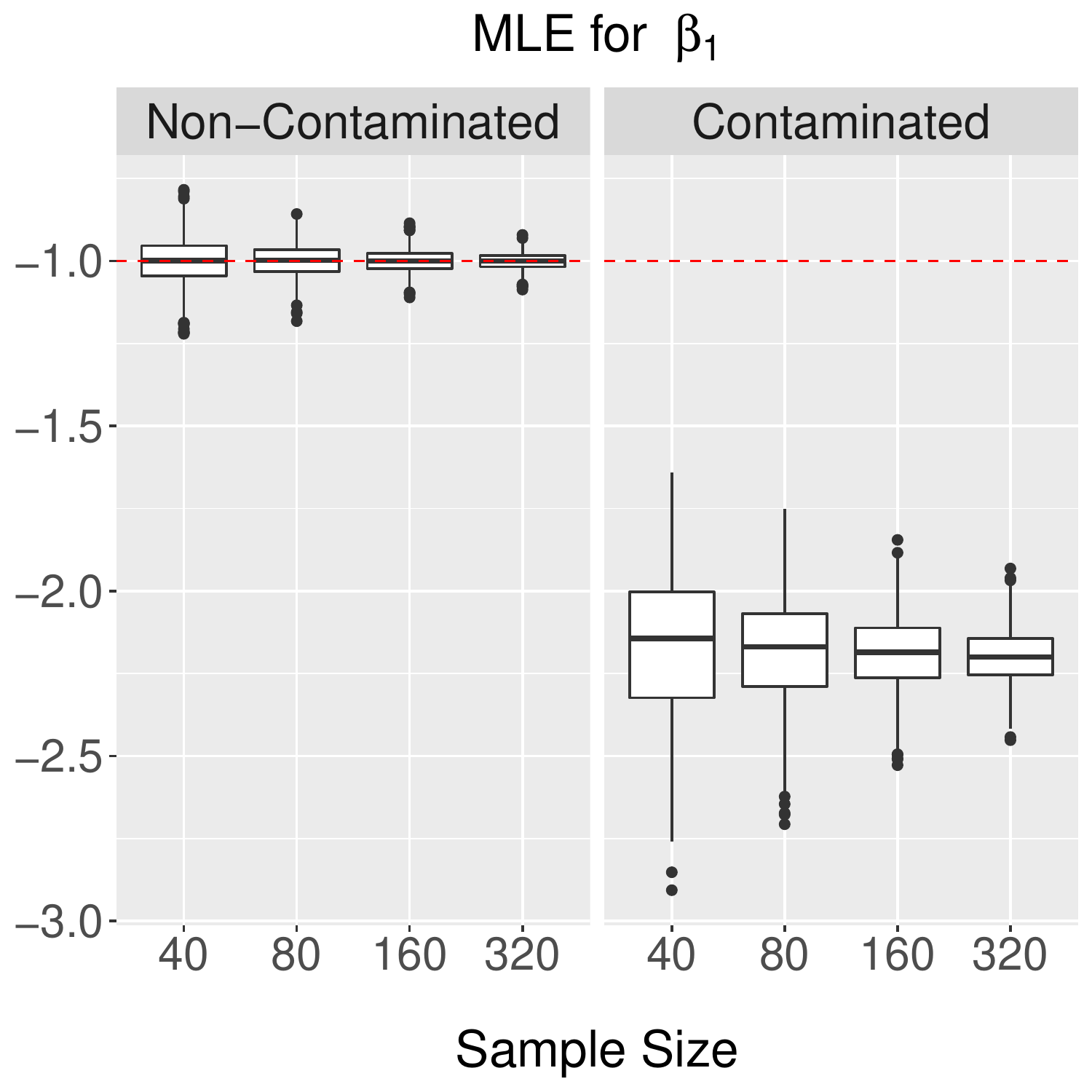}}
 \subfloat{\includegraphics[scale=0.32]{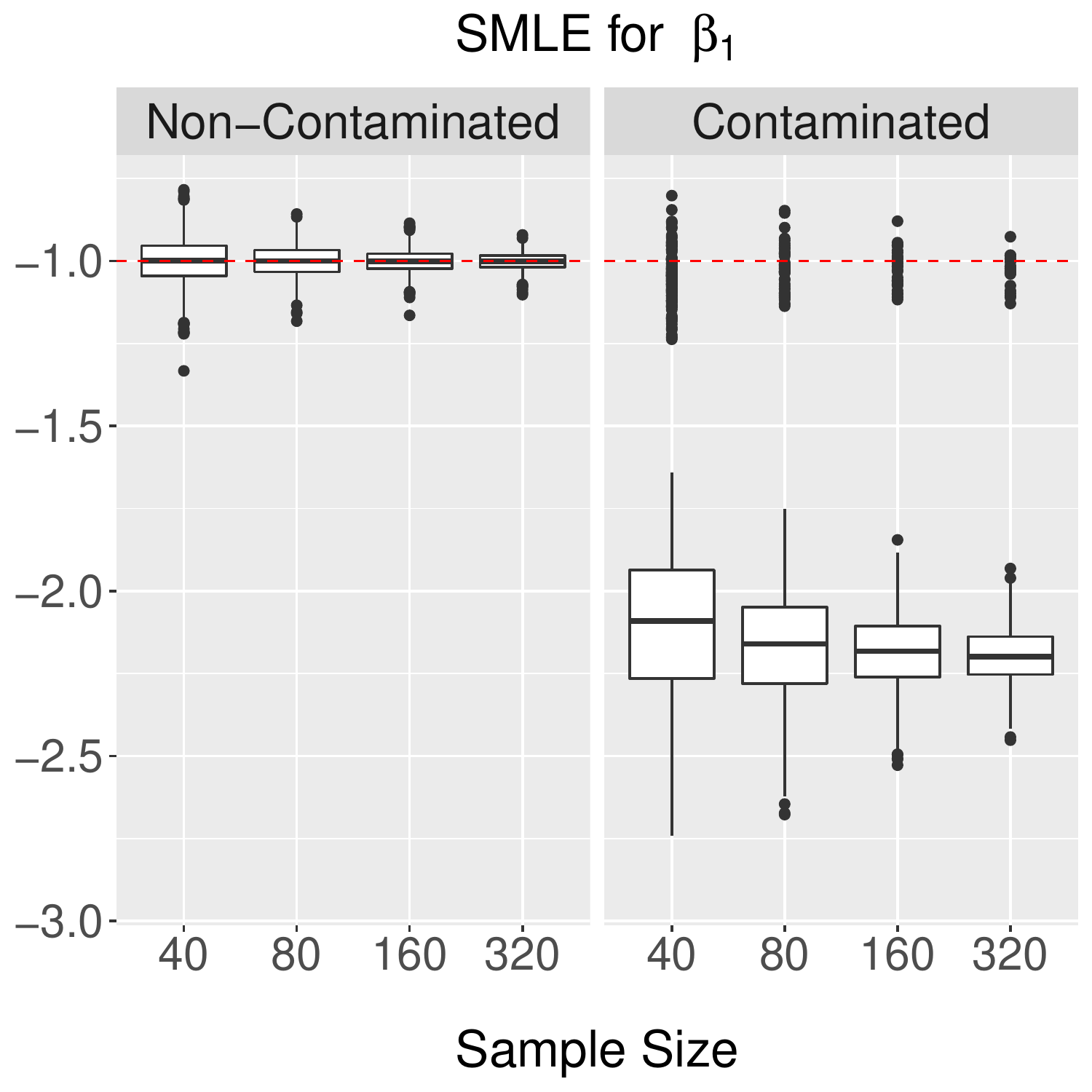}}
 \subfloat{\includegraphics[scale=0.32]{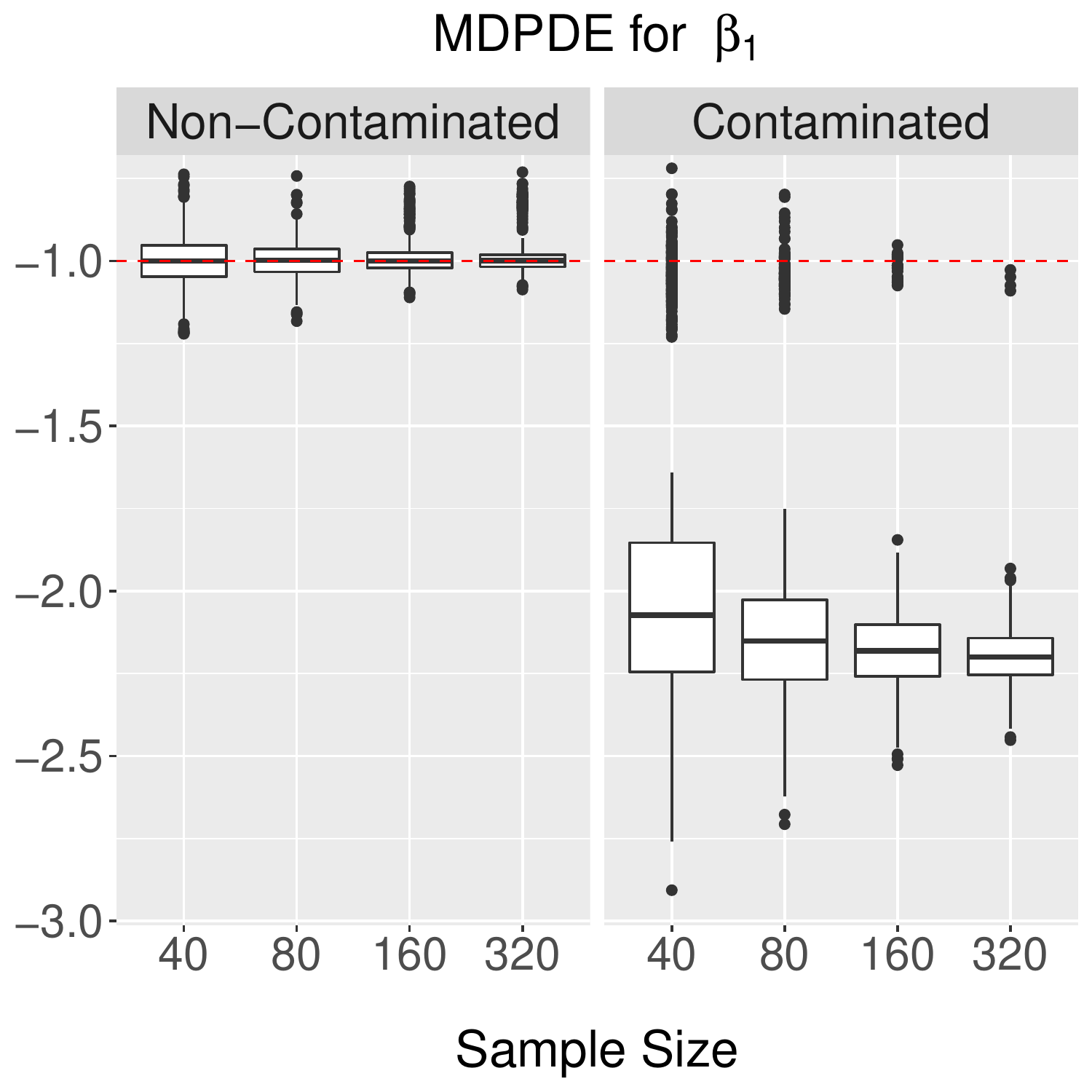}}
 \subfloat{\includegraphics[scale=0.32]{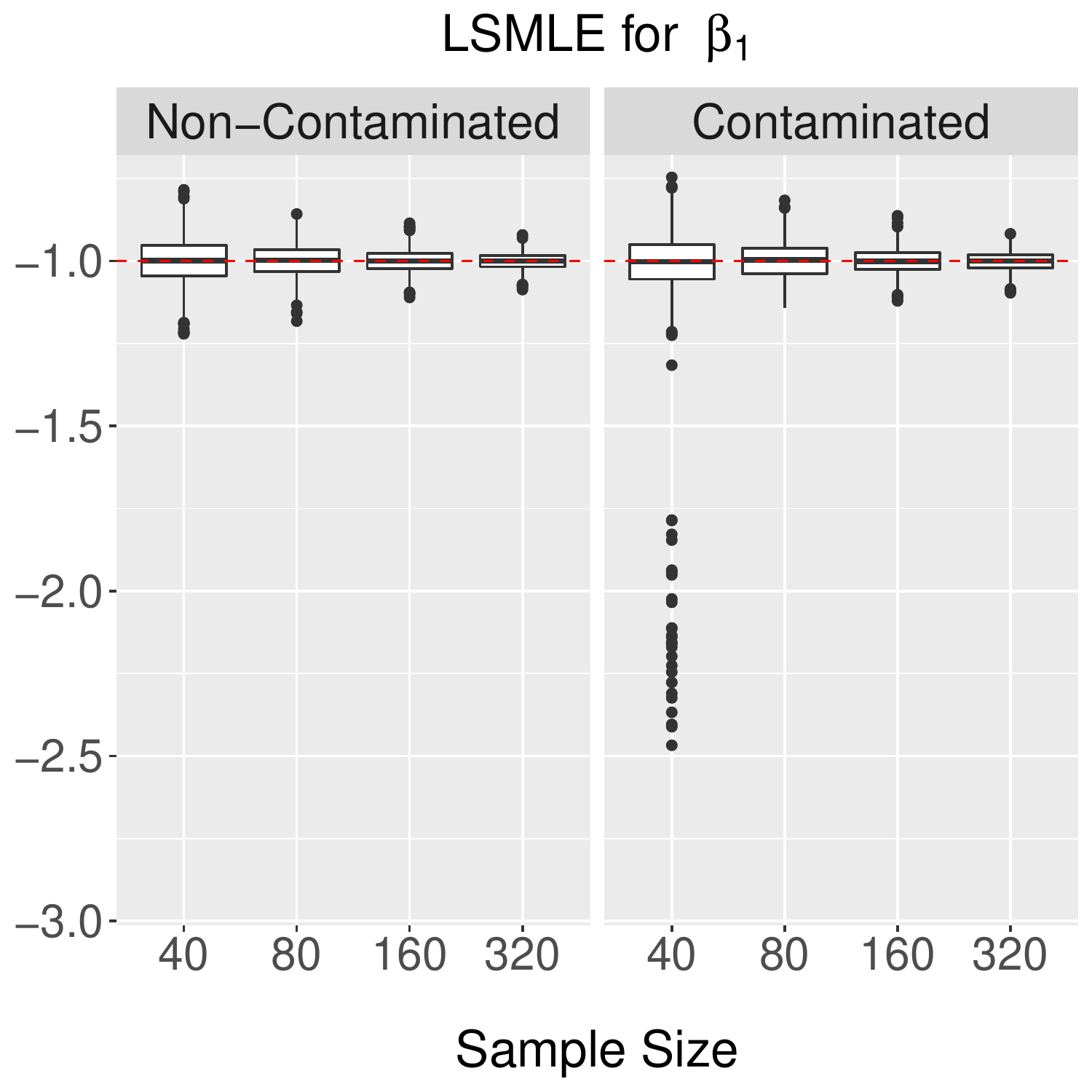}}
 \subfloat{\includegraphics[scale=0.32]{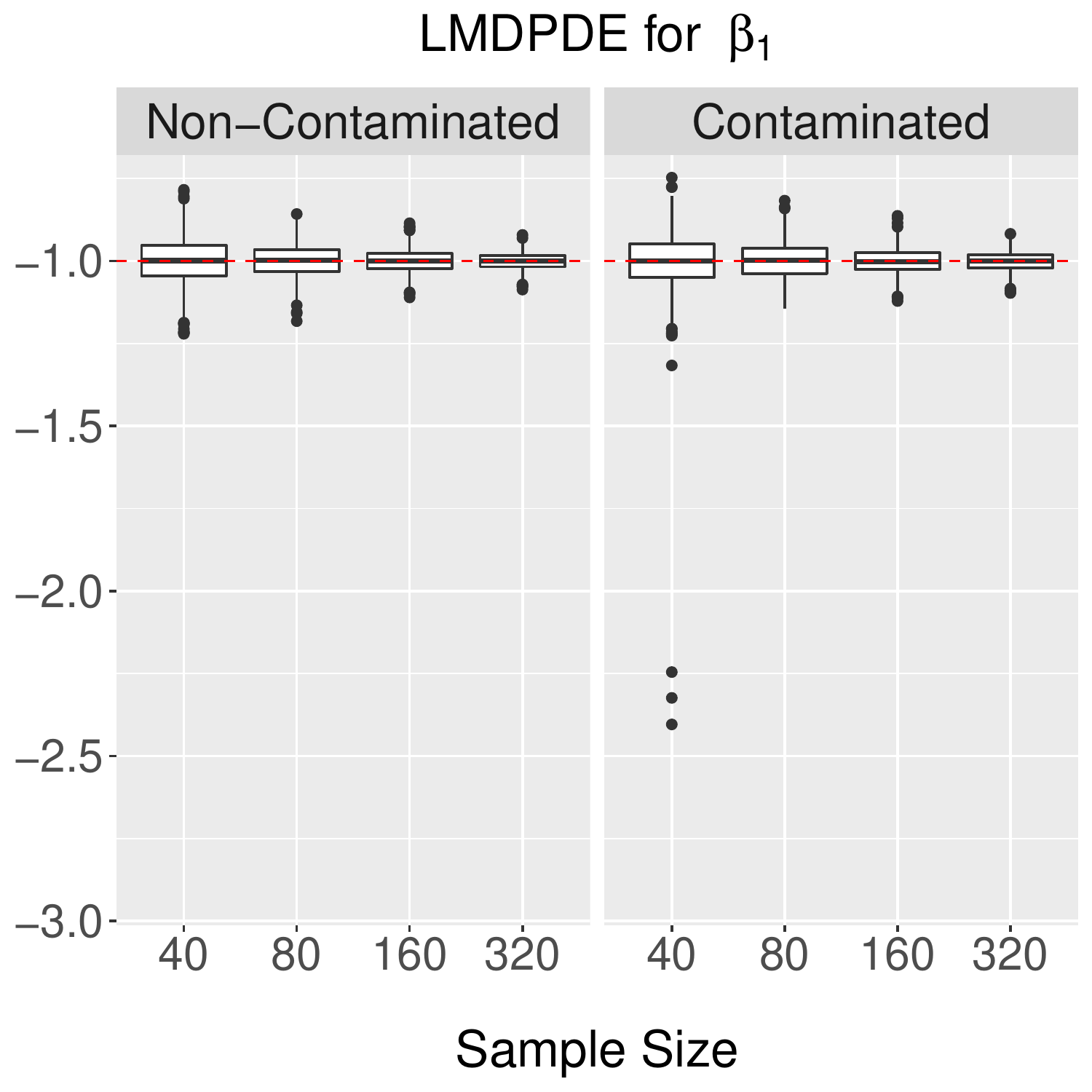}}\\
 \subfloat{\includegraphics[scale=0.32]{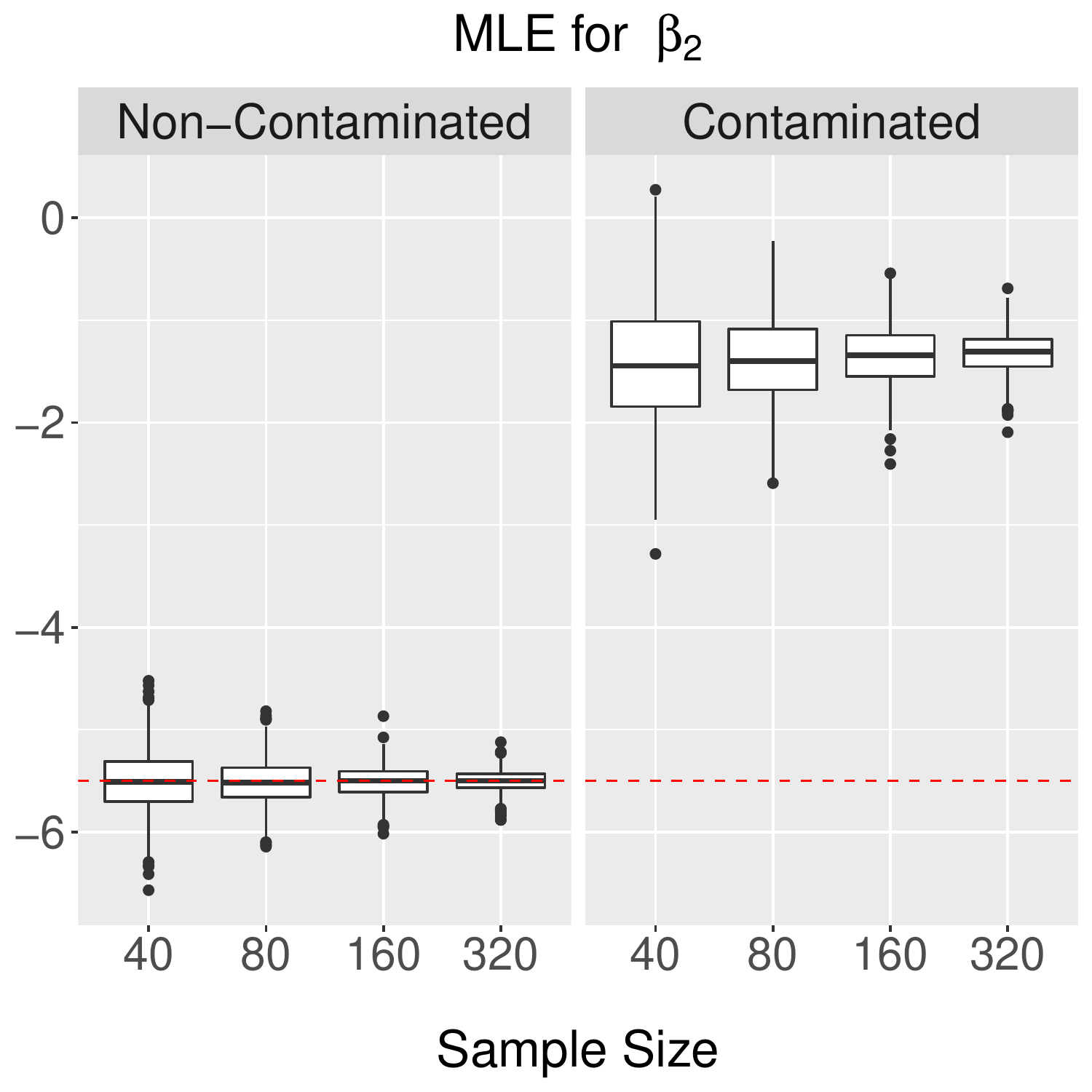}}
 \subfloat{\includegraphics[scale=0.32]{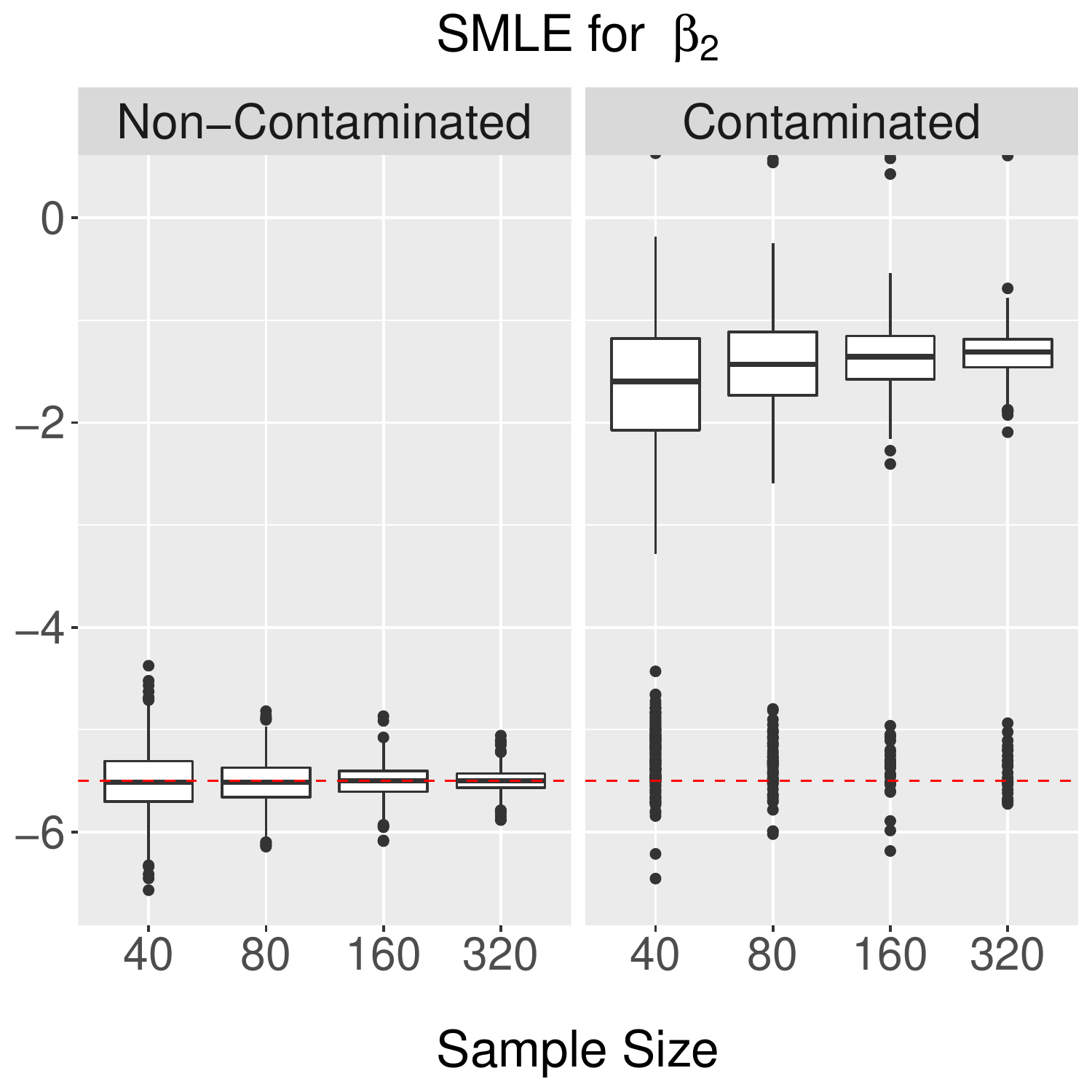}}
 \subfloat{\includegraphics[scale=0.32]{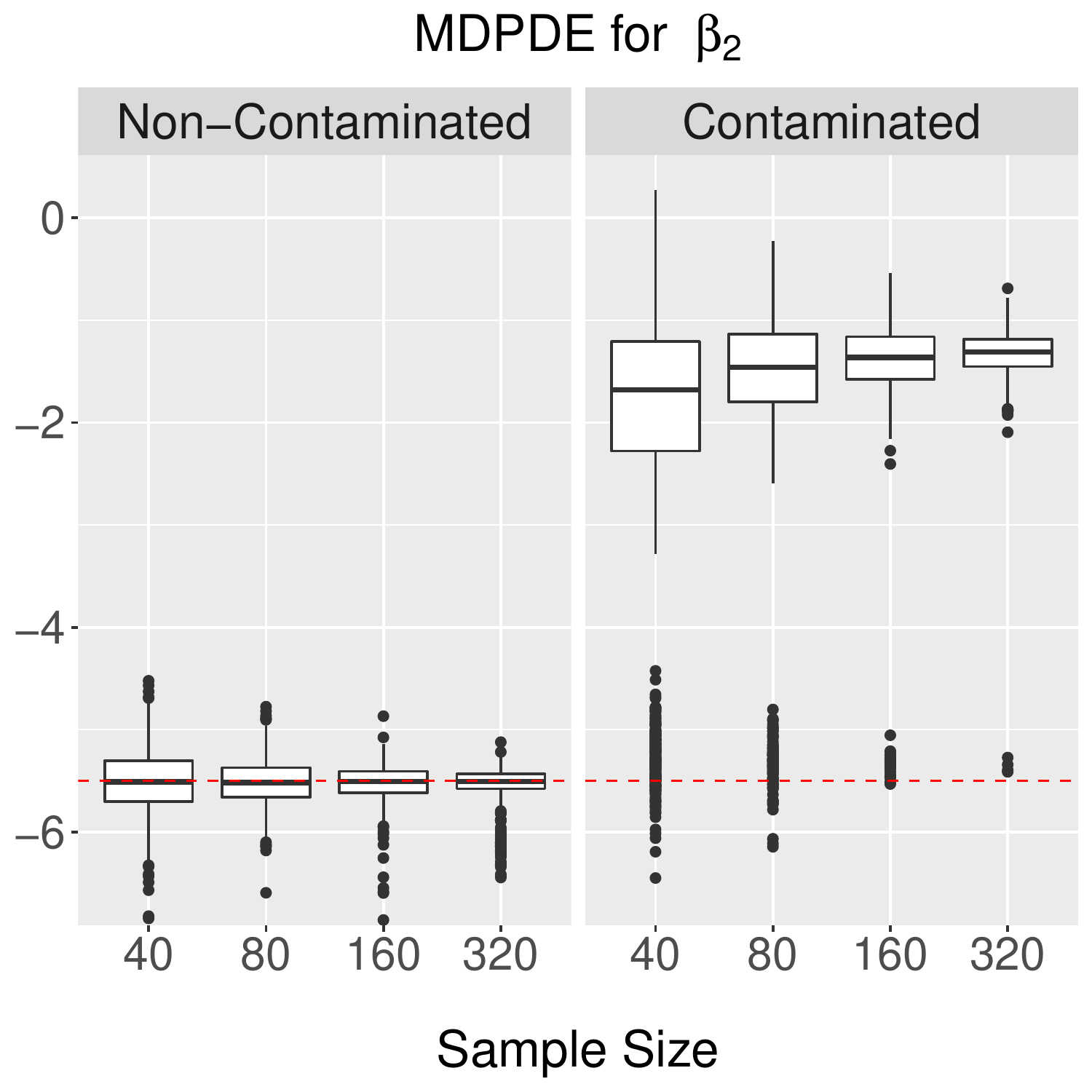}}
 \subfloat{\includegraphics[scale=0.32]{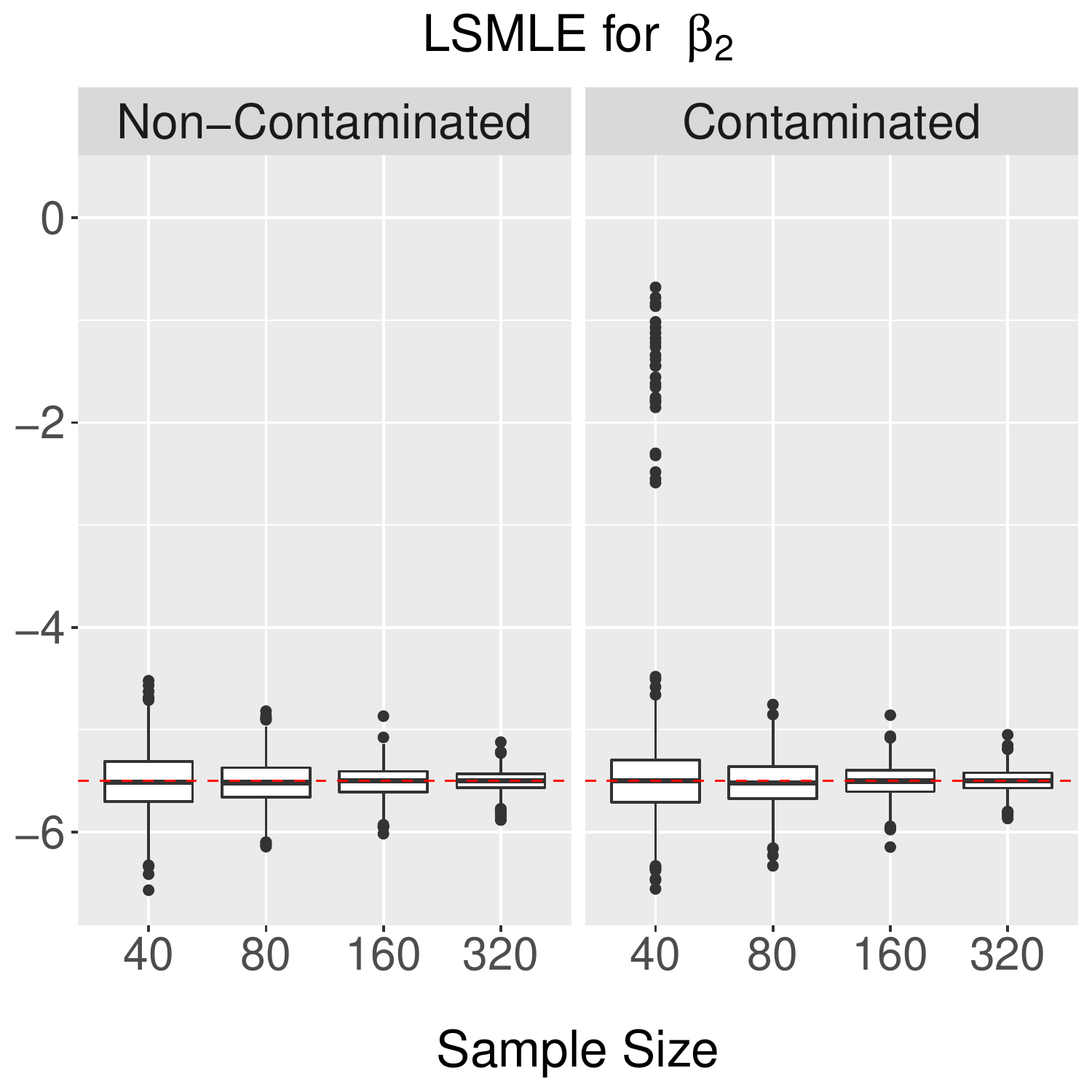}}
 \subfloat{\includegraphics[scale=0.32]{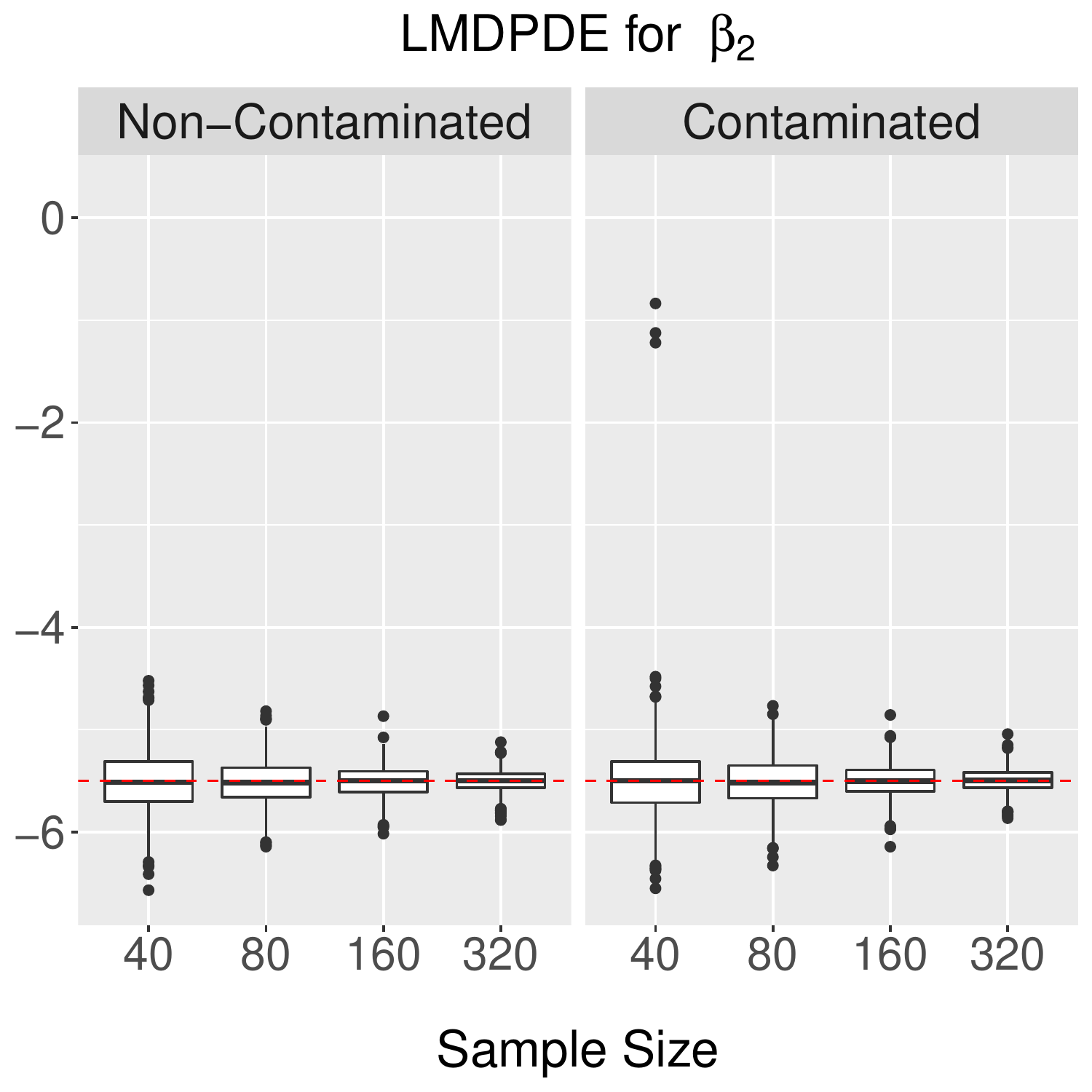}}\\
 \subfloat{\includegraphics[scale=0.32]{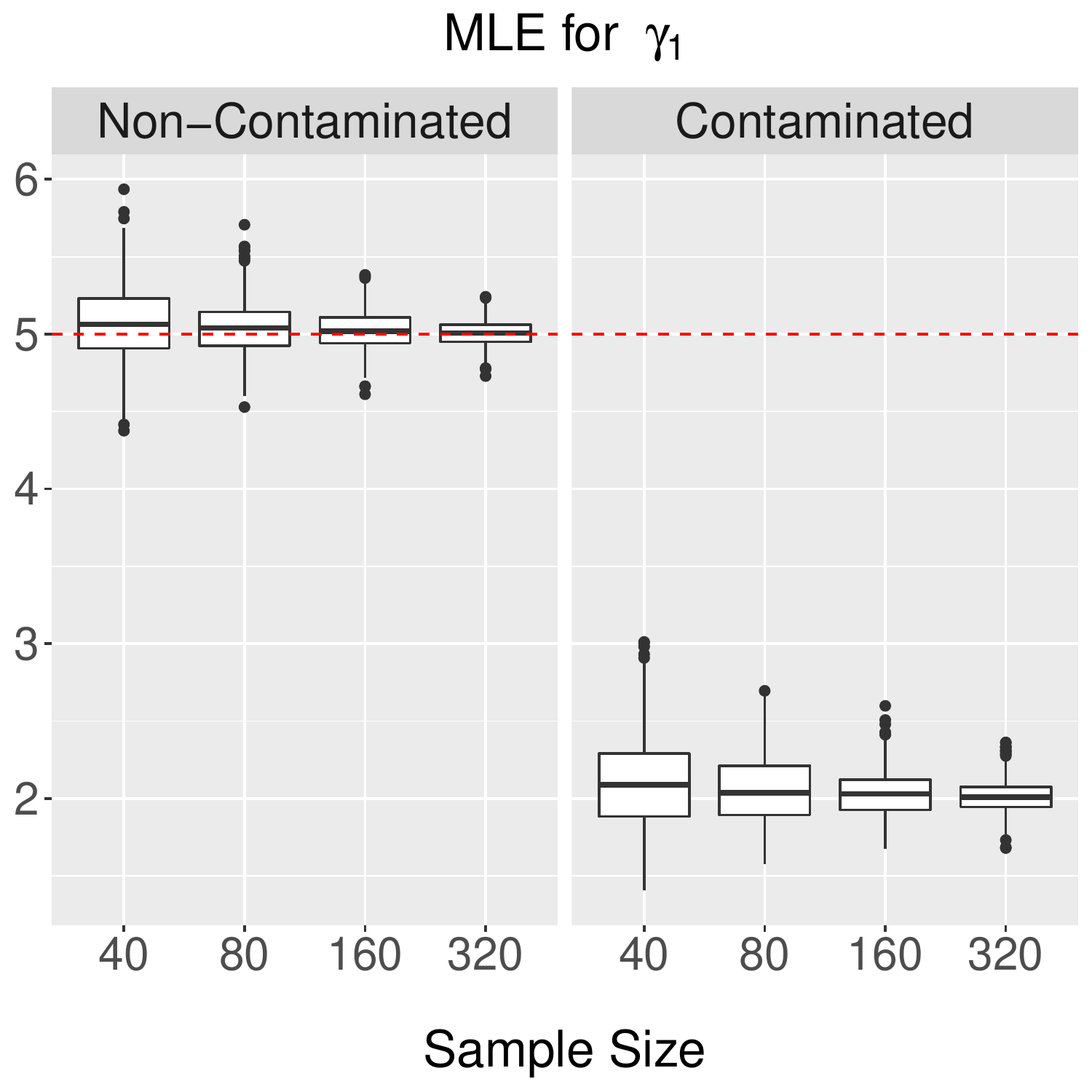}}
 \subfloat{\includegraphics[scale=0.32]{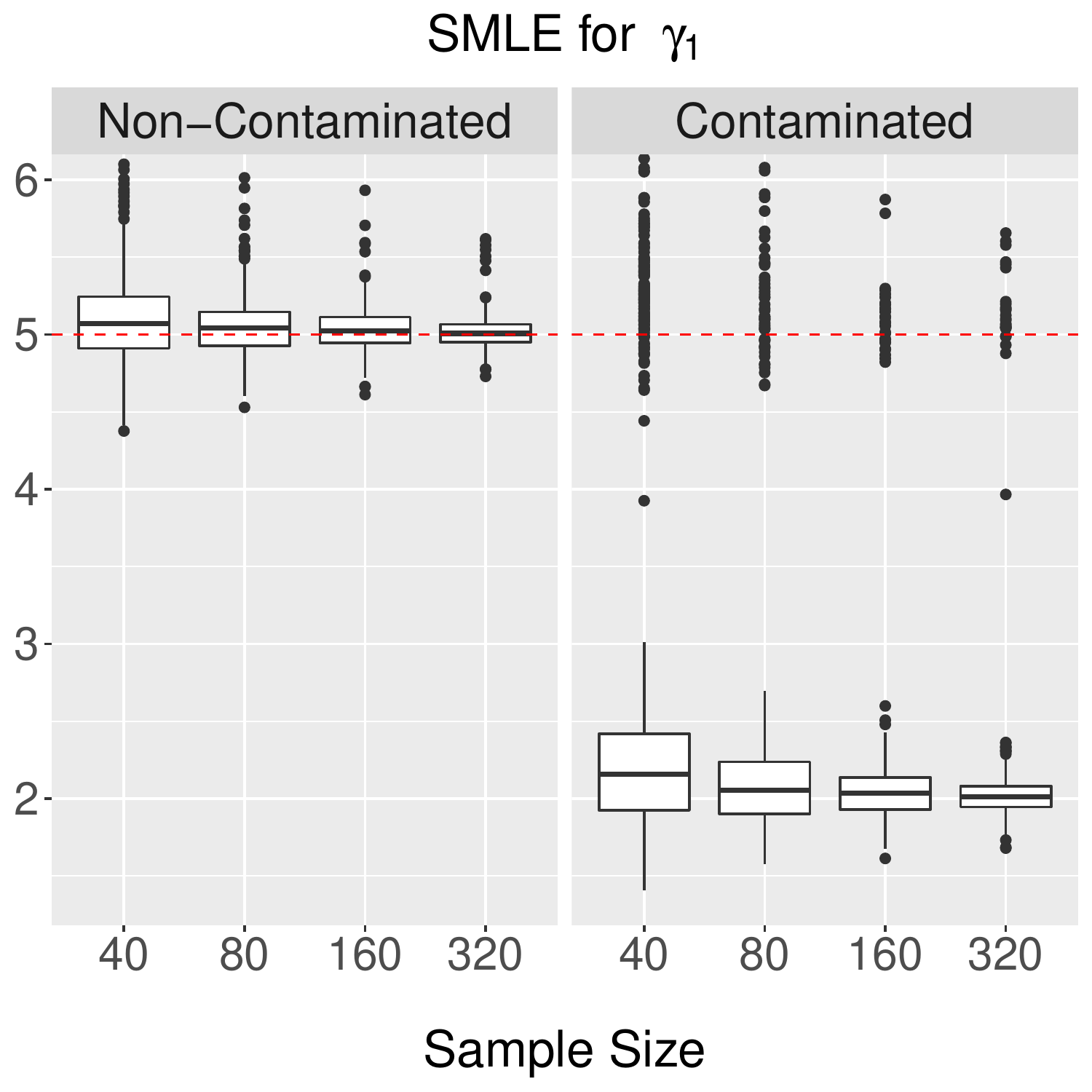}}
 \subfloat{\includegraphics[scale=0.32]{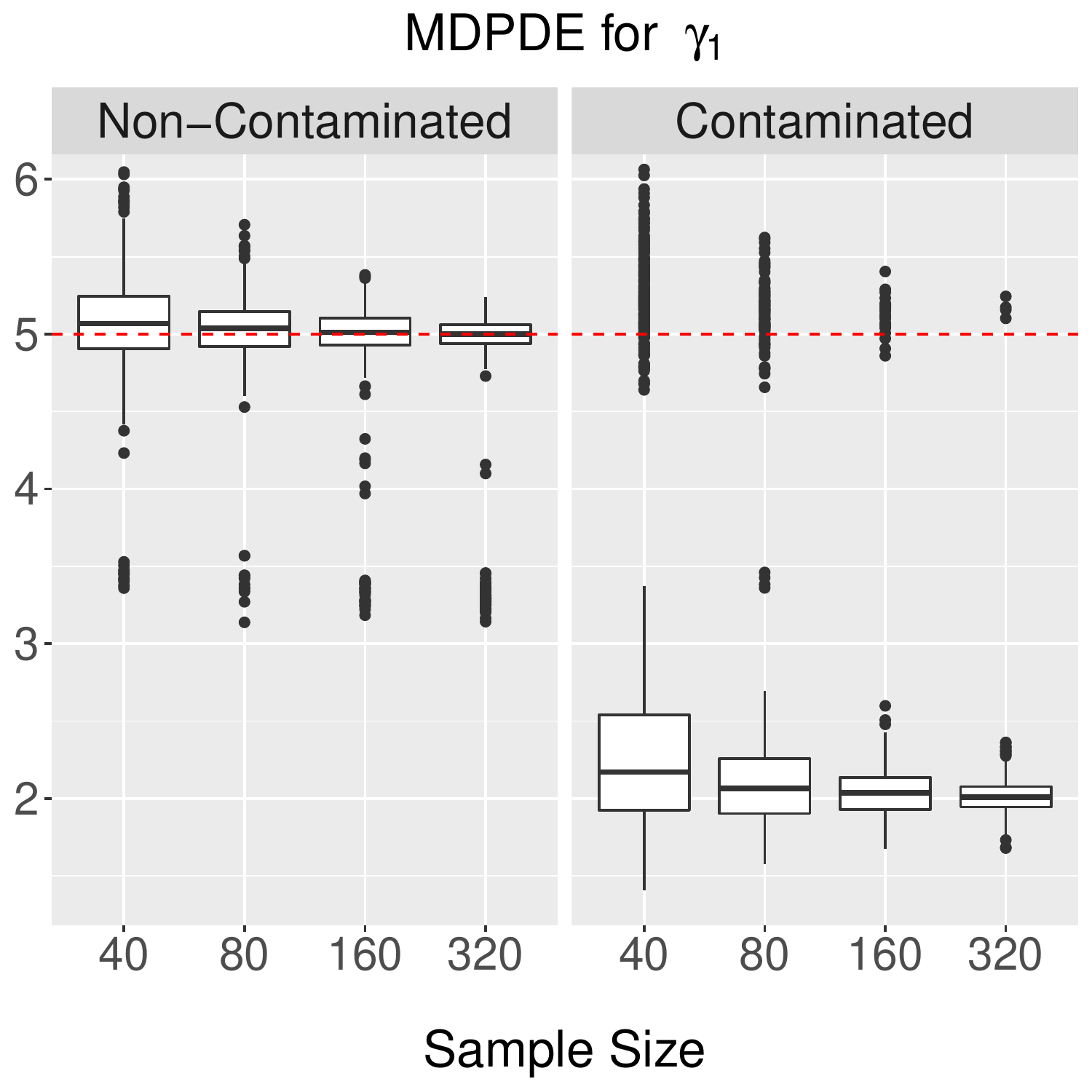}}
 \subfloat{\includegraphics[scale=0.32]{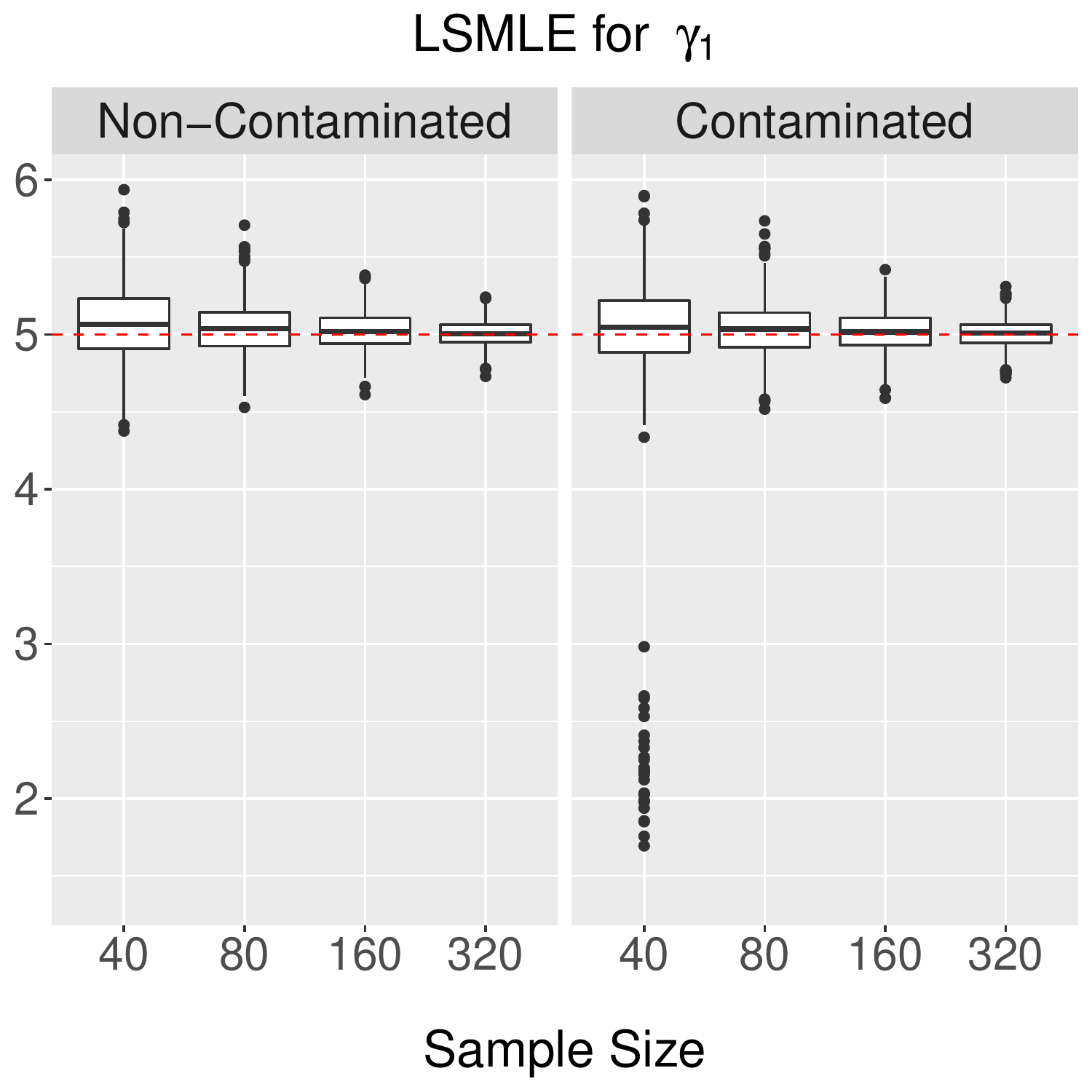}}
 \subfloat{\includegraphics[scale=0.32]{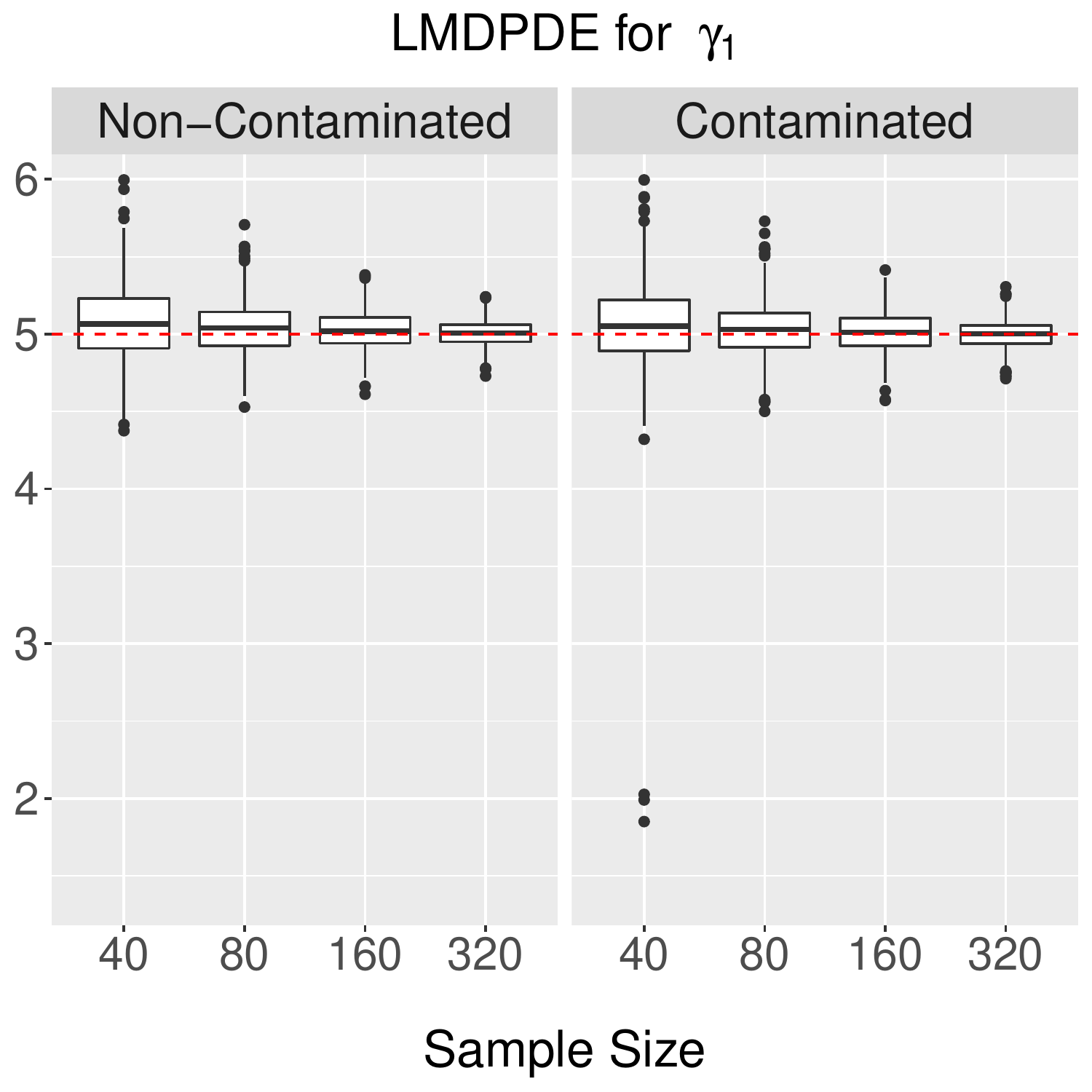}}\\
 \caption{Boxplots of estimates of $\beta_1$ (first row), $\beta_2$(second row), and $\gamma_1$(third row) for the MLE and the robust estimators. The red dashed line represents the true parameter value.}
 \label{Fig.BPB}
\end{figure}

\begin{figure}[!h]
\captionsetup[subfigure]{labelformat=empty}
\centering
 \subfloat{\includegraphics[scale=0.32]{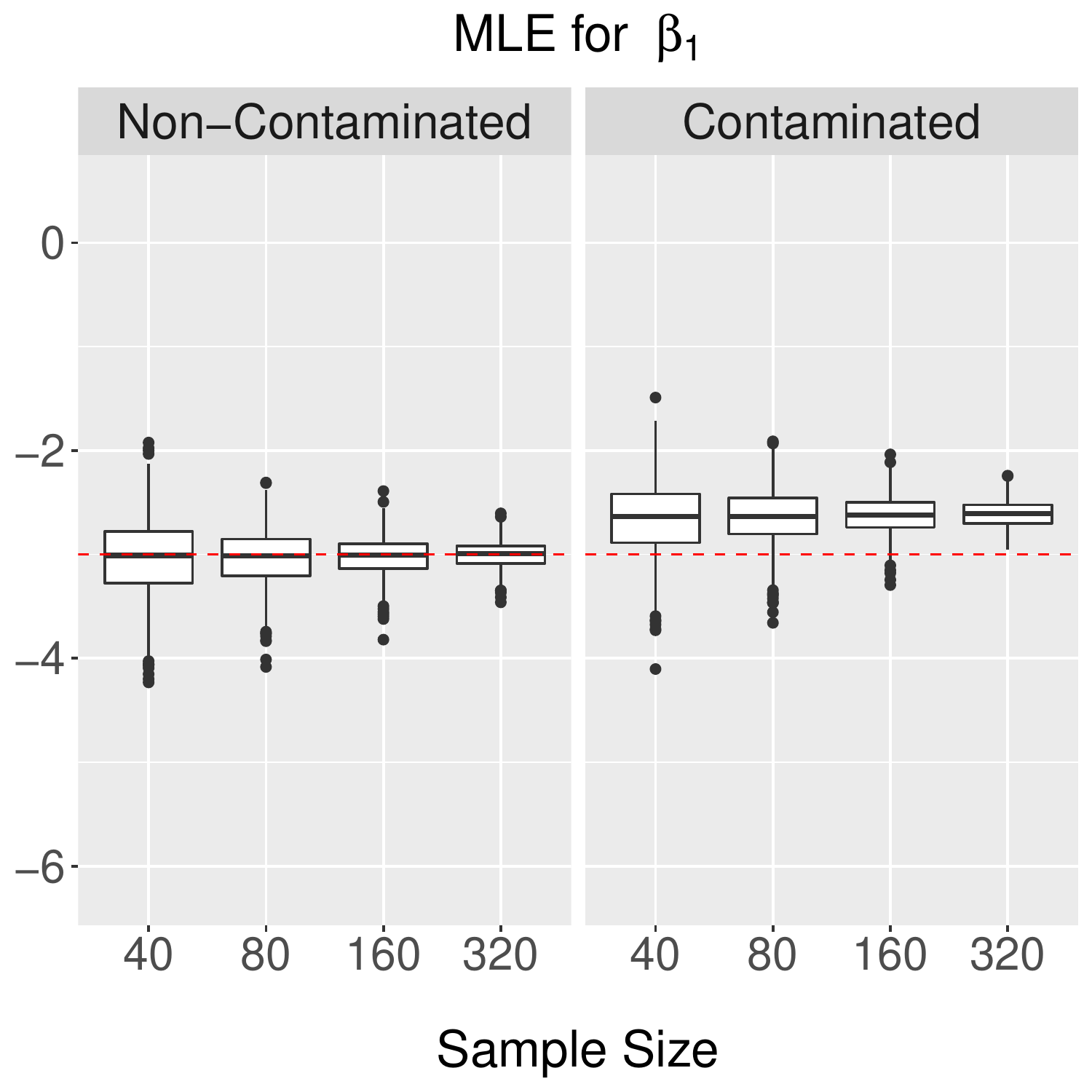}}
 \subfloat{\includegraphics[scale=0.32]{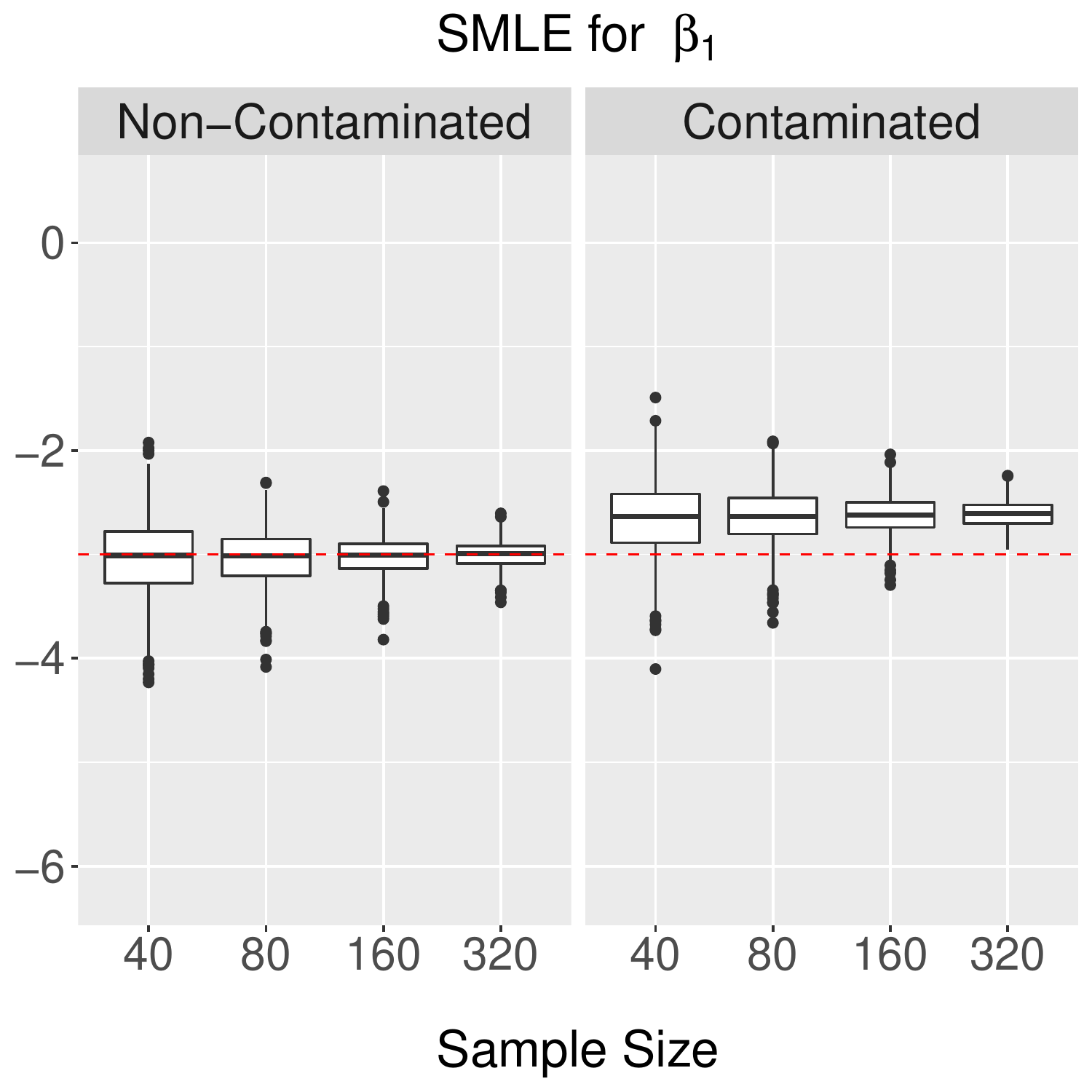}}
 \subfloat{\includegraphics[scale=0.32]{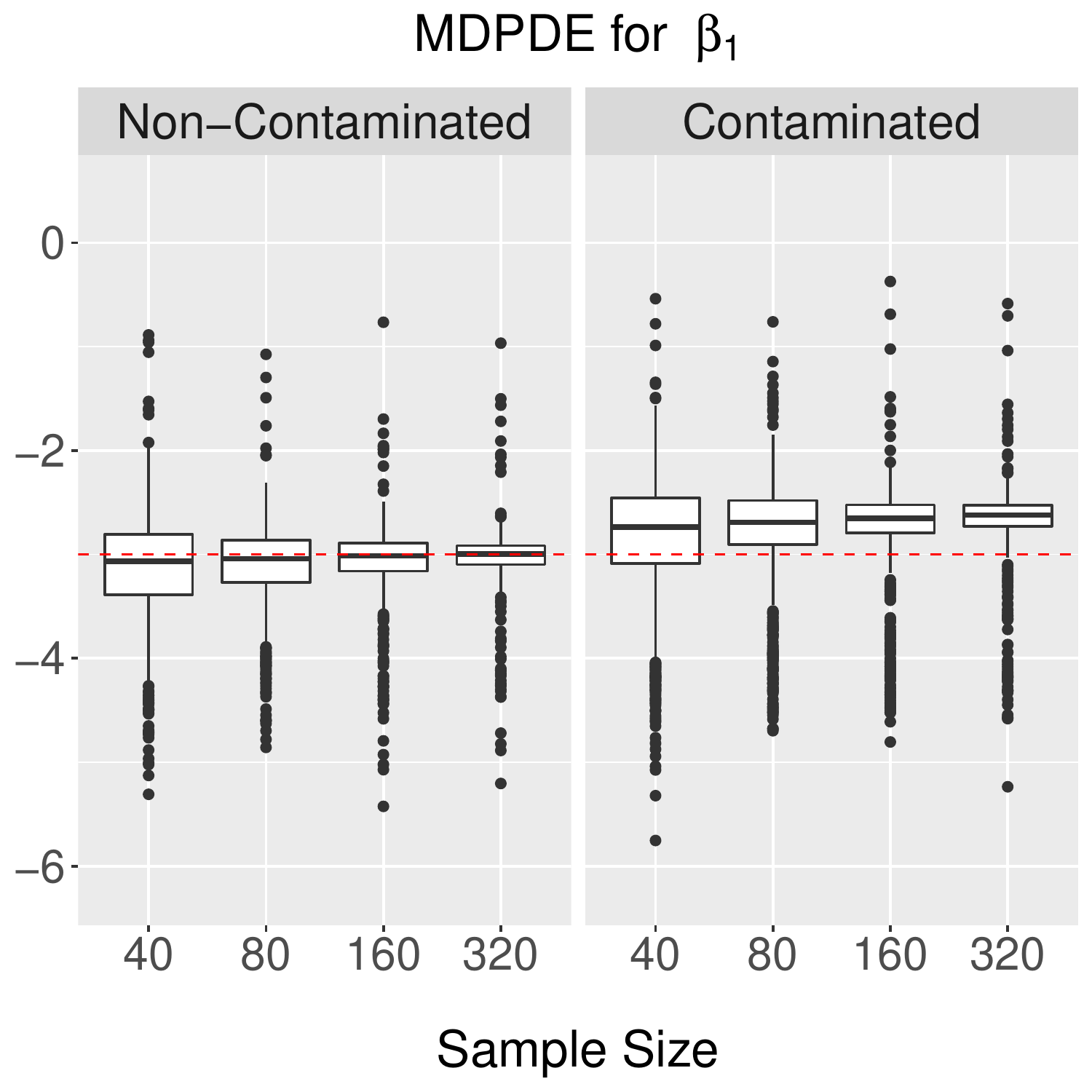}}
  \subfloat{\includegraphics[scale=0.32]{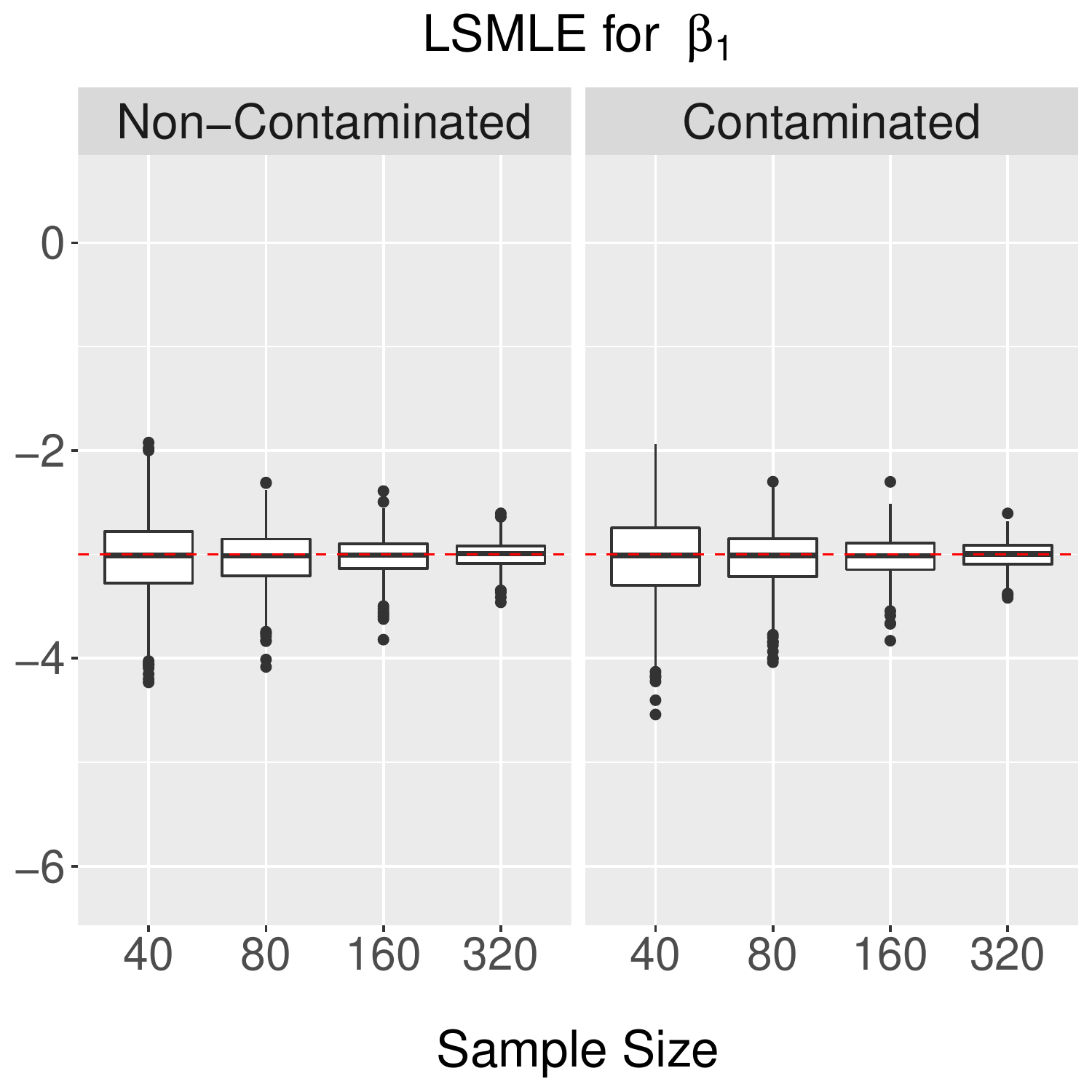}}
  \subfloat{\includegraphics[scale=0.32]{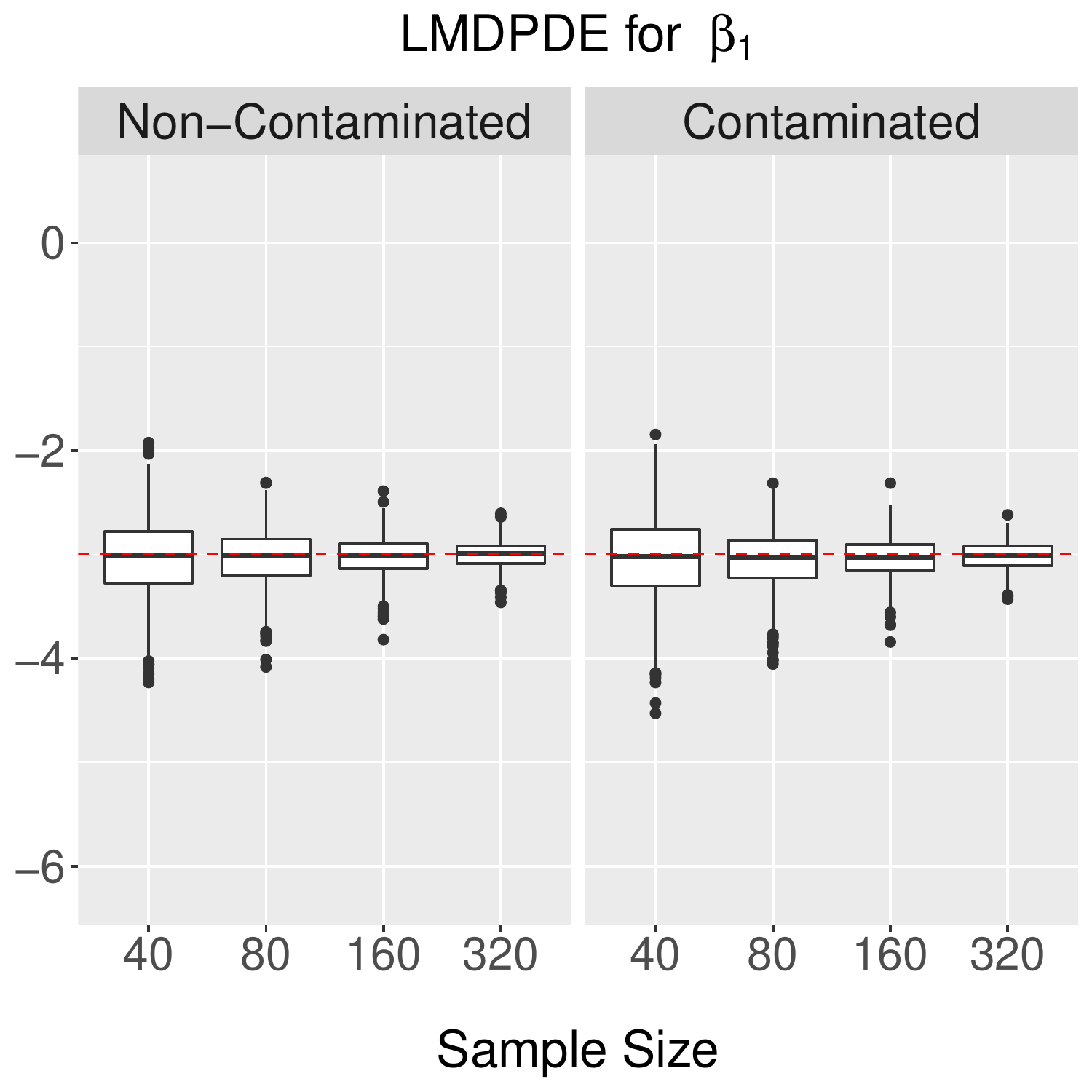}}\\
 \subfloat{\includegraphics[scale=0.32]{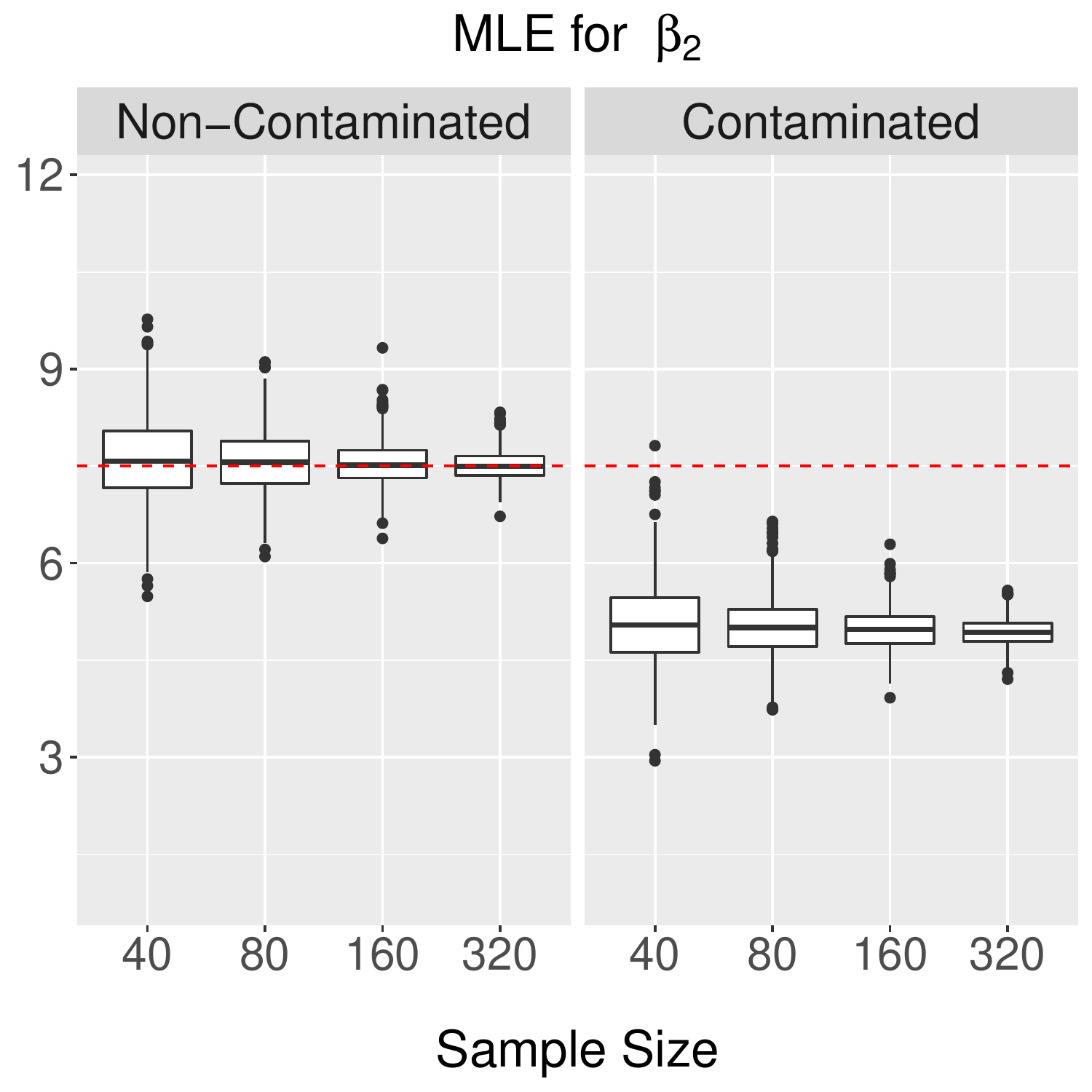}}
   \subfloat{\includegraphics[scale=0.32]{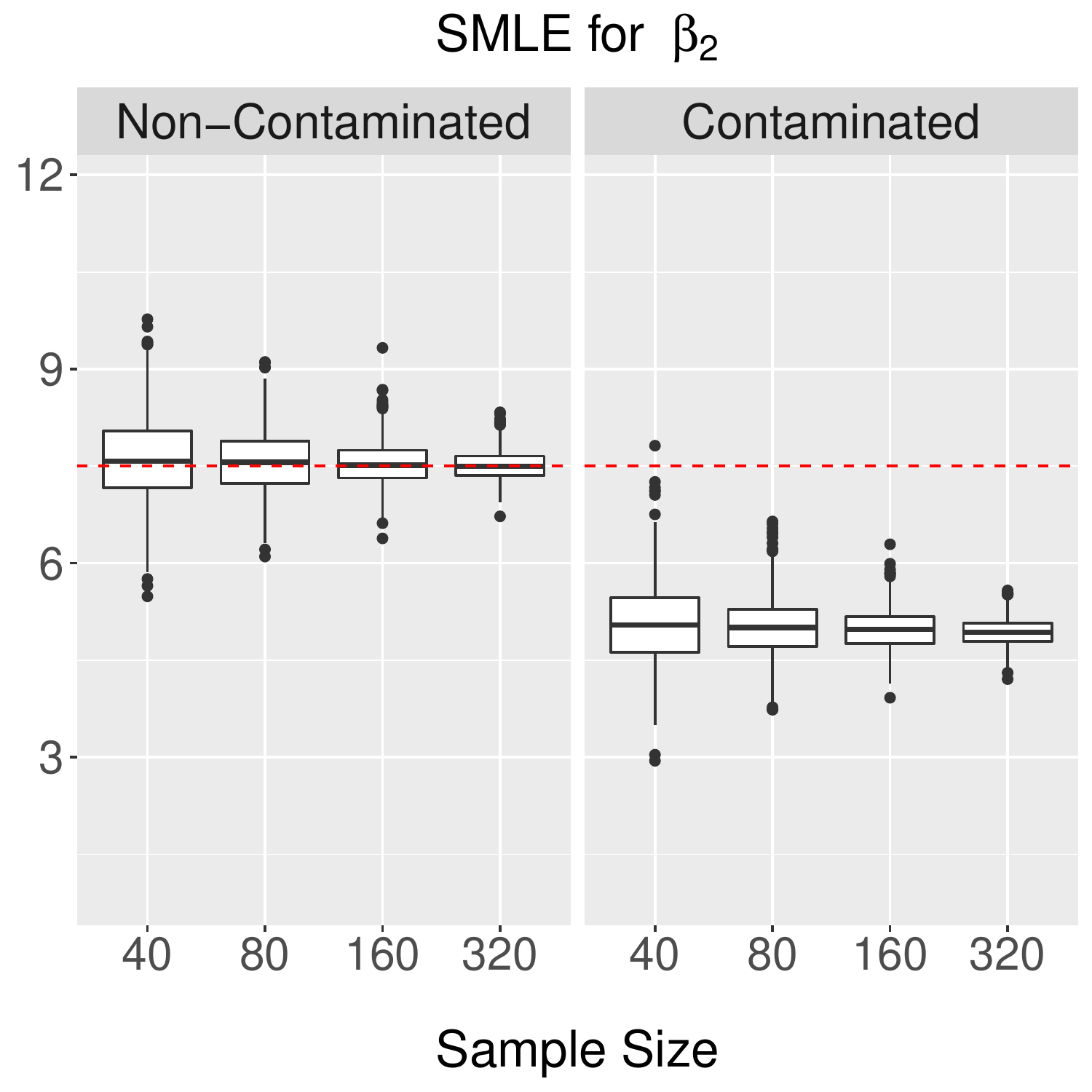}}
   \subfloat{\includegraphics[scale=0.32]{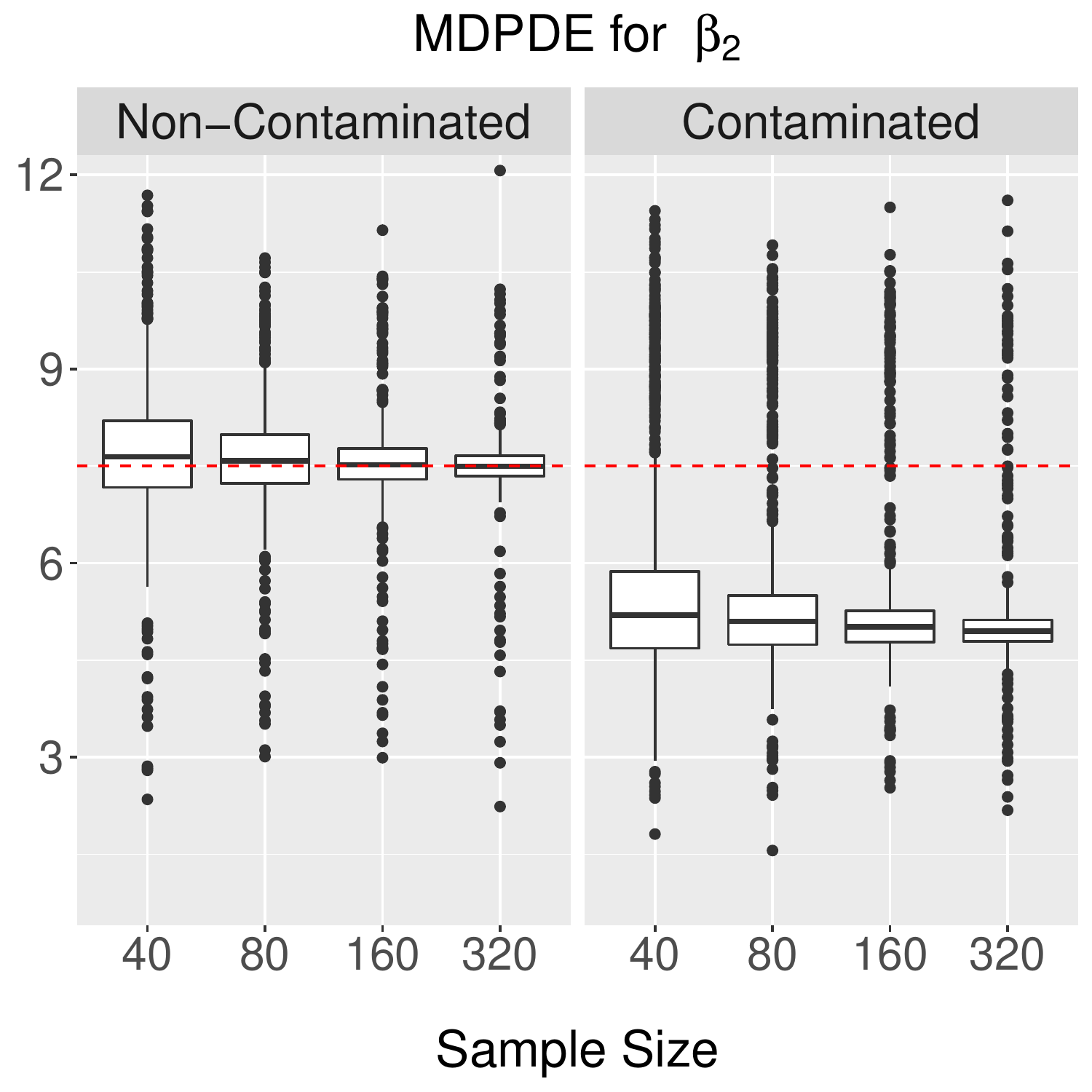}}
 \subfloat{\includegraphics[scale=0.32]{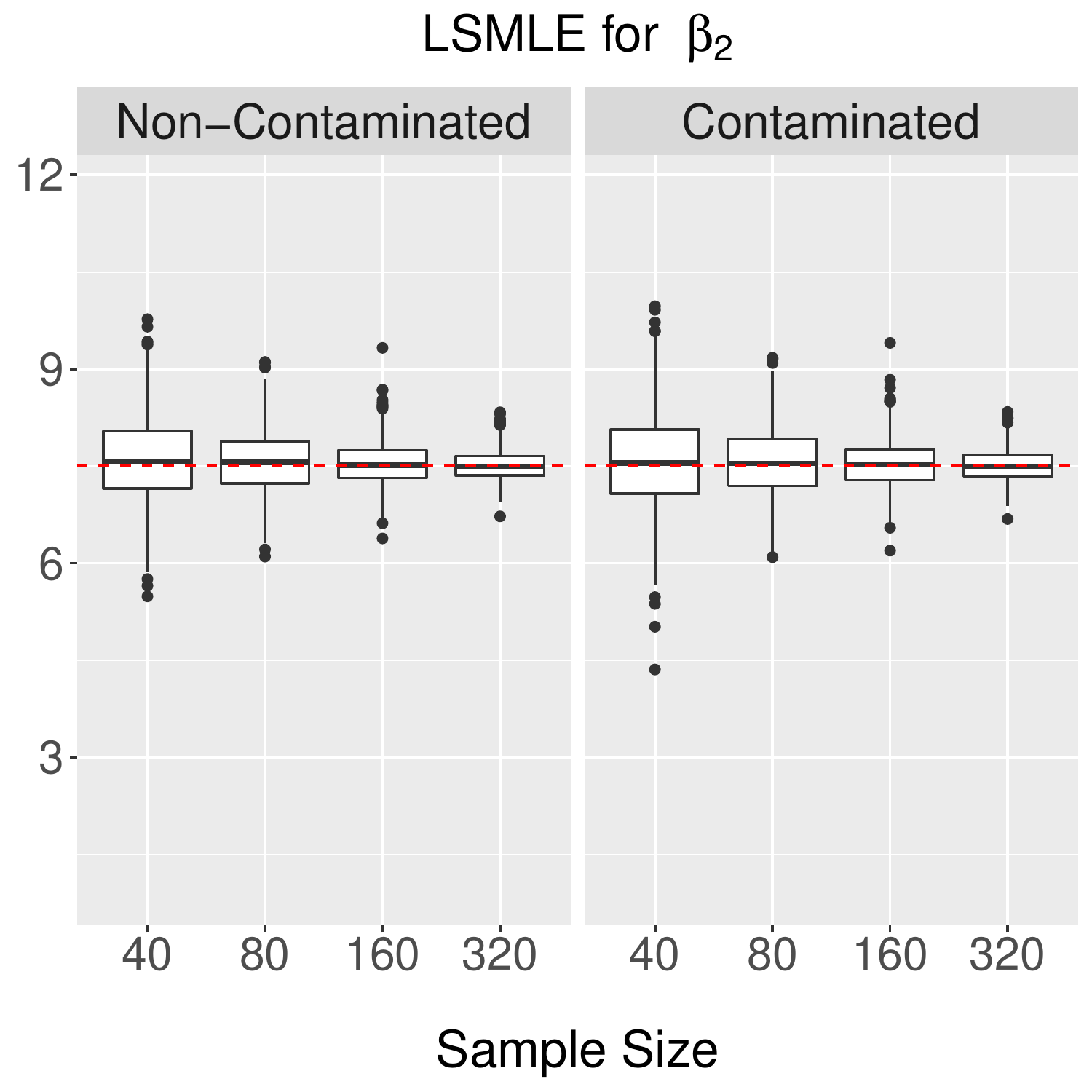}}
  \subfloat{\includegraphics[scale=0.32]{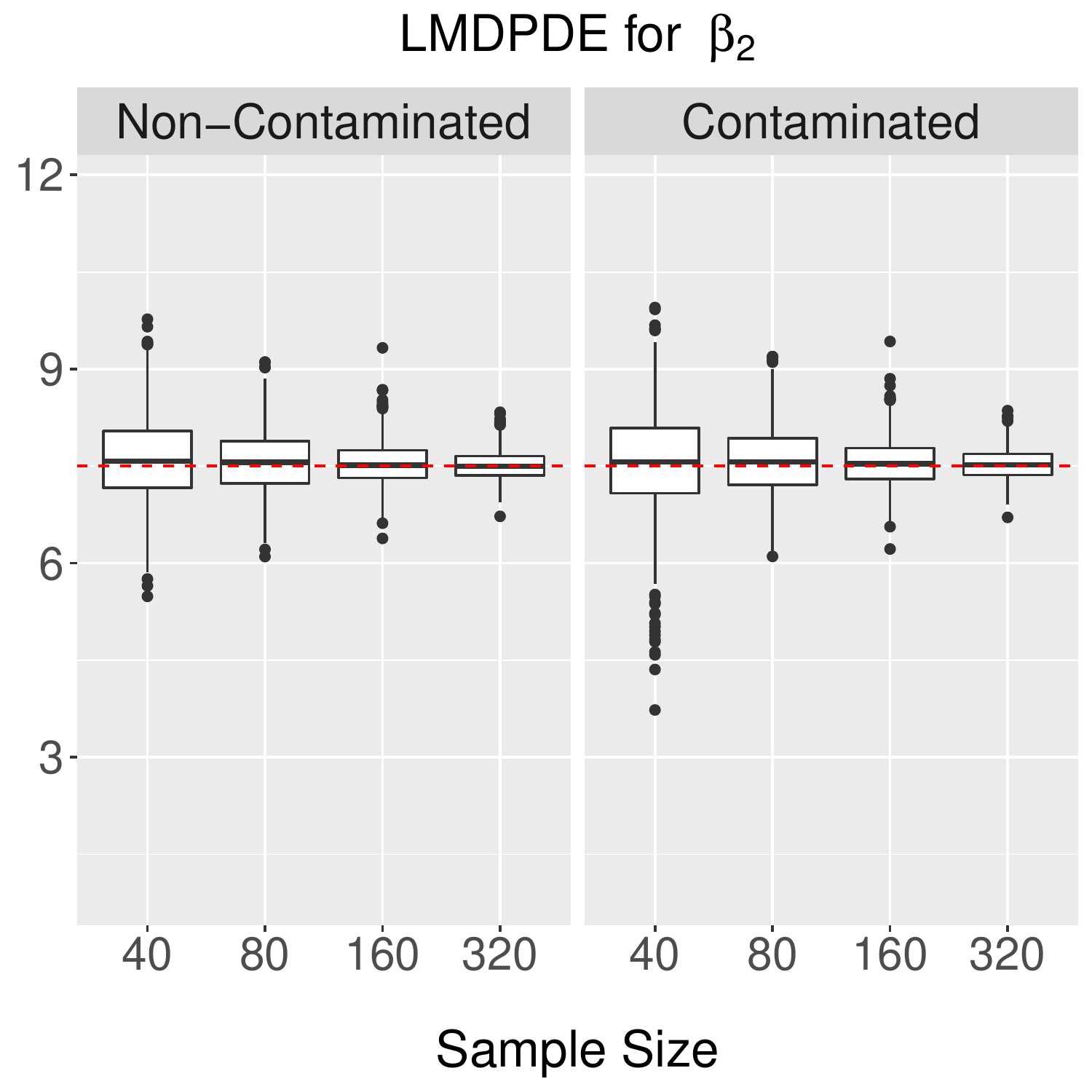}}
  \caption{Boxplots of estimates of $\beta_1$ (first row) and $\beta_2$(second row) for the MLE and the robust estimators. The red dashed line represents the true parameter value.}
 \label{Fig.BPCb}
\end{figure}

\begin{figure}[!h]
\captionsetup[subfigure]{labelformat=empty}
\centering
 \subfloat{\includegraphics[scale=0.32]{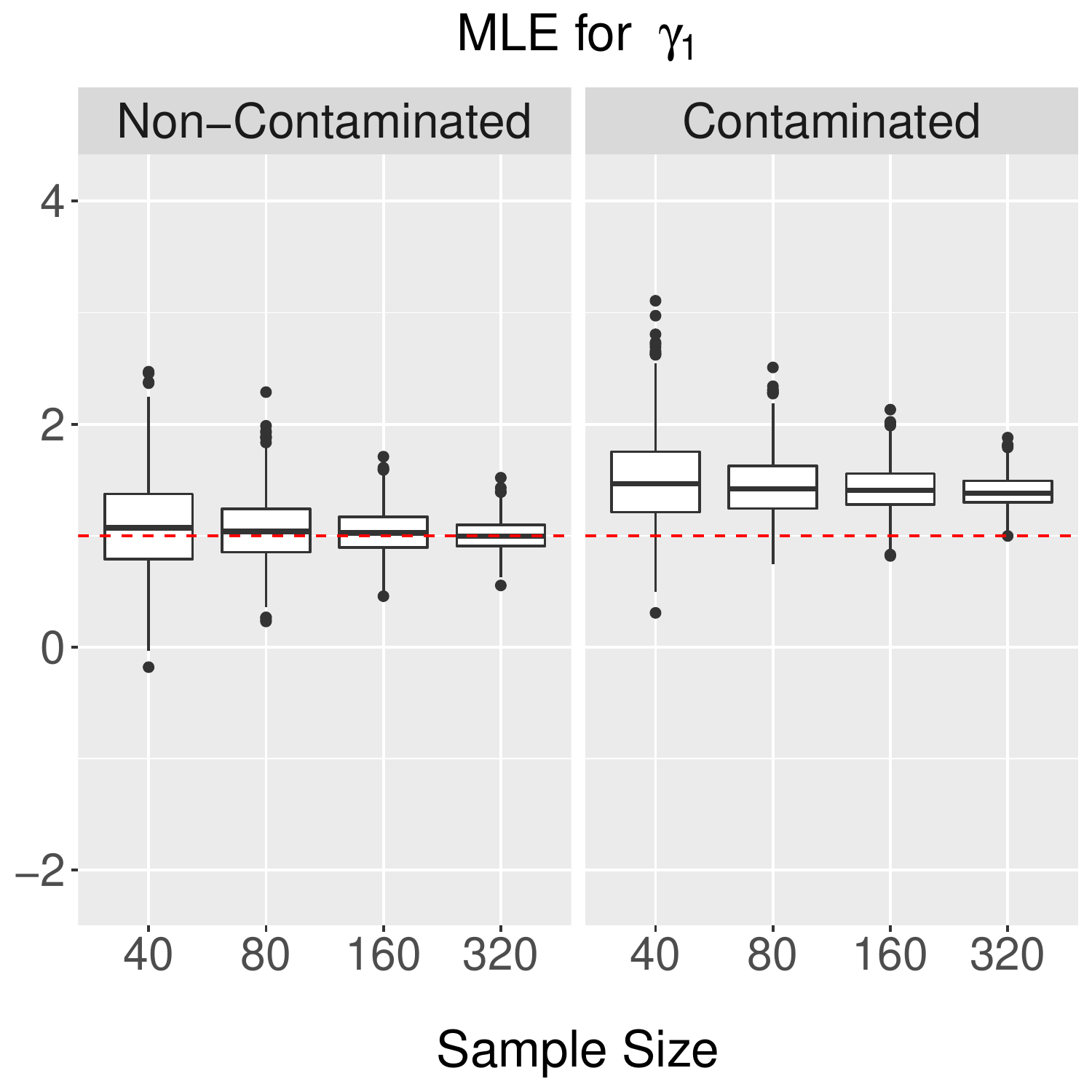}}
  \subfloat{\includegraphics[scale=0.32]{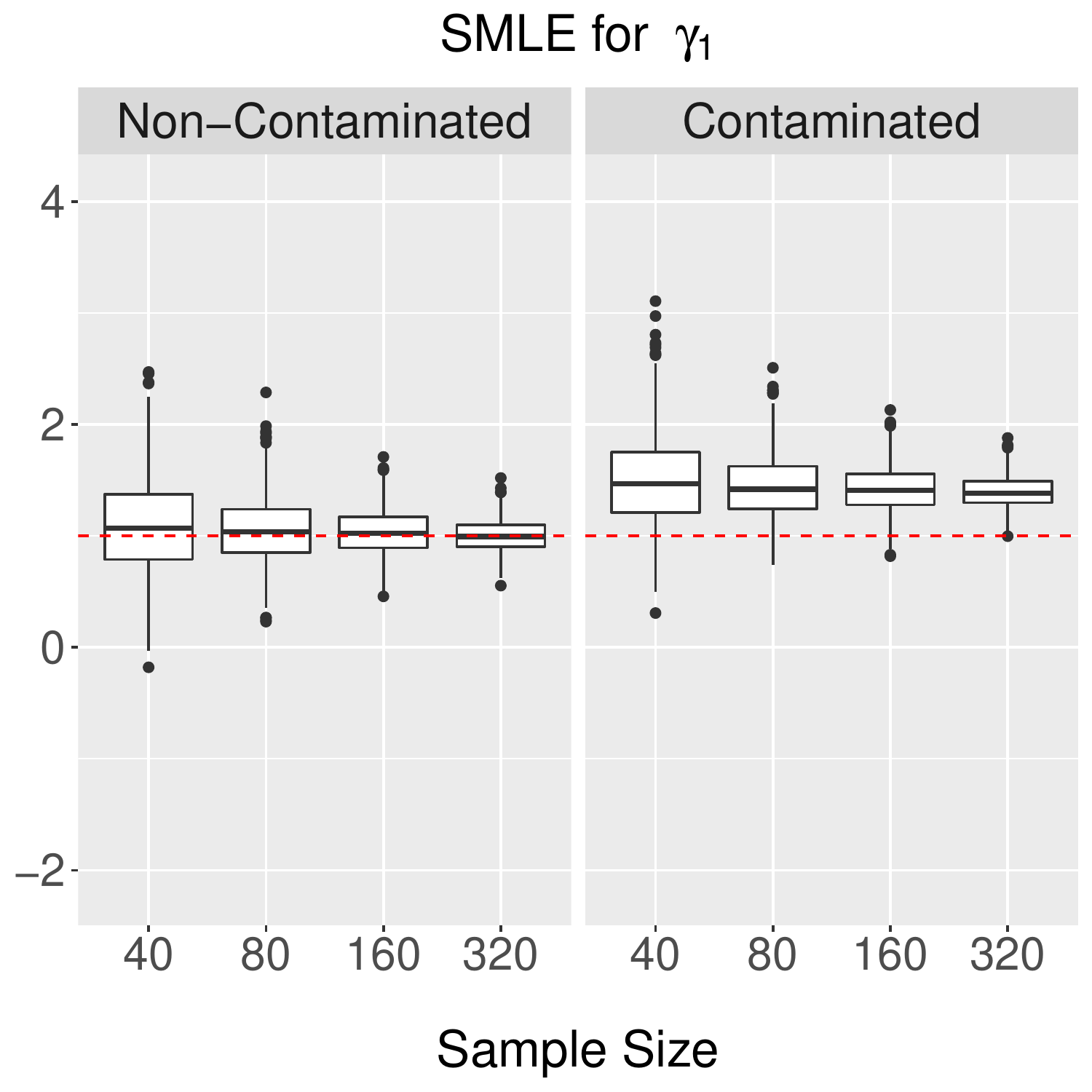}}
  \subfloat{\includegraphics[scale=0.32]{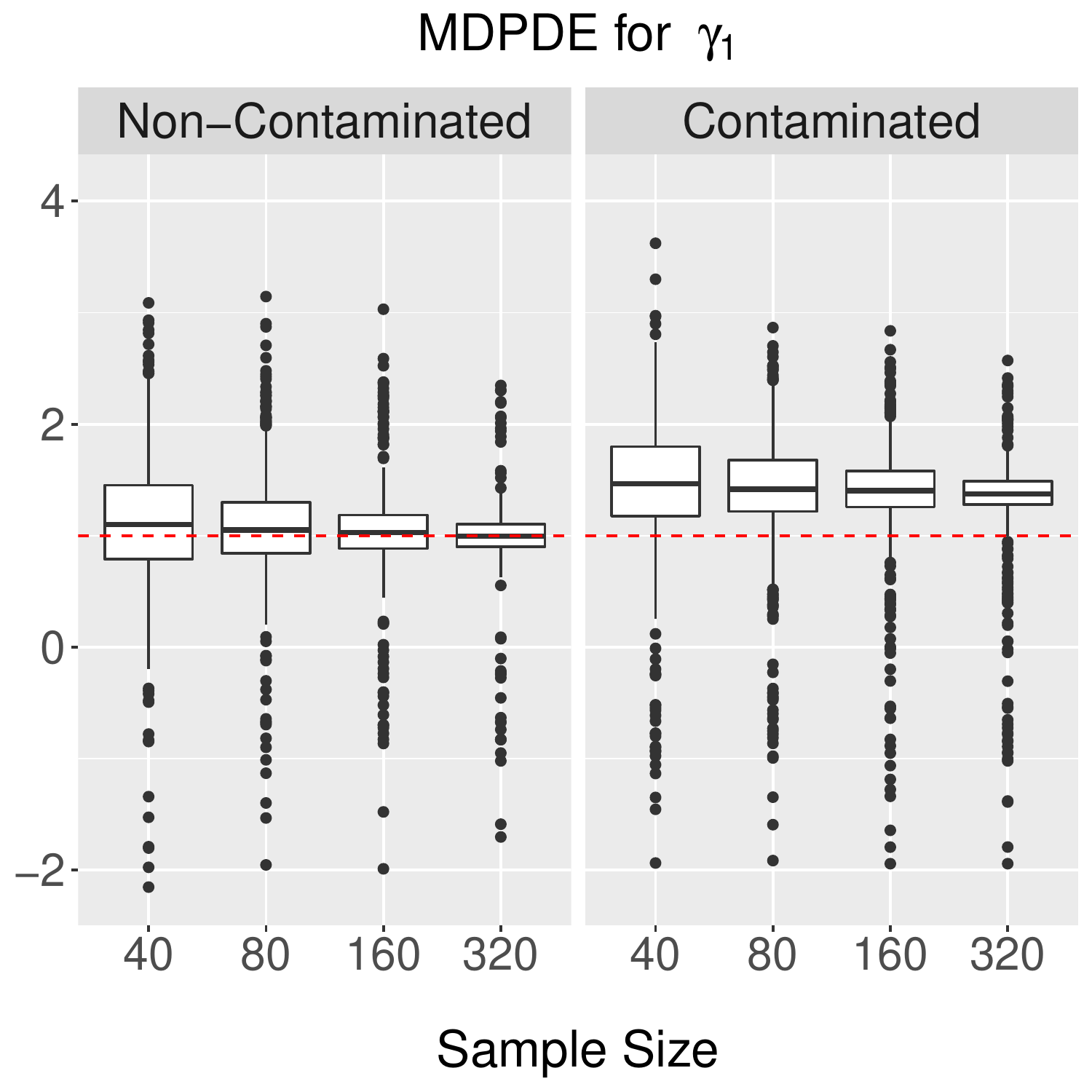}}
 \subfloat{\includegraphics[scale=0.32]{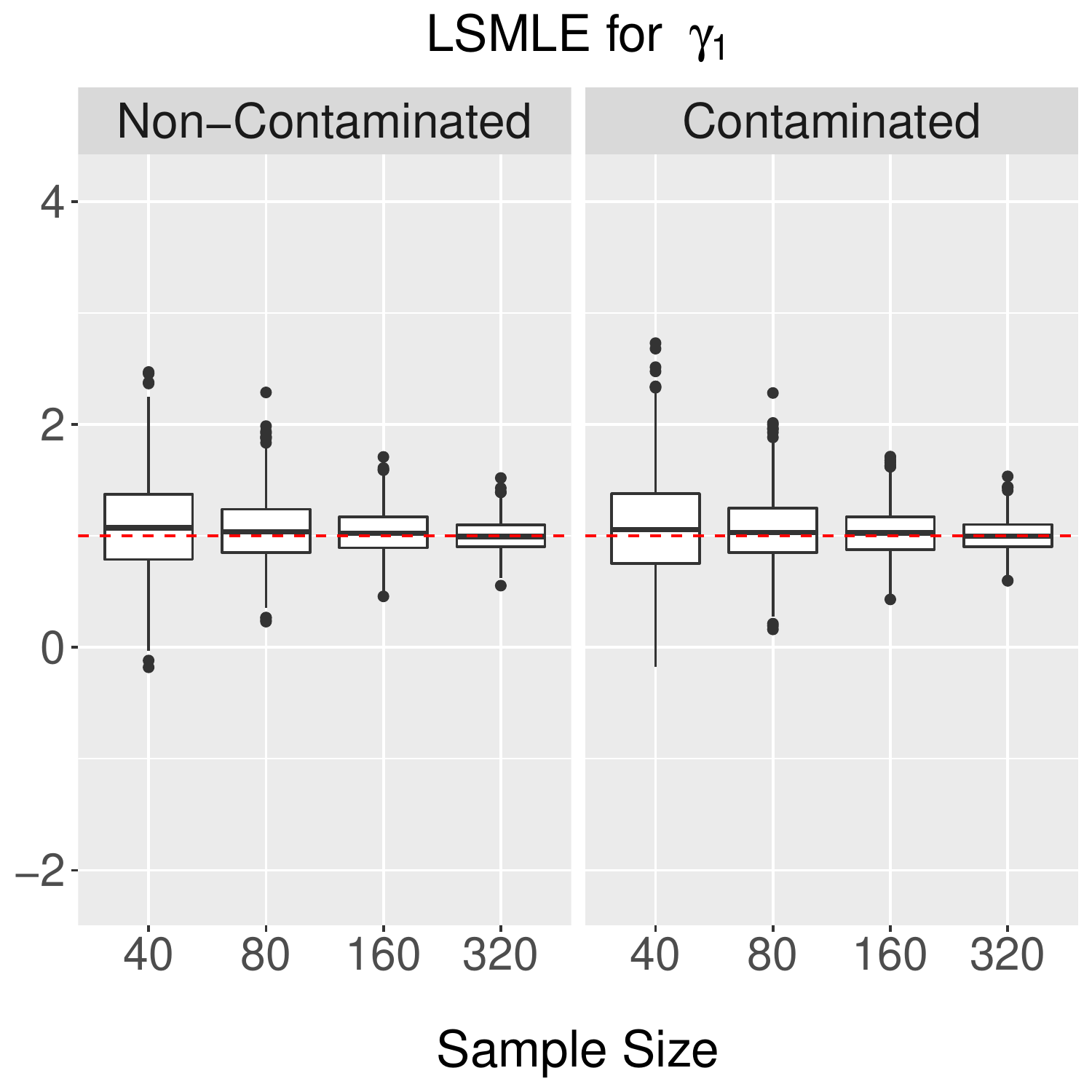}}
  \subfloat{\includegraphics[scale=0.32]{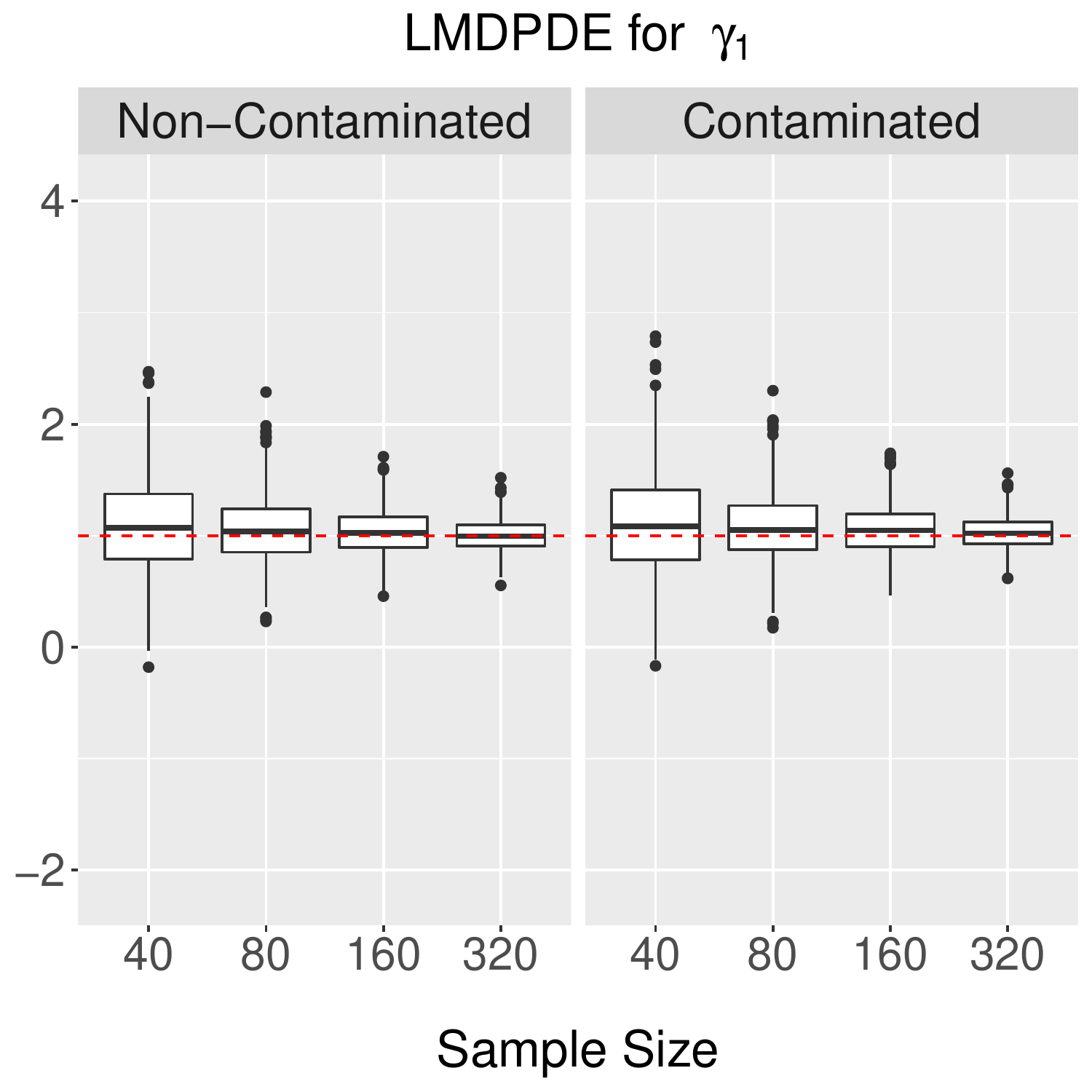}}\\
   \subfloat{\includegraphics[scale=0.32]{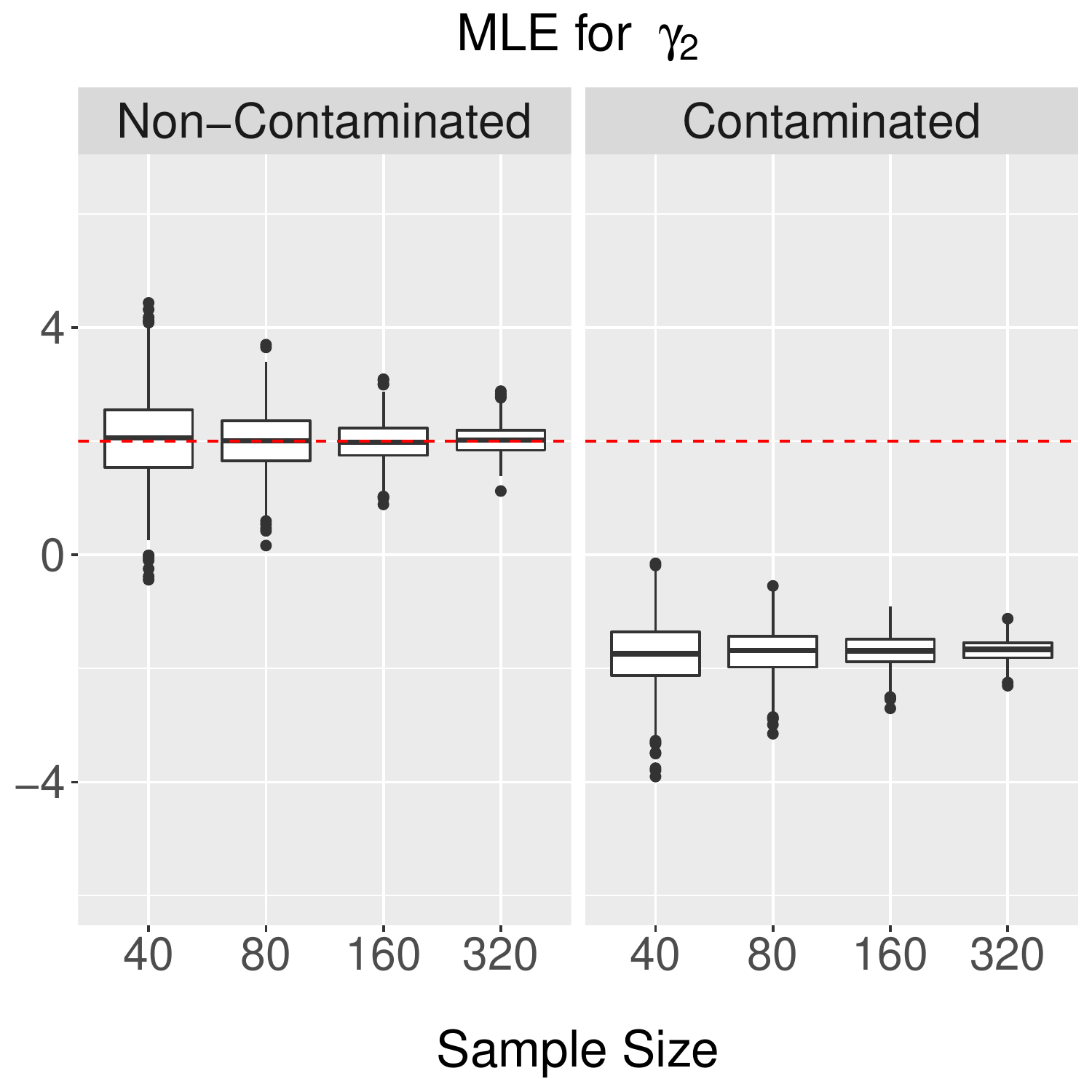}}
  \subfloat{\includegraphics[scale=0.32]{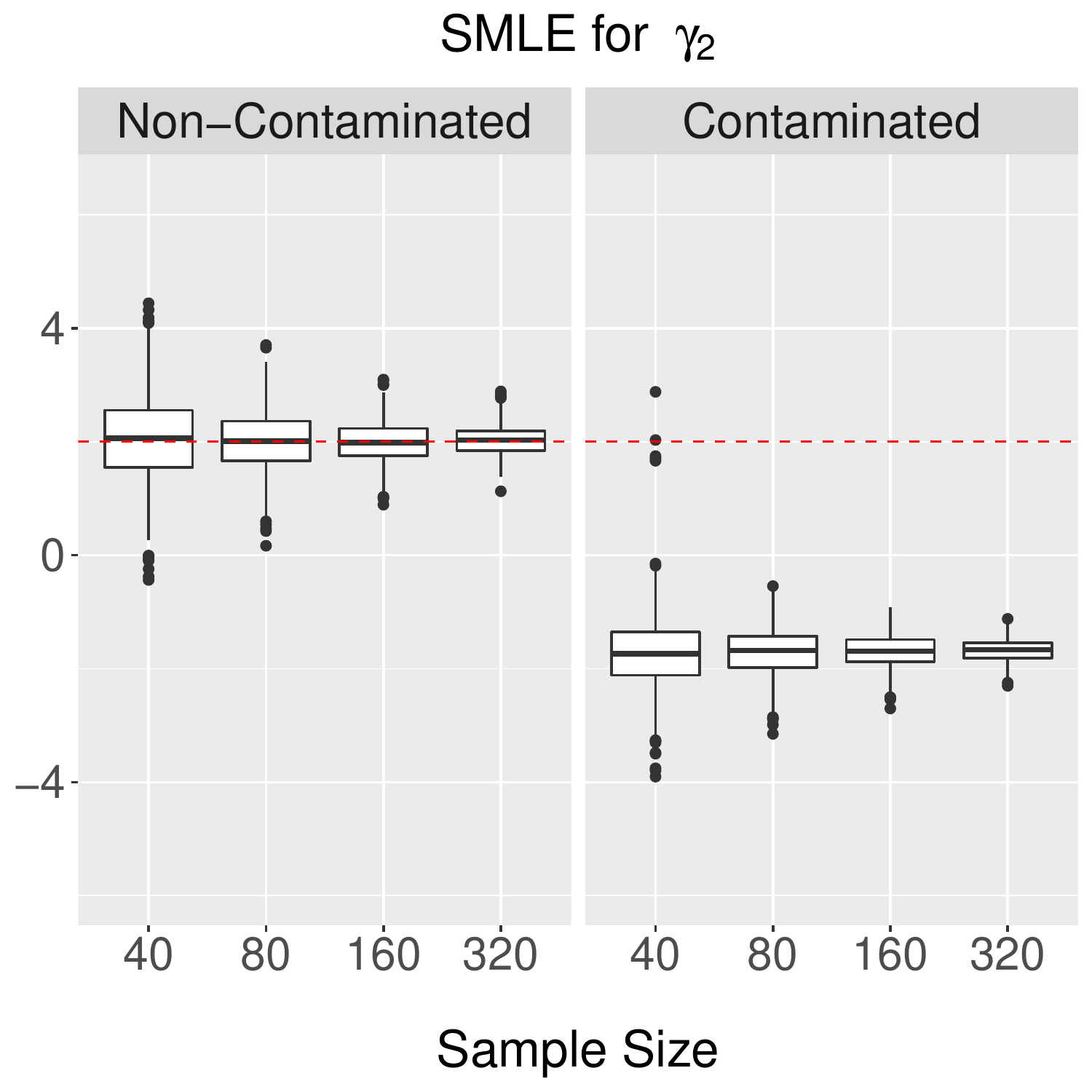}}
  \subfloat{\includegraphics[scale=0.32]{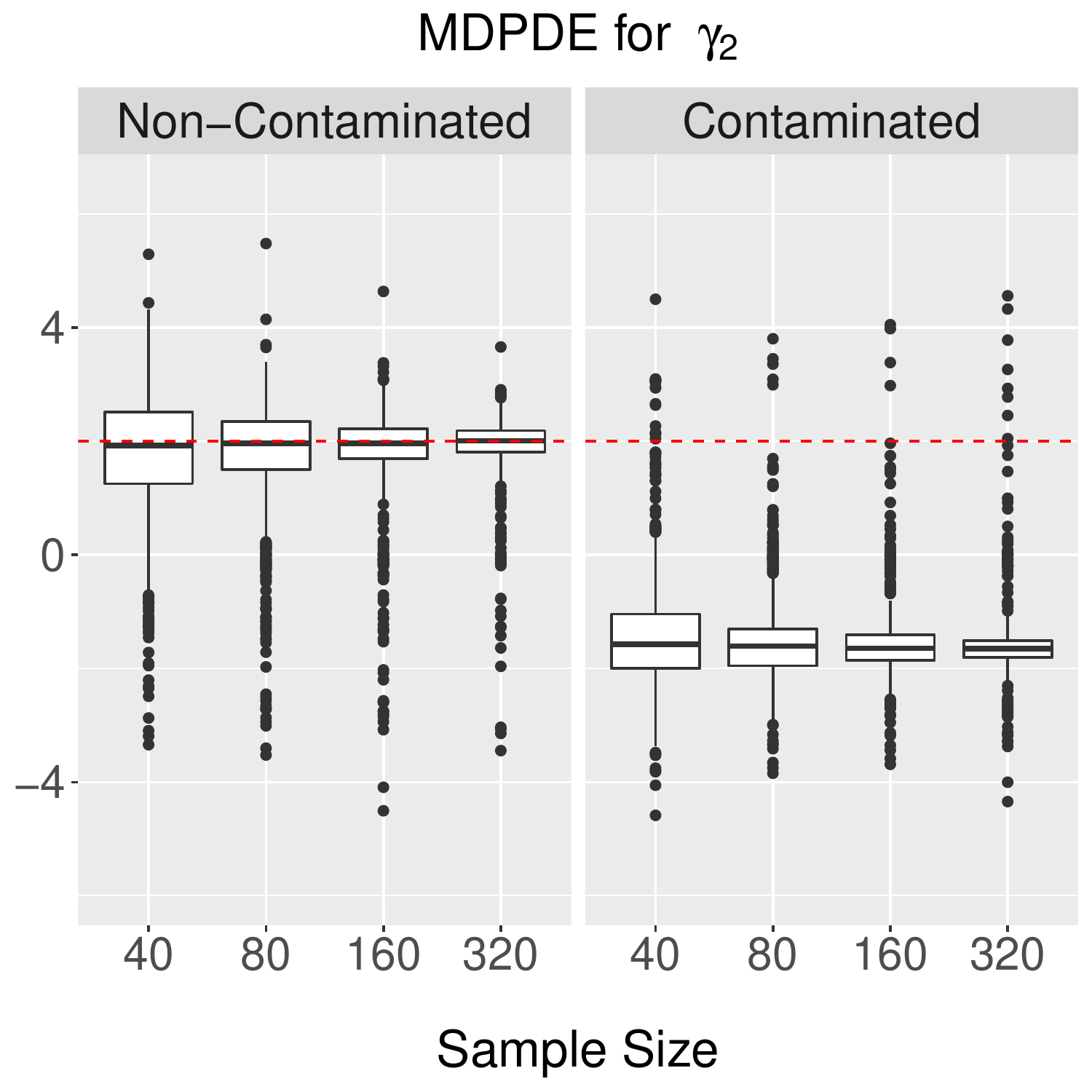}}
 \subfloat{\includegraphics[scale=0.32]{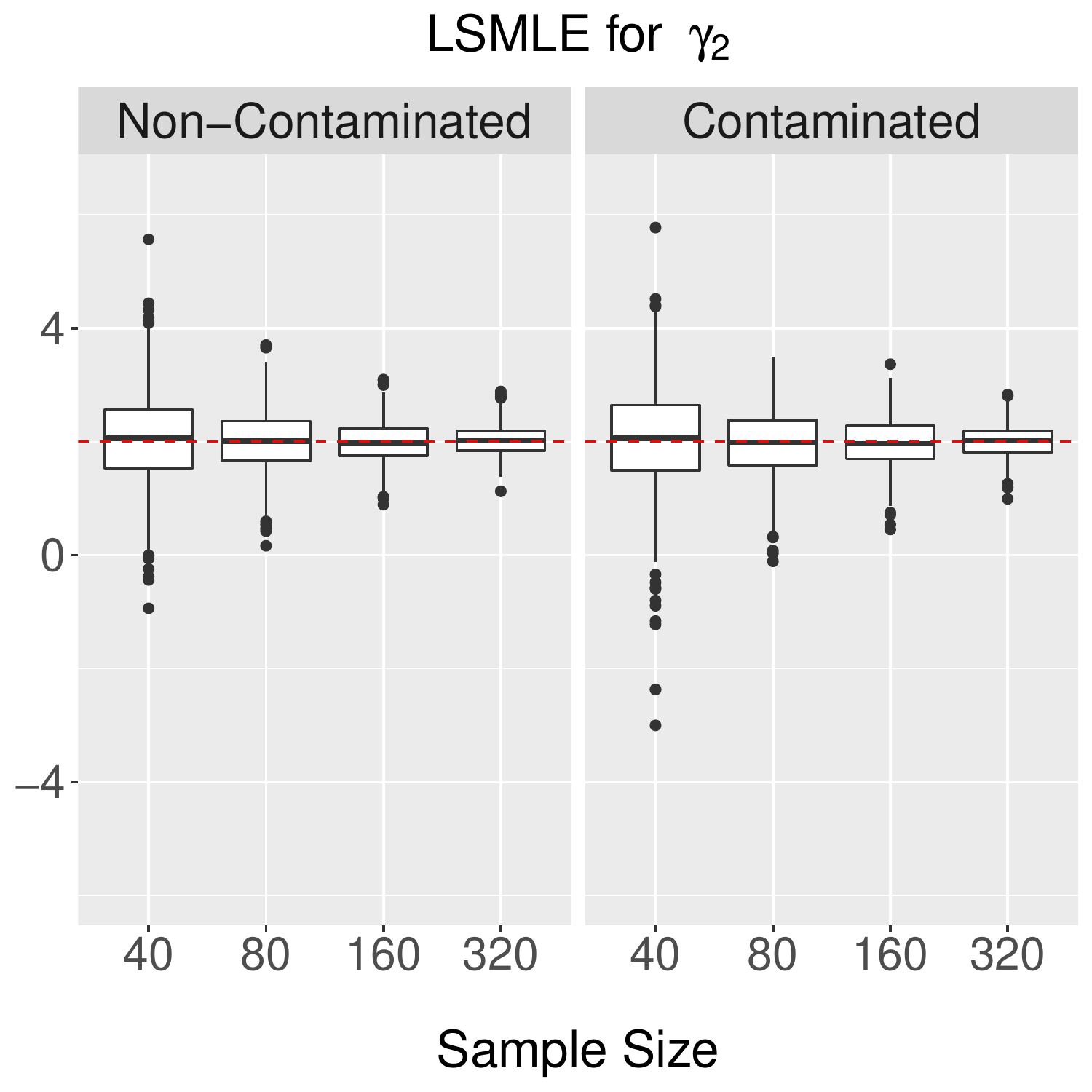}}
  \subfloat{\includegraphics[scale=0.32]{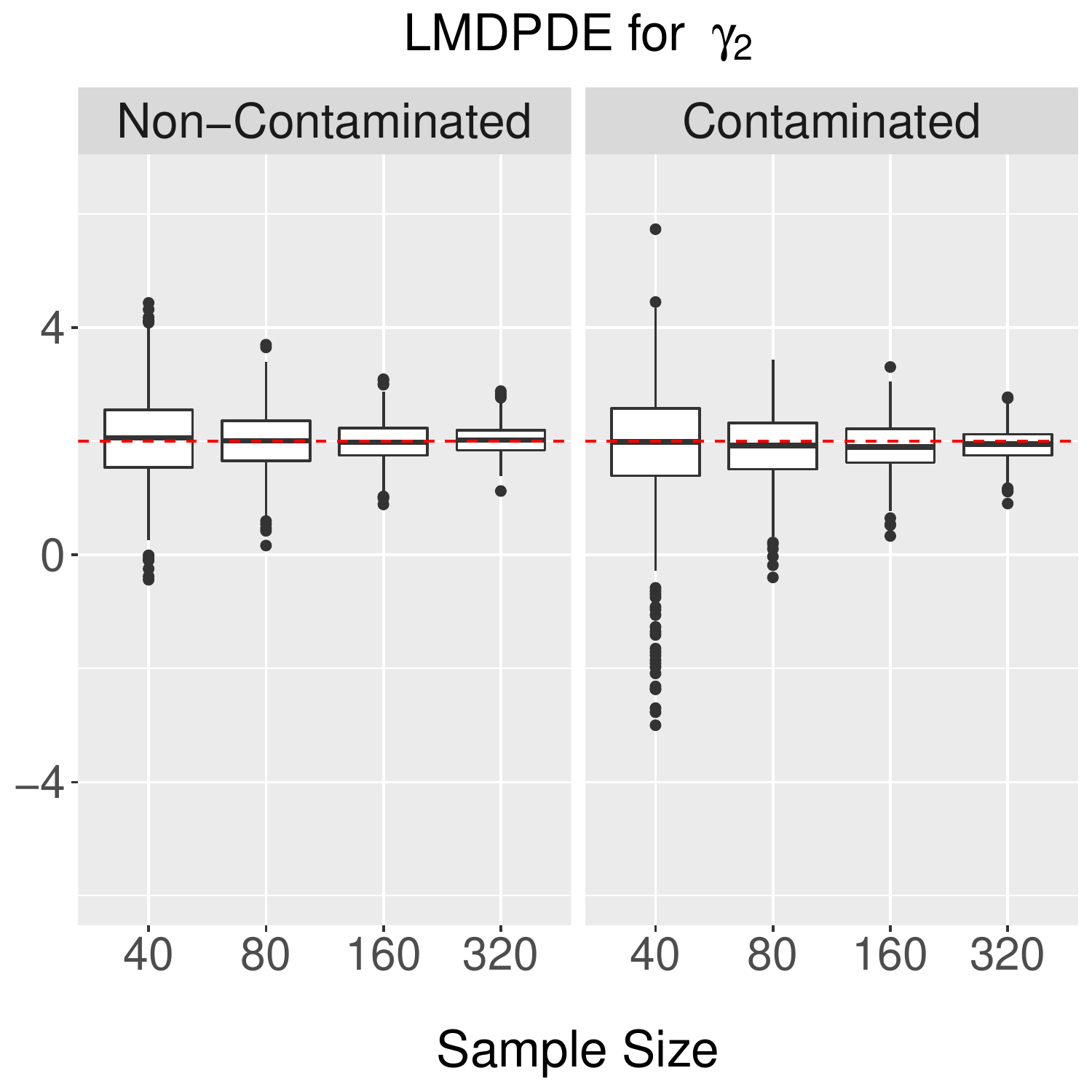}}
  \caption{Boxplots of estimates of $\gamma_1$ (first row) and $\gamma_2$(second row) for the MLE and the robust estimators. The red dashed line represents the true parameter value.}
 \label{Fig.BPCg}
\end{figure}

\end{landscape}
}
\restoregeometry

\begin{table}[!h]
\scriptsize
\centering
\def\arraystretch{1.05}
\caption{Empirical null levels of Wald-type tests under Scenarios A, B, and C at the 5\% nominal level.}
\begin{tabular}{rrrrrrrrr}
\hline
&\multicolumn{1}{c}{} & \multicolumn{7}{c}{Scenario A} \\
\cline{3-9}
& &  \multicolumn{3}{c}{Non-cont.} & & \multicolumn{3}{c}{Cont.} \\ 
\cline{3-5} \cline{7-9}
   Estimator  & $n$   & $\text{H}{0}^{1}$  & $\text{H}{0}^{2}$ & $\text{H}{0}^{3}$   & & $\text{H}{0}^{1}$  & $\text{H}{0}^{2}$ & $\text{H}{0}^{3}$   \\
                \cline{3-5} \cline{7-9}       
MLE             & $40$  & $0.05$ & $0.06$ & $0.07$ & & $0.88$ & $0.98$ & $1.00$ \\        
                & $80$  & $0.06$ & $0.06$ & $0.06$ & & $1.00$ & $1.00$ & $1.00$ \\        
                & $160$ & $0.05$ & $0.06$ & $0.06$ & & $1.00$ & $1.00$ & $1.00$ \\        
                & $320$ & $0.05$ & $0.05$ & $0.05$ & & $1.00$ & $1.00$ & $1.00$ \\   
                \cline{3-9}     
 LMDPDE	      & $40$  & $0.05$ & $0.06$ & $0.07$ & & $0.07$ & $0.07$ & $0.08$ \\        
                & $80$  & $0.06$ & $0.06$ & $0.06$ & & $0.07$ & $0.07$ & $0.07$ \\        
                & $160$ & $0.05$ & $0.06$ & $0.06$ & & $0.06$ & $0.06$ & $0.07$ \\        
                & $320$ & $0.05$ & $0.05$ & $0.05$ & & $0.06$ & $0.06$ & $0.07$ \\ 
                \cline{3-9}       
LSMLE        & $40$  & $0.05$ & $0.06$ & $0.07$ & & $0.08$ & $0.08$ & $0.09$ \\        
                & $80$  & $0.06$ & $0.06$ & $0.06$ & & $0.07$ & $0.07$ & $0.07$ \\        
                & $160$ & $0.05$ & $0.06$ & $0.06$ & & $0.06$ & $0.06$ & $0.07$ \\        
                & $320$ & $0.05$ & $0.05$ & $0.05$ & & $0.06$ & $0.06$ & $0.06$ \\        
\hline
&\multicolumn{1}{c}{} & \multicolumn{7}{c}{Scenario B} \\
\cline{3-9}
& &  \multicolumn{3}{c}{Non-cont.} & & \multicolumn{3}{c}{Cont.} \\ 
\cline{3-5} \cline{7-9}
  Estimator  & $n$   & $\text{H}{0}^{1}$  & $\text{H}{0}^{2}$ & $\text{H}{0}^{3}$   & & $\text{H}{0}^{1}$  & $\text{H}{0}^{2}$ & $\text{H}{0}^{3}$   \\
                \cline{3-5} \cline{7-9}
MLE             & $40$  & $0.07$ & $0.08$ & $0.08$ & & $1.00$ & $1.00$ & $1.00$ \\
                & $80$  & $0.05$ & $0.06$ & $0.06$ & & $1.00$ & $1.00$ & $1.00$ \\
                & $160$ & $0.05$ & $0.06$ & $0.07$ & & $1.00$ & $1.00$ & $1.00$ \\
                & $320$ & $0.06$ & $0.07$ & $0.08$ & & $1.00$ & $1.00$ & $1.00$ \\
                \cline{3-9} 
LMDPDE	      & $40$  & $0.07$ & $0.07$ & $0.08$ & & $0.07$ & $0.10$ & $0.10$ \\
                & $80$  & $0.06$ & $0.06$ & $0.06$ & & $0.07$ & $0.09$ & $0.09$ \\
                & $160$ & $0.05$ & $0.06$ & $0.07$ & & $0.06$ & $0.07$ & $0.08$ \\
                & $320$ & $0.06$ & $0.07$ & $0.08$ & & $0.08$ & $0.10$ & $0.10$ \\
                \cline{3-9} 
LSMLE      & $40$  & $0.07$ & $0.08$ & $0.08$ & & $0.09$ & $0.12$ & $0.01$ \\
                & $80$  & $0.06$ & $0.06$ & $0.06$ & & $0.08$ & $0.09$ & $0.01$ \\
                & $160$ & $0.05$ & $0.06$ & $0.07$ & & $0.06$ & $0.07$ & $0.09$ \\
                & $320$ & $0.06$ & $0.07$ & $0.08$ & & $0.08$ & $0.10$ & $0.10$ \\
\hline
&\multicolumn{1}{c}{} & \multicolumn{7}{c}{Scenario C} \\
\cline{3-9}
& &  \multicolumn{3}{c}{Non-cont.} & & \multicolumn{3}{c}{Cont.} \\ 
\cline{3-5} \cline{7-9}
Estimator    & $n$   & $\text{H}{0}^{4}$  & $\text{H}{0}^{5}$ & $\text{H}{0}^{6}$   & & $\text{H}{0}^{4}$  & $\text{H}{0}^{5}$ & $\text{H}{0}^{6}$   \\
                \cline{3-5} \cline{7-9}
MLE             & $40$  & $0.08$ & $0.08$ & $0.08$ & & $1.00$ & $1.00$ & $1.00$ \\
                & $80$  & $0.08$ & $0.07$ & $0.06$ & & $1.00$ & $1.00$ & $1.00$ \\
                & $160$ & $0.06$ & $0.06$ & $0.05$ & & $1.00$ & $1.00$ & $1.00$ \\
                & $320$ & $0.05$ & $0.05$ & $0.06$ & & $1.00$ & $1.00$ & $1.00$ \\
                \cline{3-9} 
LMDPDE	      & $40$  & $0.08$ & $0.08$ & $0.08$ & & $0.13$ & $0.13$ & $0.12$ \\
                & $80$  & $0.08$ & $0.07$ & $0.06$ & & $0.12$ & $0.10$ & $0.10$ \\
                & $160$ & $0.05$ & $0.06$ & $0.05$ & & $0.11$ & $0.12$ & $0.09$ \\
                & $320$ & $0.04$ & $0.05$ & $0.06$ & & $0.07$ & $0.08$ & $0.08$ \\
                \cline{3-9} 
LSMLE        & $40$  & $0.08$ & $0.08$ & $0.08$ & & $0.12$ & $0.12$ & $0.11$ \\
                & $80$  & $0.08$ & $0.07$ & $0.06$ & & $0.11$ & $0.10$ & $0.09$ \\
                & $160$ & $0.06$ & $0.06$ & $0.05$ & & $0.10$ & $0.09$ & $0.09$ \\
                & $320$ & $0.05$ & $0.05$ & $0.05$ & & $0.07$ & $0.07$ & $0.07$ \\
                \hline
\end{tabular} 
\label{Tab.Emp-Level-5}
\end{table}

\section{An application to health insurance coverage data}\label{Sec.App}

We shall now present and discuss an application of the new robust estimators to health insurance coverage data collected by the Institute of Applied Economic Research (Instituto de Pesquisa Econômica Aplicada, IPEA). The dataset includes information on 80 cities in the state of São Paulo, Brazil, in 2010. This application's dataset and R codes are available at \url{https://github.com/ffqueiroz/RobustBetareg}.

The response variable ($y$) is the health insurance coverage index (HIC). The covariates are the percentage of the total population who lives in the city's urban zone (Urb) and the per capita gross domestic product (GDP). We consider the beta regression model \eqref{betadensity}-\eqref{linearPredictor} with $g_\mu(\cdot)$ and $g_\phi(\cdot)$ being the logit and the log functions, respectively. First, both covariates and an intercept are included in the mean and precision submodels. We fit the model using the MLE and the robust estimators. The Wald-type tests based on all the estimators agree that the covariate Urb is not significant for the precision submodel ($p$-value greater than $0.3$ for all the tests); see the Supplementary Material, Section 4. The postulated reduced model is the beta regression model with 
\begin{equation*}
\begin{split}
  \log\left(\frac{\mu_i}{1-\mu_i}\right)&=\beta_1+\beta_2 \text{Urb}_i+\beta_3 \text{GDP}_i, \\
  \log(\phi_i)&=\gamma_1+\gamma_2 \text{GDP}_i,
\end{split}
\end{equation*}
for $i=1,...,80$. For some values of $\alpha$, the SMLE and MDPDE could not be computed and the data-driven algorithm for selecting $\alpha$ did not reach stability and returned $\alpha=0$ (MLE). The estimates and standard errors for the LSMLE and LMDPDE are similar. Here, we present the results for the LSMLE and those for the LMDPDE are shown in the Supplementary Material (Section 4). Table \ref{AppTab1} presents the estimates, asymptotic standard errors, $z$-statistics (estimate divided by the asymptotic standard error), and asymptotic $p$-values of the Wald-type tests of nullity of coefficients. It also reports the results without observation $\#1$, which is the most evident outlier. This observation corresponds to a city with an atypical value for HIC, around $0.98$. 

\begin{table}[ht]
\centering
\caption{Estimates, asymptotic standard errors (Std. error), $z$-stat, and asymptotic $p$-values for the full and reduced data.} \label{AppTab1}
\scriptsize
\def\arraystretch{1.1}
\begin{tabular}{rrrrrrrrrr}
\hline
& \multicolumn{4}{c}{MLE for the full data}    & & \multicolumn{4}{c}{LSMLE for the full data}\\
\cline{2-5} \cline{7-10}
& Estimate & Std. error & $z$-stat & $p$-value & & Estimate & Std. error & $z$-stat & $p$-value \\
\cline{2-5} \cline{7-10}
\textit{mean submodel} &&&&&&&&&\\
Intercept  & $-4.429$   & $0.629$      & $-7.037$   & $0.000$     & & $-5.992$   & $0.518$      & $-11.562$   & $0.000$     \\ 
Urb        & $3.429 $   & $0.706$      & $4.857 $   & $0.000$     & & $4.884$    & $0.581$      & $ 8.410 $   & $0.000$     \\ 
GDP        & $0.010 $   & $0.003$      & $3.535 $   & $0.000$     & & $0.013$    & $0.003$      & $ 4.652 $   & $0.000$     \\ 
\textit{precision submodel} &&&&&&&&&\\
Intercept  & $2.086 $   & $0.229$      & $9.115 $   & $0.000$     & & $3.332  $  & $0.229$      & $14.573$   & $0.000$     \\ 
GDP        & $-0.002$   & $0.005$      & $-0.456$   & $0.648$     & & $-0.013$   & $0.004$      & $-2.934$   & $0.003$     \\ 
\cline{2-5} \cline{7-10}
& \multicolumn{4}{c}{MLE without observation $\#1$}    & & \multicolumn{4}{c}{LSMLE without observation $\#1$}\\
\cline{2-5} \cline{7-10}
& Estimate & Std. error & $z$-stat & $p$-value & & Estimate & Std. error & $z$-stat & $p$-value \\
\cline{2-5} \cline{7-10}
\textit{mean submodel}&&&&&&&&&\\
Intercept  & $-5.978$   & $0.508$      & $-11.777$  & $0.000$     & & $-5.978$   & $0.508$      & $-11.777$  & $0.000$     \\ 
Urb        &$ 4.854$    & $0.569$      &$ 8.538  $  & $0.000$     & & $4.854 $   & $0.569$      &$ 8.538 $   & $0.000$     \\ 
GDP        &$ 0.013$    & $0.003$      &$ 4.721  $  & $0.000$     & & $0.013 $   & $0.003$      &$ 4.721 $   & $0.000$     \\ 
\textit{precision submodel}&&&&&&&&&\\
Intercept  & $3.391 $   & $0.229$      & $14.816$   & $0.000$     & & $3.391  $  & $0.229$      & $14.816$   & $0.000$     \\ 
GDP        & $-0.013$   & $0.004$      & $-3.003$   & $0.003$     & & $-0.013$   & $0.004$      & $-3.003$   & $0.003$     \\ 
\hline
\end{tabular}
\end{table}

For the full data, the data-driven algorithm for selecting the optimum $\alpha$ returned $\alpha=0.06$ for both LMDPDE and LSMLE, indicating that a robust fit is needed. For the data without observation $\#1$, the algorithm returned $\alpha=0$ (MLE) for both the new robust estimators. As we observe in Table \ref{AppTab1}, the MLE is highly influenced by observation $\#1$. For instance, the estimated coefficient for Urb in the mean submodel moves from $3.429$ (full data) to $4.854$ (reduced data). Additionally, the covariate GDP in the precision submodel is non-significant ($p$-value equal to $0.648$) for the full data and highly significant ($p$-value equal to $0.003$) for the reduced data. In contrast, the results based on the LSMLE are not impacted by the exclusion of the outlier observation. The results for the LSMLE for the full data are close to those for the MLE for the reduced data.

\begin{figure}[!h]
\captionsetup[subfigure]{labelformat=empty}
\centering
\includegraphics[scale=0.3]{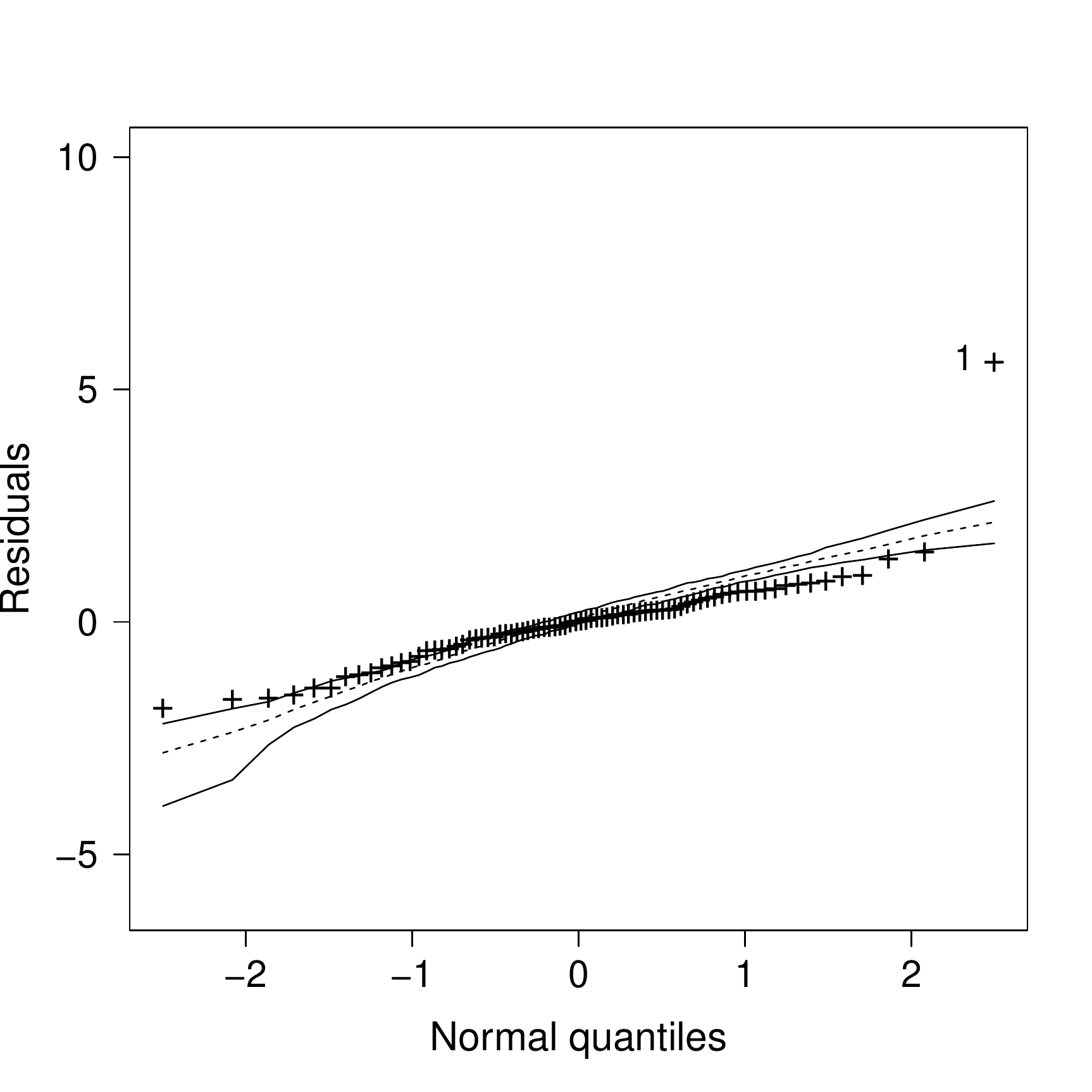}
\includegraphics[scale=0.3]{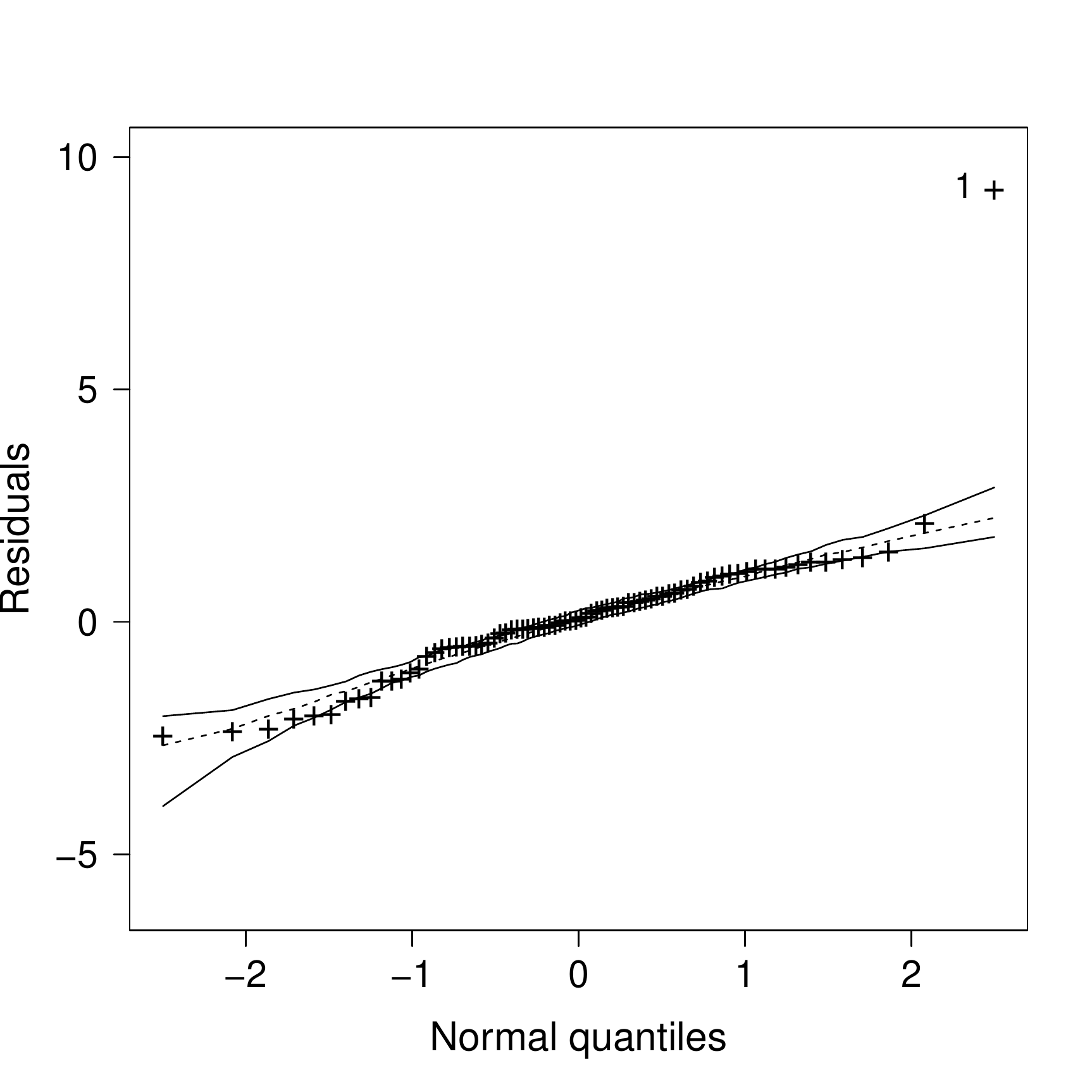}
\includegraphics[scale=0.3]{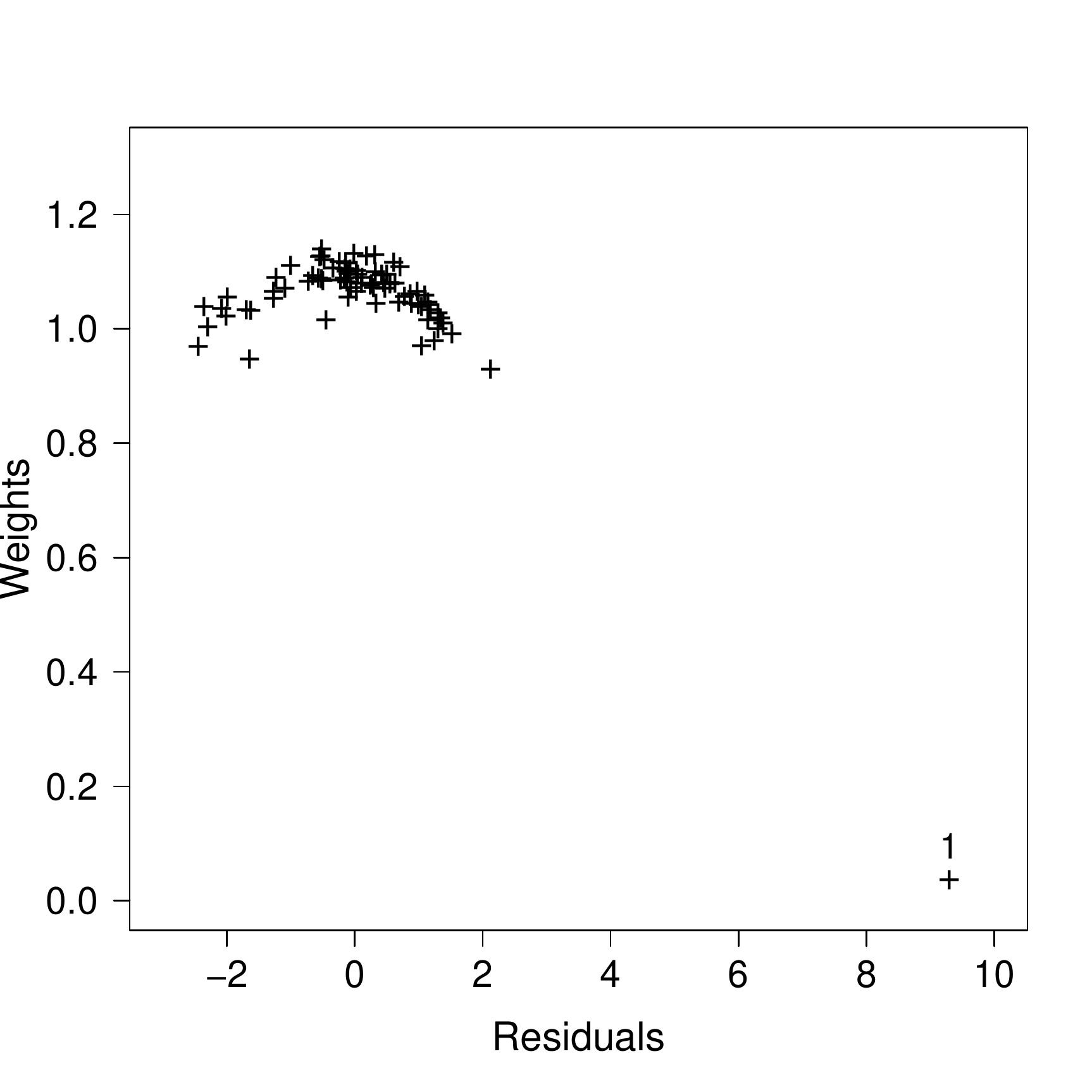}
\includegraphics[scale=0.3]{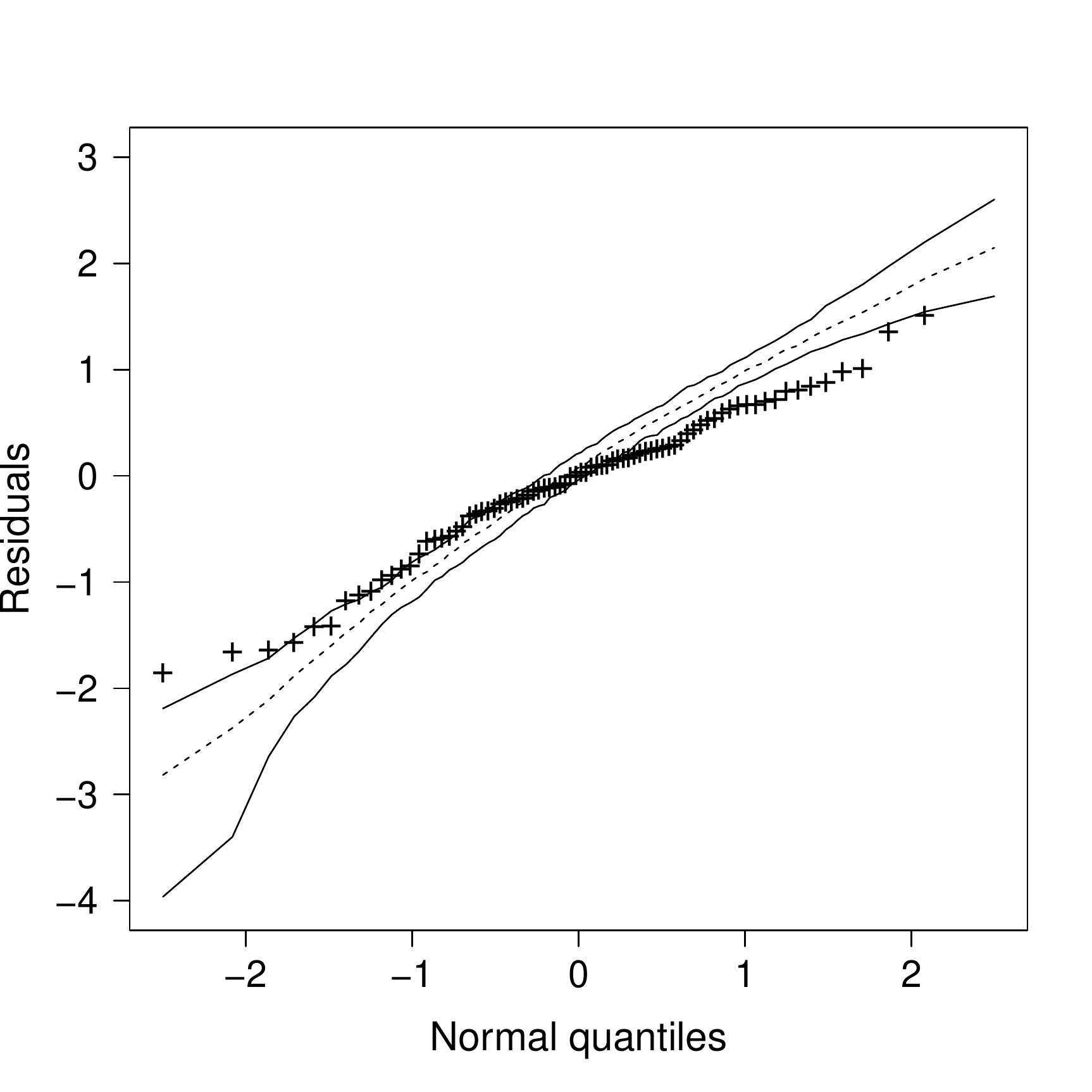}
\includegraphics[scale=0.3]{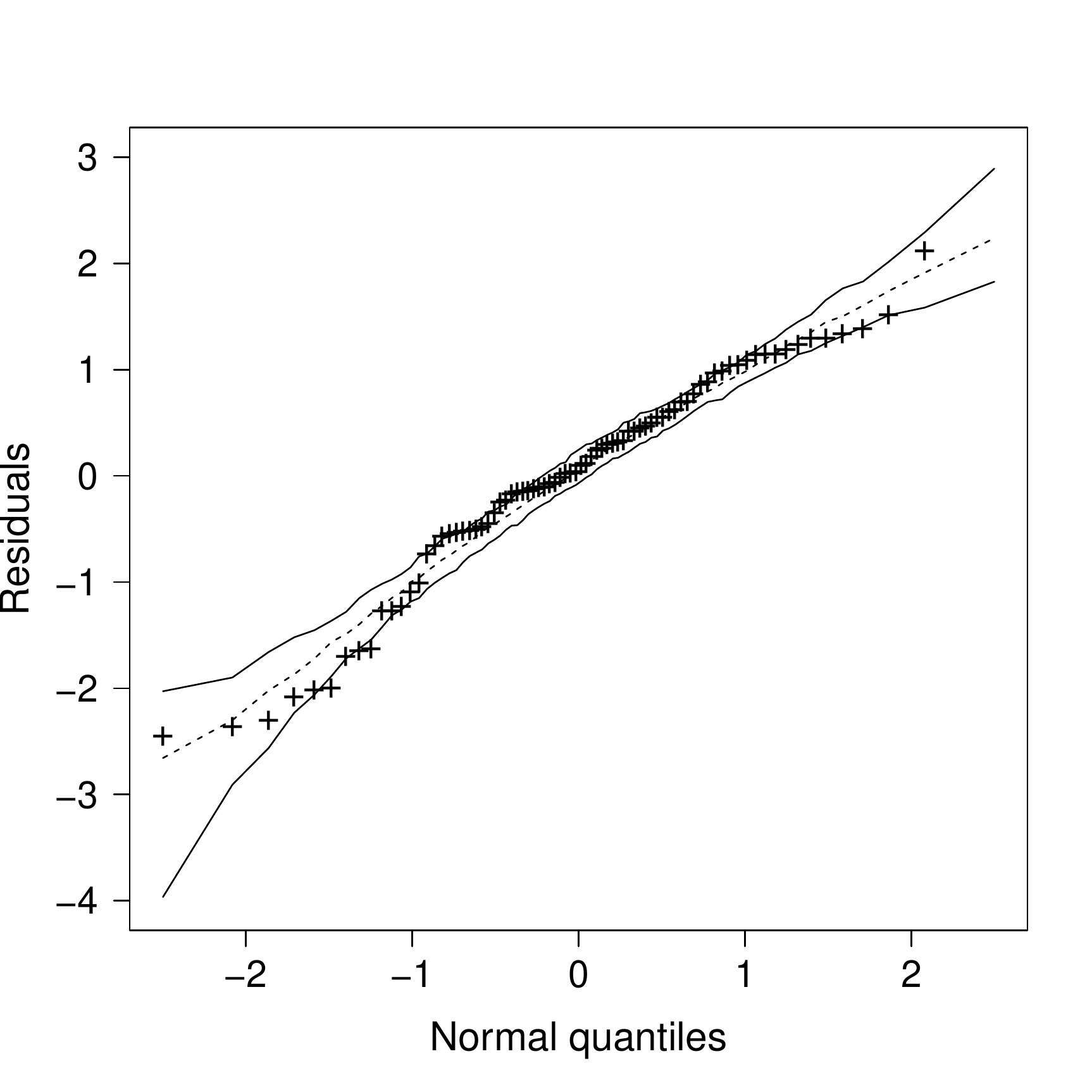}
\includegraphics[scale=0.3]{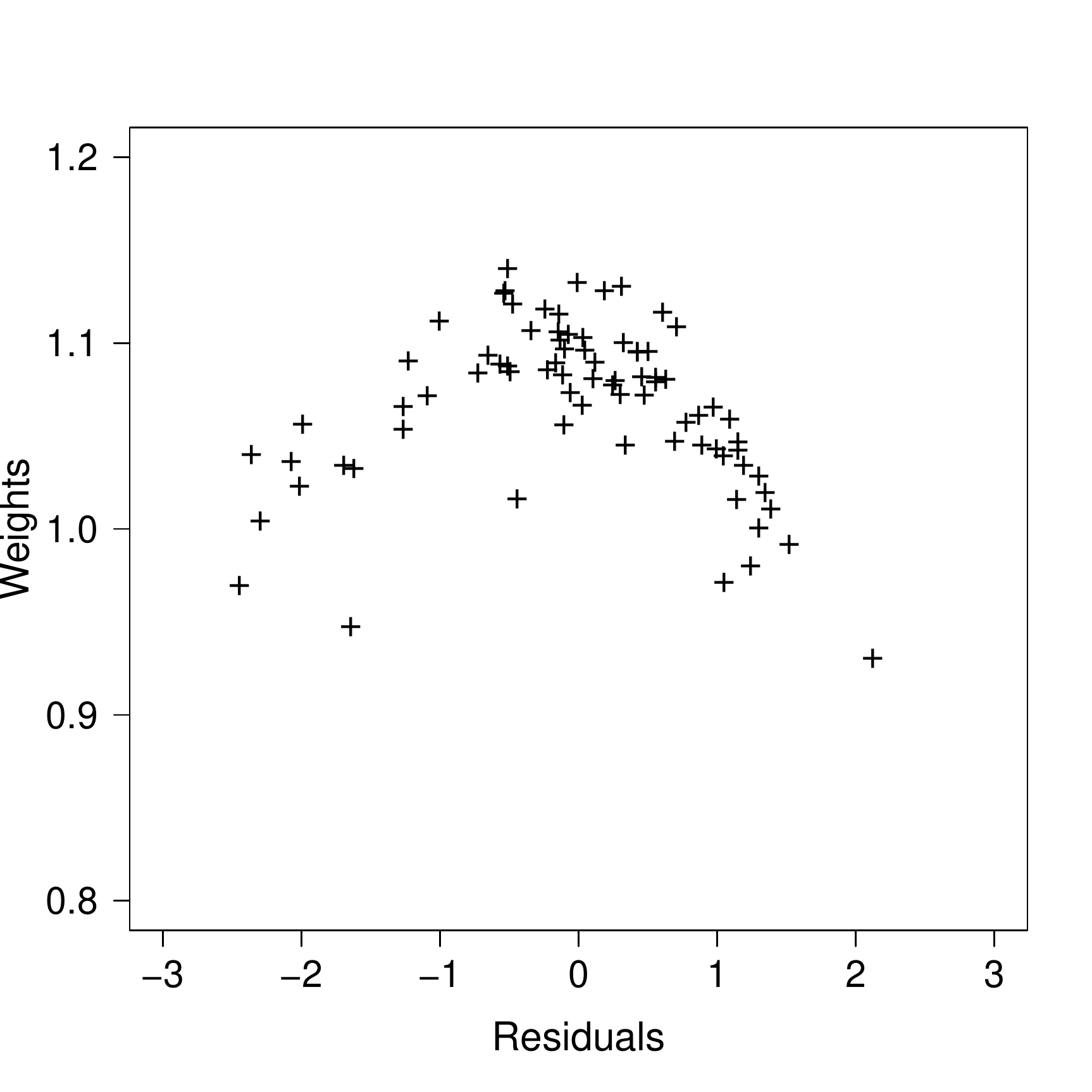}
 \caption{Normal probability plots with simulated envelope of the residuals for MLE (first column) and LSMLE (second column) and plot of estimated weights for LSMLE (third column). The plots in the second row are zoomed versions of those in the first row.}
     \label{residuals}
\end{figure}

Following \cite{RibeiroFerrari}, Figure \ref{residuals} presents the normal probability plots with simulated envelopes of residuals for the MLE and the LSMLE and the plot of estimated weights against residuals for the LSMLE. We consider the `standardized weighted residual 2' proposed by \cite{EspinheiraEtAll}. The residual plots of the MLE clearly evidence the lack of fit of the maximum likelihood estimation. As expected, observation $\#1$ is highlighted as an outlier for both the MLE and the LSMLE fits. The residual plots for the LSMLE suggest a suitable fit for all the observations except for case $\#1$. In fact, this observation receives a weight close to zero for the LSMLE fit. 

\section{Concluding remarks}\label{Sec.CR}

This paper introduces two new robust estimators for the beta regression models: the LSMLE and the LMDPDE. The proposed estimators overcome the limitations of the current robust estimators, the SMLE and the MDPDE. Simulation results and a real data application evidence the excellent performance of the new estimators even in situations where the existing estimators fail. The new robust estimators present similar behavior and are easily implemented. Practitioners may employ our R codes for their own analyses. An R package for robust beta regression inference and diagnostics is under development and will soon be available. 

The development of robust estimators for inflated beta regression models \citep{OspinaFerrari} is a natural, although non-trivial, extension of our work. The second and third authors have been working on this topic. The findings will be reported elsewhere.

\section*{Acknowledgments}
This work was financed in part by the Coordena\c{c}\~ao de Aperfei\c{c}oamento de Pessoal de N\'ivel Superior - Brazil (CAPES) - Finance Code 001 and by the Conselho Nacional de Desenvolvimento Cient\'ifico e Tecnol\'ogico - Brazil (CNPq). The second author gratefully acknowledges the funding provided by CNPq (Grant No. 305963-2018-0). A special thank goes to Terezinha T.K.A. Ribeiro for helpful discussion and for sharing her R code on which we based our implementation of the estimators developed in this paper.
	

\appendix
\section{Appendix}\label{Appendix.Int}

The matrices $\bo{\Lambda}_{1, \alpha}(\bo{\theta})$ and $\bo{\Sigma}_{1, \alpha}(\bo{\theta})$ used in the covariance matrix of the LMDPDE are
\begin{align*}
\bo{\Lambda}_{1, \alpha}(\bo{\theta}) = 
\begin{bmatrix}
\bm{{X}}^{\top}  \gamma_{11}^{(1+\alpha)} \bm{{X}} & \bm{{X}}^{\top} \gamma_{12}^{(1+\alpha)} \bm{{Z}}\\
\bm{{Z}}^{\top} \gamma_{12}^{(1+\alpha)} \bm{{X}} & \bm{{Z}}^{\top} \gamma_{22}^{(1+\alpha)} \bm{{Z}}
\end{bmatrix}
\end{align*}
and
\begin{align*}
\bo{\Sigma}_{1, \alpha}(\bo{\theta}) = 
\begin{bmatrix}
\bm{{X}}^{\top}  \left[ \gamma_{11}^{(1+2\alpha)} - {\gamma_{1}^{(1+\alpha)}}^2 \right] \bm{{X}} & \bm{{X}}^{\top} \left[ \gamma_{12}^{(1+2\alpha)} - {\gamma_{1}^{(1+\alpha)} \gamma_{2}^{(1+\alpha)}} \right] \bm{{Z}}\\
\bm{{Z}}^{\top} \left[ \gamma_{12}^{(1+2\alpha)} - {\gamma_{1}^{(1+\alpha)} \gamma_{2}^{(1+\alpha)}} \right] \bm{{X}} & \bm{{Z}}^{\top} \left[ \gamma_{22}^{(1+2\alpha)} - {\gamma_{2}^{(1+\alpha)}}^2 \right] \bm{{Z}}
\end{bmatrix},
\end{align*}
where $\gamma_{j}^{(\alpha)} = \diag \{ \gamma_{j, i}^{(\alpha)}; i=1, \ldots, n \}$, for $j=1, 2$, $\gamma_{11}^{(\alpha)} = \diag \{ \gamma_{11, i}^{(\alpha)}; i=1, \ldots, n  \}$, $\gamma_{12}^{(\alpha)} = \diag \{ \gamma_{12, i}^{(\alpha)}; i=1, \ldots, n  \}$, $\gamma_{22}^{(\alpha)} = \diag \{ \gamma_{22, i}^{(\alpha)}; i=1, \ldots, n \}$, with
\begin{align*}
\gamma_{11, i}^{(\alpha)} & = \dfrac{\phi_i^2 \mathcal{K}_{i, \alpha}(\bm{\theta}) }{[g'_\mu(\mu_i)]^2} \left[ v_{i, \alpha} + (\mu_{i,\alpha}^{\star} -\mu^\star_{i} )^2 \right], \\
\gamma_{12, i}^{(\alpha)} & =\dfrac{\phi_i \mathcal{K}_{i, \alpha}(\bm{\theta}) }{g'_\mu(\mu_i)g'_\phi(\phi_i)} \left\{ \mu_i \left[ v_{i, \alpha} + (\mu_{i, \alpha}^{ \star} -\mu^\star_{i} )^2 \right] -  \psi'((1-\mu_{i})\phi_{i, \alpha} ) + (\mu_{i, \alpha}^{ \star} -\mu^\star_{i})(\mu_{i, \alpha}^{ \dagger} -\mu^\dagger_{i}) \right\},\\
\gamma_{22, i}^{(\alpha)} & = \dfrac{ \mathcal{K}_{i, \alpha}(\bm{\theta}) }{[g'_\phi(\phi_i)]^2}\left\{ \mu_i^2  \psi'(\mu_{i}\phi_{i, \alpha}) + (1-\mu_i)^2\psi'((1-\mu_{i})\phi_{i, \alpha} ) - \psi'(\phi_{i, \alpha} ) \right. \\
& \hspace*{2.3cm} + \left. \left[\mu_i(\mu_{i, \alpha}^{ \star} -\mu^\star_{i}) + (\mu_{i, \alpha}^{ \dagger} -\mu^\dagger_{i})\right]^2 \right\},
\end{align*}
and 
$v_{i, \alpha} = \psi'(\mu_{i} {\phi}_{i, \alpha}) + \psi'((1-\mu_{i}) {\phi}_{i, \alpha})$.

For the LSMLE, the matrices $\bo{\Lambda}_{2, \alpha}(\bo{\theta})$ and $\bo{\Sigma}_{2, \alpha}(\bo{\theta})$ are given by
\begin{align*}
\bo{\Lambda}_{2, \alpha}(\bo{\theta}) = -
\begin{bmatrix}
(1-\alpha) \bm{{X}}^{\top}  \bm{B}_1 {\bm{T}}_\mu^2 \bm{\Phi}^2 \bm{V} \bm{{X}} & \bm{{X}}^{\top}  \bm{B}_1 {\bm{T}_\mu} \bm{T}_\phi^{ \ast} \bm{C} \bm{{Z}}\\
\bm{{Z}}^{\top} \bm{B}_1 {\bm{T}_\mu} \bm{T}_\phi^{ \ast} \bm{C} \bm{{X}} & (1-\alpha)^{-1}\bm{{Z}}^{\top} \bm{B}_1 {\bm{T}_\phi^{\ast}}^2 \bm{D} \bm{{Z}}
\end{bmatrix}
\end{align*}
and
\begin{align*}
\bo{\Sigma}_{2, \alpha}(\bo{\theta}) = 
\begin{bmatrix}
\bm{{X}}^{\top} \bm{B}_2 {\bm{T}}_\mu^2 \bm{\Phi}^2 \bm{V}_{1+\alpha} \bm{{X}} & (1-\alpha)^{-1} \bm{{X}}^{\top} \bm{B}_2 {\bm{T}_\mu} \bm{T}_\phi^{ \ast} \bm{C}_{1+\alpha} \bm{{Z}}\\
(1-\alpha)^{-1}\bm{{Z}}^{\top} \bm{B}_2 {\bm{T}_\mu} \bm{T}_\phi^{ \ast} \bm{C}_{1+\alpha} \bm{{X}} & (1-\alpha)^{-2}\bm{{Z}}^{\top} \bm{B}_2 {\bm{T}_\phi^{ \ast}}^2 \bm{D}_{1+\alpha} \bm{{Z}}
\end{bmatrix},
\end{align*}
where $\bm{B}_j = \diag\{b_{i, j}; i=1,\ldots, n \}$, $j=1,2$,
\[
b_{i, 1} = \dfrac{B(\mu_i \phi_i, (1-\mu_i) \phi_i)^{1-\alpha}}{B(\mu_{i} {\phi}_{i, 1-\alpha}, (1-\mu_{i}) {\phi}_{i, 1-\alpha})},
\]
\[ b_{i, 2} = \dfrac{B(\mu_{i} {\phi}_{i, 1+\alpha}, (1-\mu_{i}) {\phi}_{i, 1+\alpha})}{B(\mu_i \phi_i, (1-\mu_i) \phi_i)^{2\alpha} B(\mu_{i} {\phi}_{i, 1-\alpha}, (1-\mu_{i}) {\phi}_{i,  1-\alpha})},
\]
$\bm{T}_\mu = \diag\{t_{i, \mu}; i=1,\ldots, n \}$, $\bm{T}_\phi^{ \ast} = \diag\{t_{i, \phi}; i=1,\ldots, n \}$, 
\[
t_{i, \mu} = \left[ g'_\mu(\mu_{i}) \right]^{-1}, \quad t_{i, \phi} = \left[ g'_\phi({\phi}_{i, 1-\alpha}) \right]^{-1},
\]
$\bm{\Phi} = \diag\{\phi_{i}; i=1,\ldots, n \}$, $\bm{V} = \diag\{v_i; i=1,\ldots, n \}$, $\bm{V}_{1+\alpha} = \diag\{v_{i, 1+\alpha}; i=1,\ldots, n\}$,
$\bm{C} = \diag\{c_i; i=1,\ldots, n \}$, $\bm{C}_{1+\alpha} = \diag\{c_{i, 1+\alpha}; i=1,\ldots, n \}$,
\begin{align*}
c_i & =  \phi_{i} \left[ \mu_{i}\psi'(\mu_i \phi_i) - (1-\mu_{i}) \psi'((1-\mu_i) \phi_i) \right]  ,\\
c_{i, 1+\alpha} & = \phi_{i} \left[ \mu_{i}\psi'(\mu_{i} {\phi}_{i, 1+\alpha}) - (1-\mu_{i}) \psi'((1-\mu_{i}) {\phi}_{i, 1+\alpha}) \right],
\end{align*}
$\bm{D} = \diag\{d_i; i=1,\ldots, n \}$, $\bm{D}_{1+\alpha} = \diag\{d_{i, 1+\alpha}; i=1,\ldots, n \}$,
\begin{align*}
d_i & =   \mu_{i}^2 \psi'(\mu_i \phi_i) + (1-\mu_{i})^2 \psi'((1-\mu_i) \phi_i) -\psi'(\phi_i),\\
d_{i, 1+\alpha} & =  \mu_{i}^2\psi'(\mu_{i} {\phi}_{i, 1+\alpha}) + (1-\mu_{i})^2 \psi'((1-\mu_{i}) {\phi}_{i, 1+\alpha}) - \psi'( {\phi}_{i, 1+\alpha}).
\end{align*}

\bibliographystyle{spbasic}
\bibliography{bancoref}

\end{document}


\maketitle

The results in the next three sections follow \cite{RibeiroFerrari}.

\section{Fisher-consistency}

In order to prove the Fisher-consistency of the LSMLE we show that the individual estimating function in eq. (5) of the paper is unbiased, i.e., $\mathbb{E}\left[ \bm{U}^\ast(y^\star; \bm{\theta})h^\ast_{\bm{\theta}}(y^\star; \mu, \phi)^\alpha \right] = \bm{0}$, for all $\bm{\theta} \in \mathbb{R}^p$. In fact, 
\begin{align*}
\mathbb{E}\left[ \bm{U}^\ast(y^\star; \bm{\theta})h^\ast_{\bm{\theta}}(y^\star; \mu, \phi)^\alpha \right] & = \displaystyle\int_{-\infty}^\infty \bm{U}^\ast(y^\star; \bm{\theta})h^\ast_{\bm{\theta}}(y^\star; \mu, \phi)^\alpha h_{\bm{\theta}}(y^\star; \mu, \phi) \dd y^\star\\
&=  \displaystyle\int_{-\infty}^\infty  \left[ \nabla_{\bm{\theta}} \log(h^\ast_{\bm{\theta}}(y^\star; \mu, \phi)) \right]  h^\ast_{\bm{\theta}}(y^\star; \mu, \phi)^\alpha h_{\bm{\theta}}(y^\star; \mu, \phi) \dd y^\star\\
&=  \displaystyle\int_{-\infty}^\infty \dfrac{\nabla_{\bm{\theta}} h^\ast_{\bm{\theta}}(y^\star; \mu, \phi) }{h^\ast_{\bm{\theta}}(y^\star; \mu, \phi)}    h^\ast_{\bm{\theta}}(y^\star; \mu, \phi)^\alpha h_{\bm{\theta}}(y^\star; \mu, \phi) \dd y^\star\\
&=  \displaystyle\int_{-\infty}^\infty  \dfrac{\nabla_{\bm{\theta}} h^\ast_{\bm{\theta}}(y^\star; \mu, \phi) }{h^\ast_{\bm{\theta}}(y^\star; \mu, \phi)^{1-\alpha}}  h_{\bm{\theta}}(y^\star; \mu, \phi) \dd y^\star.
\end{align*}
Since $h_{\bm{\theta}}(y^\star; \mu, \phi) = c_\alpha(\bm{\theta}) h^\ast_{\bm{\theta}}(y^\star; \mu, \phi)^{1-\alpha}$, with $c_\alpha(\bm{\theta}) = \left[ \int_{-\infty}^\infty h_{\bm{\theta}}(y^\star; \mu, \phi)^{1/(1-\alpha)} \dd y^\star \right]^{1-\alpha}$, it follows that
\begin{align*}
\mathbb{E}\left[ \bm{U}^\ast(y^\star; \bm{\theta})h^\ast_{\bm{\theta}}(y^\star; \mu, \phi)^\alpha \right] &=  c_\alpha(\bm{\theta}) \displaystyle\int_{-\infty}^\infty  \nabla_{\bm{\theta}}  h^\ast_{\bm{\theta}}(y^\star; \mu, \phi) \dd y^\star\\
& = c_\alpha(\bm{\theta}) \nabla_{\bm{\theta}} \displaystyle\int_{-\infty}^\infty   h^\ast_{\bm{\theta}}(y^\star; \mu, \phi) \dd y^\star\\
& = 0, \quad \forall~\bm{\theta} \in \mathbb{R}^p. 
\end{align*}

The Fisher-consistency of the LMDPDE can be directly verified through eq. (4) of the paper.

\section{Asymptotic covariance matrices}

We now present the steps to obtain the asymptotic covariance matrix of the LSMLE. Those for the LMDPDE are similar. 

The covariance matrix of the LSMLE is given by $\bo{V}_{2, \alpha}(\bo{\theta}) = \bo{\Lambda}_{2, \alpha}^{-1}(\bo{\theta}) \bo{\Sigma}_{2, \alpha}(\bo{\theta})\bo{\Lambda}_{2, \alpha}^{-1}(\bo{\theta})$, where
\begin{align*}
\begin{split}
\bo{\Lambda}_{2, \alpha} &= \displaystyle\sum_{i=1}^n \mathbb{E} \left\{ \nabla_{\bm{\theta}^\top} \left[ \bm{U}^\ast(y_i^\star; \bm{\theta})h^\ast_{\bm{\theta}}(y_i^\star; \mu, \phi)^\alpha \right] \right\}
\end{split}
\end{align*}
and 
\begin{align*}
\begin{split}
\bo{\Sigma}_{2, \alpha}(\bo{\theta}) & =  \displaystyle\sum_{i=1}^n \mathbb{E} \left\{  \bm{U}^\ast(y_i^\star; \bm{\theta}) \bm{U}^\ast(y_i^\star; \bm{\theta})^\top h^\ast_{\bm{\theta}}(y_i^\star; \mu_i, \phi_i)^{2\alpha} \right\}.
\end{split}
\end{align*}
Let the following partitions $\bm{U}^\ast(y_i^\star; \bm{\theta}) = \left( \bm{U}_{\bm{\beta}}^\ast(y_i^\star; \bm{\theta})^\top, \bm{U}_{\bm{\gamma}}^\ast(y_i^\star; \bm{\theta})^\top \right)^\top$,
\[
\bo{\Lambda}_{2, \alpha} = \begin{bmatrix}
\bo{\Lambda}_{2, \alpha}^{\bm{\beta}\bm{\beta}} & \bo{\Lambda}_{2, \alpha}^{\bm{\beta}\bm{\gamma}}  \\
\bo{\Lambda}_{2, \alpha}^{\bm{\gamma}\bm{\beta}}  & \bo{\Lambda}_{2, \alpha}^{\bm{\gamma}\bm{\gamma}}  
\end{bmatrix}, \quad  \quad \bo{\Sigma}_{2, \alpha} = \begin{bmatrix}
\bo{\Sigma}_{2, \alpha}^{\bm{\beta}\bm{\beta}} & \bo{\Sigma}_{2, \alpha}^{\bm{\beta}\bm{\gamma}}  \\
\bo{\Sigma}_{2, \alpha}^{\bm{\gamma}\bm{\beta}}  & \bo{\Sigma}_{2, \alpha}'^{\bm{\gamma}\bm{\gamma}} 
\end{bmatrix}.
\]

First, we obtain $\bo{\Lambda}_{2, \alpha}(\bm{\theta})$. Note that
\begin{align}\label{derivativescore}
\begin{split}
\nabla_{\bm{\theta}^\top} \left[ \bm{U}^\ast(y_i^\star; \bm{\theta})h^\ast_{\bm{\theta}}(y_i^\star; \mu_i, \phi_i)^\alpha \right] & =  \left[ \nabla_{\bm{\theta}^\top} \bm{U}^\ast(y_i^\star; \bm{\theta})  \right]h^\ast_{\bm{\theta}}(y_i^\star; \mu_i, \phi_i)^\alpha \\
& \hspace*{0.5cm}+  \bm{U}^\ast(y_i^\star; \bm{\theta}) \nabla_{\bm{\theta}^\top} \exp\{ \alpha \log(h^\ast_{\bm{\theta}}(y_i^\star; \mu_i, \phi_i))  \}\\
& = \left[ \nabla_{\bm{\theta}^\top} \bm{U}^\ast(y_i^\star; \bm{\theta})  \right]h^\ast_{\bm{\theta}}(y_i^\star; \mu_i, \phi_i)^\alpha \\
& \hspace*{0.5cm} + \alpha \bm{U}^\ast(y_i^\star; \bm{\theta}) \bm{U}^\ast(y_i^\star; \bm{\theta})^\top h^\ast_{\bm{\theta}}(y_i^\star; \mu_i, \phi_i)^\alpha.
\end{split}
\end{align}
Then, 
\begin{align*}
\mathbb{E} \left\{  \nabla_{\bm{\theta}^\top} \left[ \bm{U}^\ast(y_i^\star; \bm{\theta})h^\ast_{\bm{\theta}}(y_i^\star; \mu_i, \phi_i)^\alpha \right]    \right\}  & = \int_{-\infty}^\infty \left[ \nabla_{\bm{\theta}^\top}  \bm{U}^\ast(y_i^\star; \bm{\theta})\right] h^\ast_{\bm{\theta}}(y_i^\star; \mu_i, \phi_i)^\alpha h_{\bm{\theta}}(y_i^\star; \mu_i, \phi_i) \dd y_i^\star \\
& + \alpha \int_{-\infty}^\infty  \bm{U}^\ast(y_i^\star; \bm{\theta}) \bm{U}^\ast(y_i^\star; \bm{\theta})^\top h^\ast_{\bm{\theta}}(y_i^\star; \mu_i, \phi_i)^\alpha h_{\bm{\theta}}(y_i^\star; \mu_i, \phi_i) \dd y_i^\star\\
& = b_{i, 1} \int_{-\infty}^\infty \left[ \nabla_{\bm{\theta}^\top}  \bm{U}^\ast(y_i^\star; \bm{\theta})\right] h^\ast_{\bm{\theta}}(y_i^\star; \mu_i, \phi_i) \dd y_i^\star \\
& \hspace*{0.5cm}+ \alpha b_{i, 1} \int_{-\infty}^\infty  \bm{U}^\ast(y_i^\star; \bm{\theta}) \bm{U}^\ast(y_i^\star; \bm{\theta})^\top h^\ast_{\bm{\theta}}(y_i^\star; \mu_i, \phi_i) \dd y_i^\star\\
& = b_{i, 1} \mathbb{E} \left\{  \nabla_{\bm{\theta}^\top} \bm{U}^\ast(y_i^\star; \bm{\theta})    \right\} + b_{i, 1} \alpha \mathbb{E} \left\{  \bm{U}^\ast(y_i^\star; \bm{\theta}) \bm{U}^\ast(y_i^\star; \bm{\theta})^\top     \right\},
\end{align*}
in which the second equality comes from the fact that
\[
h^\ast_{\bm{\theta}}(y_i^\star; \mu_i, \phi_i)^\alpha h_{\bm{\theta}}(y_i^\star; \mu_i, \phi_i) = b_{i, 1} h^\ast_{\bm{\theta}}(y_i^\star; \mu_i, \phi_{i}).
\]

After some algebra, it follows that
\begin{align*}
\nabla_{\bm{\beta}^\top} \bm{U}_{\bm{\beta}}^\ast(y_i^\star; \bm{\theta}) &= -\dfrac{\phi_i}{[g'_\mu(\mu_i)]^2} \left[ \phi v_i + (y_i^\star - \mu_i^\star) \dfrac{g^{''}_\mu(\mu_i)}{g'_\mu(\mu_i)}  \right] \bm{X}_i^\top \bm{X}_i,\\
\nabla_{\bm{\gamma}^\top} \bm{U}_{\bm{\gamma}}^\ast(y_i^\star; \bm{\theta}) &= -\dfrac{(1-\alpha)^{-1}}{[g'_\phi({\phi}_{i, 1-\alpha})]^2} \left\{    (1-\alpha)^{-1} d_i + \dfrac{g^{''}_\phi({\phi}_{i, 1+\alpha})}{g'_\phi({\phi}_{i, 1-\alpha})} [\mu_i(y_i^\star - \mu_i^\star) + (y_i^\dagger - \mu_i^\dagger)] \right\}  \bm{Z}_i^\top \bm{Z}_i,\\
\nabla_{\bm{\beta}^\top} \bm{U}_{\bm{\gamma}}^\ast(y_i^\star; \bm{\theta}) & = -\dfrac{(1-\alpha)^{-1}}{g'_\mu(\mu_i) g'_\phi({\phi}_{i, 1-\alpha})} f_i \bm{Z}_i^\top \bm{X}_i,
\end{align*}
where $f_i = c_i - (y_i^\star - \mu_i^\star)$. Also, 
\begin{align*}
\bm{U}_{\bm{\beta}}^\ast(y_i^\star; \bm{\theta}) \bm{U}_{\bm{\beta}}^\ast(y_i^\star; \bm{\theta})^\top & =\phi_i^2 \dfrac{(y_i^\star - \mu_i^\star)^2}{[g'_\mu(\mu_i)]^2} \bm{X}_i^\top \bm{X}_i,\\
\bm{U}_{\bm{\gamma}}^\ast(y_i^\star; \bm{\theta}) \bm{U}_{\bm{\gamma}}^\ast(y_i^\star; \bm{\theta})^\top & =\dfrac{(1-\alpha)^{-2}}{[g'_\phi({\phi}_{i, 1-\alpha})]^2} \left[  \mu_i(y_i^\star - \mu_i^\star) + (y_i^\dagger - \mu_i^\dagger) \right]^2  \bm{Z}_i^\top \bm{Z}_i,\\
\bm{U}_{\bm{\gamma}}^\ast(y_i^\star; \bm{\theta}) \bm{U}_{\bm{\beta}}^\ast(y_i^\star; \bm{\theta})^\top & =\dfrac{\phi_i(1-\alpha)^{-1}}{g'_\mu(\mu_i) g'_\phi({\phi}_{i, 1-\alpha})} (y_i^\star - \mu_i^\star) \left[  \mu_i(y_i^\star - \mu_i^\star) + (y_i^\dagger - \mu_i^\dagger)\right]  \bm{Z}_i^\top \bm{X}_i.
\end{align*}

To obtain $\bo{\Lambda}_{2, \alpha}^{\bm{\beta}\bm{\beta}}$, we have that
\begin{align*}
\mathbb{E} \left\{ \nabla_{\bm{\beta}^\top} \bm{U}_{\bm{\beta}}^\ast(y_i^\star; \bm{\theta})    \right\} & = \mathbb{E} \left\{ \phi_i^2 \dfrac{(y_i^\star - \mu_i^\star)^2}{[g'_\mu(\mu_i)]^2} \bm{X}_i^\top \bm{X}_i    \right\}\\
& = - \dfrac{\phi_i^2 v_i}{[g'_\mu(\mu_i)]^2} \bm{X}_i^\top \bm{X}_i
\end{align*}
and
\begin{align*}
\mathbb{E} \left\{  \bm{U}_{\bm{\beta}}^\ast(y_i^\star; \bm{\theta}) \bm{U}_{\bm{\beta}}^\ast(y_i^\star; \bm{\theta})^\top     \right\} & = \mathbb{E} \left\{ \phi_i^2 \dfrac{(y_i^\star - \mu_i^\star)^2}{[g'_\mu(\mu_i)]^2} \bm{X}_i^\top \bm{X}_i    \right\}\\
& = \dfrac{\phi_i^2 v_i}{[g'_\mu(\mu_i)]^2} \bm{X}_i^\top \bm{X}_i.
\end{align*}
Hence,
\begin{align*}
\mathbb{E} \left\{  \nabla_{\bm{\beta}^\top} \left[ \bm{U}_{\bm{\beta}}^\ast(y_i^\star; \bm{\theta})h^\ast_{\bm{\theta}}(y_i^\star; \mu_i, \phi_i)^\alpha \right]     \right\}  & = b_{i, 1} \mathbb{E} \left\{  \nabla_{\bm{\beta}^\top} \bm{U}_{\bm{\beta}}^\ast(y_i^\star; \bm{\theta})    \right\} + b_{i, 1} \alpha \mathbb{E} \left\{  \bm{U}_{\bm{\beta}}^\ast(y_i^\star; \bm{\theta}) \bm{U}_{\bm{\beta}}^\ast(y_i^\star; \bm{\theta})^\top     \right\} \\
&= -(1-\alpha) \dfrac{ b_{i, 1} \phi_i^2 v_i}{[g'_\mu(\mu_i)]^2} \bm{X}_i^\top \bm{X}_i.
\end{align*}

The matrix $\bo{\Lambda}_{2, \alpha}^{\bm{\gamma}\bm{\beta}}$ can be obtained using
\begin{align*}
\mathbb{E} \left\{  \nabla_{\bm{\beta}^\top} \bm{U}_{\bm{\gamma}}^\ast(y_i^\star; \bm{\theta})    \right\} & = \mathbb{E} \left\{   -\dfrac{(1-\alpha)^{-1}}{g'_\mu(\mu_i) g'_\phi({\phi}_{i, 1-\alpha})} f_i \bm{Z}_i^\top \bm{X}_i       \right\} = - \dfrac{(1-\alpha)^{-1}}{g'_\mu(\mu_i) g'_\phi({\phi}_{i, 1-\alpha})} c_i \bm{Z}_i^\top \bm{X}_i
\end{align*}
and
\begin{align*}
\mathbb{E} \left\{  \bm{U}_{\bm{\gamma}}^\ast(y_i^\star; \bm{\theta}) \bm{U}_{\bm{\beta}}^\ast(y_i^\star; \bm{\theta})^\top     \right\} & = \mathbb{E} \left\{  \dfrac{\phi_i(1-\alpha)^{-1}}{g'_\mu(\mu_i) g'_\phi({\phi}_{i, 1-\alpha})} (y_i^\star - \mu_i^\star) \left[  \vphantom{y_i^\dagger} \mu_i(y_i^\star - \mu_i^\star) + (y_i^\dagger - \mu_i^\dagger)\right]  \bm{Z}_i^\top \bm{X}_i    \right\}\\
& =  \dfrac{(1-\alpha)^{-1}}{g'_\mu(\mu_i) g'_\phi({\phi}_{i, 1-\alpha})} c_i \bm{Z}_i^\top \bm{X}_i.
\end{align*}
Thus,
\begin{align*}
\mathbb{E} \left\{  \nabla_{\bm{\beta}^\top} \left[ \bm{U}_{\bm{\gamma}}^\ast(y_i^\star; \bm{\theta})h^\ast_{\bm{\theta}}(y_i^\star; \mu_i, \phi_i)^\alpha \right]     \right\}  & = b_{i, 1} \mathbb{E} \left\{  \nabla_{\bm{\beta}^\top} \bm{U}_{\bm{\gamma}}^\ast(y_i^\star; \bm{\theta})    \right\}  + b_{i, 1} \alpha \mathbb{E} \left\{  \bm{U}_{\bm{\beta}}^\ast(y_i^\star; \bm{\theta}) \bm{U}_{\bm{\gamma}}^\ast(y_i^\star; \bm{\theta})^\top     \right\} \\
&=- \dfrac{ b_{i, 1} c_i}{g'_\mu(\mu_i) g'_\phi({\phi}_{i, 1-\alpha})} \bm{Z}_i^\top \bm{X}_i.
\end{align*}
Finally, we can obtain $\bo{\Lambda}_{2, \alpha}^{\bm{\gamma}\bm{\gamma}}$ using 
\begin{align*}
\mathbb{E} \left\{  \nabla_{\bm{\gamma}^\top} \bm{U}_{\bm{\gamma}}^\ast(y_i^\star; \bm{\theta})    \right\} & = -\dfrac{(1-\alpha)^{-1}}{[g'_\phi({\phi}_{i, 1-\alpha})]^2} \mathbb{E} \left\{     (1-\alpha)^{-1} d_i + \dfrac{g^{''}_\phi({\phi}_{i, 1+\alpha})}{g'_\phi({\phi}_{i, 1-\alpha})} \left[ \vphantom{y_i^\dagger} \mu_i(y_i^\star - \mu_i^\star)  \right. \right.\\
& \hspace*{0.5cm} \left. \left.+ (y_i^\dagger - \mu_i^\dagger)\right]    \right\} \bm{Z}_i^\top \bm{Z}_i\\
& = - \dfrac{(1-\alpha)^{-2}}{[g'_\phi({\phi}_{i, 1-\alpha})]^2} d_i \bm{Z}_i^\top \bm{Z}_i
\end{align*}
and
\begin{align*}
\mathbb{E} \left\{  \bm{U}_{\bm{\gamma}}^\ast(y_i^\star; \bm{\theta}) \bm{U}_{\bm{\gamma}}^\ast(y_i^\star; \bm{\theta})^\top     \right\} & = \mathbb{E} \left\{   \dfrac{(1-\alpha)^{-2}}{[g'_\phi({\phi}_{i, 1-\alpha})]^2} \left[ \vphantom{y_i^\dagger} \mu_i(y_i^\star - \mu_i^\star)  + (y_i^\dagger - \mu_i^\dagger) \right]^2  \bm{Z}_i^\top \bm{Z}_i    \right\}\\
& =  \dfrac{(1-\alpha)^{-2}}{[g'_\phi({\phi}_{i, 1-\alpha})]^2} d_i \bm{Z}_i^\top \bm{Z}_i.
\end{align*}
Then,
\begin{align*}
\mathbb{E} \left\{  \nabla_{\bm{\gamma}^\top} \left[ \bm{U}_{\bm{\gamma}}^\ast(y_i^\star; \bm{\theta})h^\ast_{\bm{\theta}}(y_i^\star; \mu_i, \phi_i)^\alpha \right]     \right\}  & = b_{i, 1} \mathbb{E} \left\{  \nabla_{\bm{\gamma}^\top} \bm{U}_{\bm{\gamma}}^\ast(y_i^\star; \bm{\theta})    \right\}   + b_{i, 1} \alpha \mathbb{E} \left\{  \bm{U}_{\bm{\gamma}}^\ast(y_i^\star; \bm{\theta}) \bm{U}_{\bm{\gamma}}^\ast(y_i^\star; \bm{\theta})^\top     \right\} \\
&= - \dfrac{(1-\alpha)^{-1}}{[g'_\phi({\phi}_{i, 1-\alpha})]^2} b_{i, 1} d_i \bm{Z}_i^\top \bm{Z}_i.
\end{align*}

To obtain $\bo{\Sigma}_{2, \alpha}(\bo{\theta})$, we have that
\begin{align*}
\mathbb{E} \left\{  \bm{U}^\ast(y_i^\star; \bm{\theta}) \bm{U}^\ast(y_i^\star; \bm{\theta})^\top h^\ast_{\bm{\theta}}\right. & \left. (y_i^\star; \mu_i, \phi_i)^{2\alpha}    \right\}\\
& = \displaystyle \int_{-\infty}^\infty \bm{U}^\ast(y_i^\star; \bm{\theta}) \bm{U}^\ast(y_i^\star; \bm{\theta})^\top h^\ast_{\bm{\theta}}(y_i^\star; \mu_i, \phi_i)^{2\alpha} h_{\bm{\theta}}(y_i^\star; \mu_i, \phi_i) \dd y_i^\star\\
& = b_{i, 2} \displaystyle\int_{-\infty}^\infty \bm{U}^\ast(y_i^\star; \bm{\theta}) \bm{U}^\ast(y_i^\star; \bm{\theta})^\top h^\ast_{\bm{\theta}}(y_i^\star; \mu_i, {\phi}_{i, 1+\alpha}) \dd y_i^\star
\end{align*}
because
\[
h^\ast_{\bm{\theta}}(y_i^\star; \mu_i, \phi_i)^{2\alpha} h_{\bm{\theta}}(y_i^\star; \mu_i, \phi_i) = b_{i, 2} h^\ast_{\bm{\theta}}(y_i^\star; \mu_i, {\phi}_{i, 1+\alpha}).
\]

After some algebra, it follows that
\begin{align*}
\mathbb{E} \left\{  \bm{U}_{\bm{\beta}}^\ast(y_i^\star; \bm{\theta}) \bm{U}_{\bm{\beta}}^\ast(y_i^\star; \bm{\theta})^\top \right. & \left. h^\ast_{\bm{\theta}}(y_i^\star; \mu_i, \phi_i)^{2\alpha}    \right\}\\
& = b_{i, 2} \displaystyle\int_{-\infty}^\infty \bm{U}_{\bm{\beta}}^\ast(y_i^\star; \bm{\theta}) \bm{U}_{\bm{\beta}}^\ast(y_i^\star; \bm{\theta})^\top h^\ast_{\bm{\theta}}(y_i^\star; \mu_i, {\phi}_{i, 1+\alpha}) \dd y_i^\star\\
& =  \dfrac{b_{i, 2} \phi_i^2}{[g'_\mu(\mu_i)]^2} \bm{X}_i^\top \bm{X}_i \displaystyle\int_{-\infty}^\infty (y_i^\star - \mu_i^\star)^2  h^\ast_{\bm{\theta}}(y_i^\star; \mu_i, {\phi}_{i, 1+\alpha}) \dd y_i^\star\\
& =  \dfrac{b_{i, 2} \phi_i^2 v_{i, 1+\alpha}}{[g'_\mu(\mu_i)]^2} \bm{X}_i^\top \bm{X}_i ,
\end{align*}
\begin{align*}
\mathbb{E} \left\{ \right. & \left. \bm{U}_{\bm{\gamma}}^\ast(y_i^\star; \bm{\theta})  \bm{U}_{\bm{\beta}}^\ast(y_i^\star; \bm{\theta})^\top  h^\ast_{\bm{\theta}}(y_i^\star; \mu_i, \phi_i)^{2\alpha}    \right\}\\
& = b_{i, 2} \displaystyle\int_{-\infty}^\infty \bm{U}_{\bm{\gamma}}^\ast(y_i^\star; \bm{\theta}) \bm{U}_{\bm{\beta}}^\ast(y_i^\star; \bm{\theta})^\top h^\ast_{\bm{\theta}}(y_i^\star; \mu_i, {\phi}_{i, 1+\alpha}) \dd y_i^\star\\
& =  \dfrac{b_{i, 2} \phi_i(1-\alpha)^{-1}}{g'_\mu(\mu_i) g'_\phi({\phi}_{i, 1-\alpha})} \bm{Z}_i^\top \bm{X}_i \displaystyle\int_{-\infty}^\infty 
(y_i^\star - \mu_i^\star) \left[  \mu_i(y_i^\star - \mu_i^\star) + (y_i^\dagger - \mu_i^\dagger)\right]^2  
h^\ast_{\bm{\theta}}(y_i^\star; \mu_i, {\phi}_{i, 1+\alpha}) \dd y_i^\star\\
& =  \dfrac{b_{i, 2} \phi_i(1-\alpha)^{-1}}{g'_\mu(\mu_i) g'_\phi({\phi}_{i, 1-\alpha})} \bm{Z}_i^\top \bm{X}_i \left\{ \mu_i \displaystyle\int_{-\infty}^\infty (y_i^\star - \mu_i^\star)^2
h^\ast_{\bm{\theta}}(y_i^\star; \mu_i, {\phi}_{i, 1+\alpha}) \dd y_i^\star \right.\\
 & \hspace*{4.5cm}+ \left.\displaystyle\int_{-\infty}^\infty (y_i^\star - \mu_i^\star)(y_i^\dagger - \mu_i^\dagger) h^\ast_{\bm{\theta}}(y_i^\star; \mu_i, {\phi}_{i, 1+\alpha}) \dd y_i^\star\right\}\\
& = \dfrac{b_{i, 2} \phi_i(1-\alpha)^{-1}}{g'_\mu(\mu_i) g'_\phi({\phi}_{i, 1-\alpha})} \bm{Z}_i^\top \bm{X}_i \left[ \mu_i v'_{i, 1+\alpha} - \psi'((1-\mu_i){\phi}_{i, 1+\alpha})  \right]\\
& =  \dfrac{b_{i, 2}(1-\alpha)^{-1} c_{i, 1+\alpha}}{g'_\mu(\mu_i) g'_\phi({\phi}_{i, 1-\alpha})} \bm{Z}_i^\top \bm{X}_i,
\end{align*}
and
\begin{align*}
\mathbb{E} \left\{  \bm{U}_{\bm{\gamma}}^\ast(y_i^\star; \bm{\theta})\right. & \left. \bm{U}_{\bm{\gamma}}^\ast(y_i^\star; \bm{\theta})^\top  h^\ast_{\bm{\theta}}(y_i^\star; \mu_i, \phi_i)^{2\alpha}    \right\}\\
& = b_{i, 2} \displaystyle\int_{-\infty}^\infty \bm{U}_{\bm{\gamma}}^\ast(y_i^\star; \bm{\theta}) \bm{U}_{\bm{\gamma}}^\ast(y_i^\star; \bm{\theta})^\top h^\ast(y_i^\star; \mu_i, {\phi}_{i, 1+\alpha}) \dd y_i^\star\\
& =  \dfrac{b_{i, 2}(1-\alpha)^{-2}}{[g'_\phi({\phi}_{i, 1-\alpha})]^2} \bm{Z}_i^\top \bm{Z}_i \displaystyle\int_{-\infty}^\infty  \left[  \mu_i(y_i^\star - \mu_i^\star) + (y_i^\dagger - \mu_i^\dagger) \right]^2  h^\ast_{\bm{\theta}}(y_i^\star; \mu_i, {\phi}_{i, 1+\alpha}) \dd y_i^\star\\
& =  \dfrac{b_{i, 2} (1-\alpha)^{-2}d_{i, 1+\alpha}}{[g'_\phi({\phi}_{i, 1-\alpha})]^2} \bm{Z}_i^\top \bm{Z}_i.
\end{align*}

Hence, the matrices  $\bo{\Lambda}_{2, \alpha}(\bm{\theta})$ and $\bo{\Lambda}_{2, \alpha}(\bm{\theta})$ can now be written as in the Appendix of the paper.

\section{Robustness analysis}

In this section, we will show the steps to prove the robustness properties for the LSMLE. For the LMDPDE the steps are similar.

\subsubsection*{Bounded influence function}

The influence function of the LSMLE is 
$\text{IF}(y^\star; \widetilde{\bo{\theta}}_\alpha) = \bo{\Lambda}_{2, \alpha}^{-1}(\bo{\theta})\bm{U}^\ast(y^\star; \bm{\theta})h^\ast_{\bm{\theta}}(y^\star; \mu, \phi)^\alpha$. 
Now, we will show that this function is bounded. Since $\bm{U}^\ast(y^\star; \bm{\theta})h^\ast_{\bm{\theta}}(y^\star; \mu, \phi)^\alpha$ is a continuous function of $y^\star$, it is sufficient to show that $\bm{U}^\ast(y^\star; \bm{\theta})h^\ast_{\bm{\theta}}(y^\star; \mu, \phi)^\alpha$ is finite when $y^\star \rightarrow -\infty$ or $y^\star \rightarrow \infty$. First, note that $h^\ast_{\bm{\theta}}(y^\star; \mu, \phi)^\alpha \rightarrow 0$ when $y^\star$ goes to $-\infty$ or $\infty$. In addition, using the L'Hôspital rule, we have
\begin{align}\label{lim1}
\lim_{y^\star \to \pm \infty} (y^\star - {\mu}^\star) h^\ast_{\bm{\theta}}(y^\star; \mu, \phi)^\alpha = 0
\end{align}
and
\begin{align}\label{lim2}
\lim_{y^\star \to \pm \infty} (y^\dagger - {\mu}^\dagger) h^\ast_{\bm{\theta}}(y^\star; \mu, \phi)^\alpha = \lim_{y^\star \to \pm \infty} (-\log(1+\text{e}^{y^\star}) - {\mu}^\dagger) h^\ast_{\bm{\theta}}(y^\star; \mu, \phi)^\alpha = 0.
\end{align}

Therefore, it is easy to verify that $\bm{U}_{\bm{\beta}}^\ast(y_i^\star; \bm{\theta})h^\ast_{\bm{\theta}}(y^\star; \mu, \phi)^\alpha \rightarrow 0$ and $\bm{U}_{\bm{\gamma}}^\ast(y_i^\star; \bm{\theta})h^\ast_{\bm{\theta}}(y^\star; \mu, \phi)^\alpha \rightarrow 0$  when $y^\star \rightarrow \pm \infty$.

\subsubsection*{Bounded change-of-variance function}

The change-of-variance function of the LSMLE is bounded if the weighted modified score
vector and its first derivative are bounded. Then, it remains to show that the derivative of $\bm{U}^\ast(y^\star; \bm{\theta})h^\ast_{\bm{\theta}}(y^\star; \mu, \phi)^\alpha$, given in \eqref{derivativescore}, is bounded when $y^\star$ goes to $-\infty$ or $\infty$. Using the results in \eqref{lim1}-\eqref{lim2}, we can verify the boundedness of the first term of \eqref{derivativescore}. The boundedness of the second term follows using the L'Hôspital rule.

\section{Additional results for the HIC application}

\begin{table}[ht]
\caption{Estimates, asymptotic standard errors (Std. error), $z$-stat, and asymptotic $p$-values for the MLE and the robust estimators considering the model with covariates Urb and GDP in both mean and precision submodels.}
\centering
\scriptsize
\def\arraystretch{1.1}
\begin{tabular}{rrrrrrrrrr}
\hline
& \multicolumn{4}{c}{MLE}    & & \multicolumn{4}{c}{SMLE ($\alpha  = 0.06$)}\\
\cline{2-5} \cline{7-10}
& Estimate & Std. error & $z$-stat & $p$-value & & Estimate & Std. error & $z$-stat & $p$-value \\
\cline{2-5} \cline{7-10}
\textit{mean submodel} &&&&&&&&&\\
Intercept & $-4.516 $ & $0.681$ & $-6.629 $ & $ 0.000$  & & $ -5.732$  & $0.545$  & $ -10.521$  & $0.000$  \\ 
Urb       & $3.475  $ & $0.767$ & $4.531  $ & $ 0.000$  & & $ 4.581 $  & $0.617$  & $ 7.428  $  & $0.000$  \\ 
GDP       & $0.011  $ & $0.003$ & $3.967  $ & $ 0.000$  & & $ 0.013 $  & $0.003$  & $ 4.758  $  & $0.000$  \\ 
\textit{precision submodel} &&&&&&&&&\\
Intercept & $2.987  $ & $1.019$ & $2.930  $ & $ 0.003$  & & $ 3.991 $  & $1.036$  & $ 3.854  $  & $0.000$  \\ 
Urb       & $-1.040 $ & $1.169$ & $-0.889 $ & $ 0.374$  & & $ -0.838$  & $1.190$  & $ -0.704 $  & $0.481$  \\ 
GDP       & $-0.003 $ & $0.005$ & $-0.595 $ & $ 0.552$  & & $ -0.012$  & $0.004$  & $ -2.788 $  & $0.005$  \\ 
\cline{2-5} \cline{7-10}
& \multicolumn{4}{c}{MDPDE ($\alpha  = 0.42$)}    & & \multicolumn{4}{c}{LSMLE ($\alpha  = 0.06$)}\\
\cline{2-5} \cline{7-10}
& Estimate & Std. error & $z$-stat & $p$-value & & Estimate & Std. error & $z$-stat & $p$-value \\
\cline{2-5} \cline{7-10}
\textit{mean submodel}&&&&&&&&&\\
Intercept & $-6.051 $ & $0.649$ & $-9.319 $ & $ 0.000$  & & $ -5.907$  & $0.552$  & $ -10.693$  & $0.000$  \\ 
Urb       & $4.973  $ & $0.721$ & $6.895  $ & $ 0.000$  & & $ 4.776 $  & $0.624$  & $ 7.658  $  & $0.000$  \\ 
GDP       & $0.014  $ & $0.003$ & $4.675  $ & $ 0.000$  & & $ 0.013 $  & $0.003$  & $ 4.736  $  & $0.000$  \\ 
\textit{precision submodel} &&&&&&&&&\\
Intercept & $2.779  $ & $0.486$ & $5.717  $ & $ 0.000$  & & $ 3.845 $  & $1.010$  & $ 3.806  $  & $0.000$  \\ 
Urb       & $0.604  $ & $0.608$ & $0.994  $ & $ 0.320$  & & $ -0.622$  & $1.161$  & $ -0.536 $  & $0.592$  \\ 
GDP       & $-0.011 $ & $0.006$ & $-1.892 $ & $ 0.059$  & & $ -0.013$  & $0.004$  & $ -2.906 $  & $0.004$  \\ 
\cline{2-5} \cline{7-10}
& \multicolumn{4}{c}{LMDPDE ($\alpha  = 0.06$)}    & &  &   &    & \\
\cline{2-5} 
& Estimate & Std. error & $z$-stat & $p$-value & &   &   &   &   \\
\cline{2-5} 
\textit{mean submodel}&&&&&&&&&\\
Intercept & $-5.886$ & $0.555$ & $-10.614$ &$ 0.000$  & &    &   &    &    \\ 
Urb       & $4.755  $& $0.626$ & $7.594   $& $0.000 $ & &    &   &    &    \\ 
GDP       & $0.013  $& $0.003$ & $4.724   $& $0.000 $ & &    &   &    &    \\ 
\textit{precision submodel} &&&&&&&&&\\
Intercept & $3.839 $ & $1.011$ & $3.797 $ & $0.000$  & &    &   &    &    \\ 
Urb       & $-0.639$ & $1.162$ & $-0.550$ & $0.582$  & &    &   &    &    \\ 
GDP       & $-0.013$ & $0.004$ & $-2.862$ & $0.004$  & &    &   &    &    \\ 
\cline{2-5} \cline{7-10}

\hline
\end{tabular}
\end{table}

\begin{table}[!ht]
\centering
\caption{Estimates, asymptotic standard errors (Std. error), $z$-stat, and asymptotic $p$-values for the full and reduced data for the LMDPDE.}
\scriptsize
\def\arraystretch{1.1}
\begin{tabular}{rrrrrrrrrr}
\hline
& \multicolumn{4}{c}{LMDPDE for the full data}    & & \multicolumn{4}{c}{LMDPDE without observation $\#1$}\\
\cline{2-5} \cline{7-10}
& Estimate & Std. error & $z$-stat & $p$-value & & Estimate & Std. error & $z$-stat & $p$-value \\
\cline{2-5} \cline{7-10}
\textit{mean submodel} &&&&&&&&&\\
Intercept & $-5.973$ & $0.520$ & $-11.479$ & $0.000$ & & $-5.978$ & $0.508$ & $-11.777$ & $0.000$ \\ 
Urb       & $4.865 $ & $0.583$ & $8.345  $ & $0.000$ & & $4.854 $ & $0.569$ & $8.538  $ & $0.000$ \\ 
GDP       & $0.013 $ & $0.003$ & $4.634  $ & $0.000$ & & $0.013 $ & $0.003$ & $4.721  $ & $0.000$ \\ 
\textit{precision submodel} &&&&&&&&&\\
Intercept & $3.312 $ & $0.229$ & $14.476 $ & $0.000$ & & $3.391 $ & $0.229$ & $14.816 $ & $0.000$ \\ 
GDP       & $-0.013$ & $0.004$ & $-2.889 $ & $0.004$ & & $-0.013$ & $0.004$ & $-3.003 $ & $0.003$ \\ 
\hline
\end{tabular}
\end{table}

\newpage
\bibliographystyle{spbasic}
\bibliography{bancoref}